\documentclass[prd,,aps,nofootinbib,amsmath,amssymb]{revtex4}
\usepackage{amssymb,epsfig}
\usepackage{mathrsfs}
\usepackage[usenames,dvipsnames]{color}
\usepackage[bookmarks]{hyperref}

\newcommand{\Natural}{\mathbb{N}}
\newcommand{\Integer}{\mathbb{Z}}
\newcommand{\Real}{\mathbb{R}}
\newcommand{\Complex}{\mathbb{C}}

\newcommand{\diag}{\mbox{diag}}
\newcommand{\id}{\mbox{id}}
\newcommand{\sign}{\mbox{sign}}
\newcommand{\supp}{\mbox{supp}}
\newcommand{\identy}{1 \! \! 1}

\newcommand{\ve}[1]{{\bf{#1}}}

\newcommand{\DD}{\mathbb{D}}
\newcommand{\EE}{\mathbb{E}}
\newcommand{\FF}{\mathbb{F}}
\newcommand{\GG}{\mathbb{G}}
\newcommand{\HH}{\mathbb{H}}
\newcommand{\II}{\mathbb{I}}
\newcommand{\KK}{\mathbb{K}}

\newcommand{\proof}{\noindent {\bf Proof. }}
\newcommand{\proofof}[1]{\noindent {\bf Proof of #1. }}
\newcommand{\qed}{\hfill \fbox{} \vspace{.3cm}}

\newtheorem{definition}{Definition}
\newtheorem{lemma}{Lemma}
\newtheorem{proposition}{Proposition}
\newtheorem{corollary}{Corollary}
\newtheorem{theorem}{Theorem}
\newtheorem{remark}{Remark}

\begin{document}

\title{Phase space mixing of a Vlasov gas in the exterior of a Kerr black hole}

\author{Paola Rioseco}
\affiliation{Departamento de Ingenier\'ia Matem\'atica and Centro de Modelamiento Matem\'atico, Universidad de Chile, Beauchef 851, Santiago, Chile}

\author{Olivier Sarbach}
\affiliation{Instituto de F\'isica y Matem\'aticas,
Universidad Michoacana de San Nicol\'as de Hidalgo,
Edificio C-3, Ciudad Universitaria, 58040 Morelia, Michoac\'an, M\'exico}

\begin{abstract}
We study the dynamics of a collisionless kinetic gas whose particles follow future-directed timelike and spatially bound geodesics in the exterior of a sub-extremal Kerr black hole spacetime. Based on the use of generalized action-angle variables, we analyze the large time asymptotic behavior of macroscopic observables associated with the gas. We show that, as long as the fundamental frequencies of the system satisfy a suitable non-degeneracy condition, these macroscopic observables converge in time to the corresponding observables determined from an averaged distribution function. In particular, this implies that the final state is characterized by a distribution function which is invariant with respect to the full symmetry group of the system, that is, it is stationary, axisymmetric and Poisson-commutes with the integral of motion associated with the Carter constant.

As a corollary of our result, we demonstrate the validity of the strong Jeans theorem in our setting, stating that the distribution function belonging to a stationary state must be a function which is independent of the generalized angle variables. An analogous theorem in which the assumption of stationarity is replaced with the requirement of invariance with respect to the Carter flow is also proven.

Finally, we prove that the aforementioned non-degeneracy condition holds. This is achieved by providing suitable asymptotic expansions for the energy and Carter constant in terms of action variables for orbits having sufficiently large radii, and by exploiting the analytic dependency of the fundamental frequencies on the integrals of motion.
\end{abstract}

\date{\today}

\maketitle

\tableofcontents

\section{Introduction}

In recent years there has been much interest on the properties of solutions of the Einstein-Vlasov system (see~\cite{hA11} for a review). Even neglecting its self-gravity, studying the properties of a collisionless gas on a fixed (but curved) background leads to interesting problems and phenomena. In particular, it has been argued that, as a consequence of the  integrability properties of the underlying geodesic flow~\cite{bC68,mWrP70}, the Vlasov equation on a Kerr background admits an explicit solution representation in terms of generalized action-angle variables~\cite{oStZ14b}. Based on this representation, it has been possible to study accretion problems which constitute the kinetic analogue of the popular hydrodynamic Bondi-Michel~\cite{hB52,fM72} and Bondi-Hoyle-Lyttleton~\cite{fHrL39,hBfH44} models by analytic methods~\cite{pRoS16,pRoS17,pMoA21a,pMoA21b,pMaO22,aGetal21}. These accretion models are based on a careful analysis of the relativistic phase space corresponding to future-directed timelike geodesics of the Schwarzschild spacetime which are unbounded in the spatial directions. By imposing suitable boundary conditions at spatial infinity (describing, typically, a state in thermodynamic equilibrium) one can show that under reasonable assumptions on the initial data, the gas configuration settles down to a steady-state configuration whose one-particle distribution function (DF) depends only on the integrals of motion. This is not surprising, since a gas particle following an unbound orbit either falls into the black hole in finite proper time or disperses to infinity, implying that a stationary observer perceives a steady-state configuration after long enough time. At the mathematical level, this translates into the fact that the DF describing this scenario converges pointwise in time to the DF specified at spatial infinity~\cite{pRoS16}. For recent related work analyzing the steady-state accretion of a Vlasov gas in the equatorial plane of a Kerr black hole, see~\cite{aCpMaO22}.

The complementary region of phase space, describing spatially bound future-directed timelike geodesics in a Schwarzschild or Kerr black hole exterior spacetime, leads to (non-accreting) gas clouds surrounding the black hole. As long as the gas is collisionless, it is sufficient to focus ones attention to the free-particle flow in the bound region of phase space to understand such configurations. Unlike the unbounded case, a DF whose support lies inside the bound region does not converge pointwise since the geodesic motion is quasi-periodic in this region. Interestingly however, the macroscopic observables associated with the gas, which we define by ``smearing out" the DF by multiplying it with a suitable test function and integrating over phase space, may nevertheless converge as time goes to infinity, due to phase space mixing~\cite{jLoP73,CornfeldFominSinai-Book}. This effect, which plays a fundamental role in diverse fields of physics, including stellar dynamics (see e.g.~\cite{dL62,dL67,sTmHdL86,dM99,sT99}) nonlinear Landau damping in plasma physics~\cite{cMcV11,bY16} and quantum physics~\cite{rMeT17,tDaKeKyS02,tDaKnRyS02}, is based on the fact that orbits belonging to neighboring invariant tori have, in general, slightly different fundamental frequencies associated with them, implying that the flow spreads in the angle directions. As a consequence, from the point of view of the macroscopic observables, the DF can be replaced with the angle-averaged DF at large times; in other words, the DF converges weakly to its angle-average~\cite{cM19,pRoS20}. Recently, based on the explicit solution representation of the DF in terms of action-angle variables, we have analyzed this mixing phenomena for the restricted case in which the gas particles were confined to equatorial orbits in a Kerr black hole exterior, and we provided evidence that the macroscopic observables do relax in time to a stationary and axisymmetric state~\cite{pRoS18}. In the present article, we extend the results of our previous work to the case in which the gas particles are free to follow any spatially bound, future-directed timelike geodesic trajectory in the exterior of a Kerr black hole, and we prove that as long as the black hole is rotating, phase space mixing also occurs in this more general case.

In the next subsection we provide an overview of the different steps and intermediate results (many of them having an interest on their own) involved in establishing our result. We emphasize that we restrict ourselves to a kinetic gas consisting of massive particles; the behavior of a massless Vlasov gas propagating on a fixed black hole background has been studied in Refs.~\cite{lApBjS18,lB20} and leads to decay. See also~\cite{hA21} for self-gravitating static solutions describing a Schwarzschild black hole surrounded by a finite shell of massless Vlasov matter.

\subsection{Organization of the article and description of the main results}

In section~\ref{Sec:KerrGeodesics} we start with a compilation of relevant results (most of them which are known) regarding the Hamiltonian flow associated with the geodesic motion in the Kerr spacetime. For definiteness, we restrict our attention to the sub-extremal case in which the rotation parameter's magnitude is strictly smaller than the black hole mass. For generality and in view of potential future applications to the accretion problem, we describe these results in terms of a coordinate chart based on Kerr coordinates, which covers the future event horizon and a part of the black hole interior in addition to the exterior region. In particular, we list the four integrals of motion, show that they satisfy (almost everywhere) the hypotheses for an integrable system and characterize the region $\Gamma_{bound}$ of phase space corresponding to spatially bound future-directed timelike geodesics which are confined to the exterior region.\footnote{A similar characterization has recently been worked out in Ref.~\cite{fJ22}.} This region $\Gamma_{bound}$ is naturally foliated by the invariant subsets on which all integrals of motion are constant, and we show that (after subtracting a zero measure set from $\Gamma_{bound}$), these subsets are smooth manifolds with topology $\Real\times T^3$. Here, the presence of the non-compact factor $\Real$ is a consequence of the fact that we work on (the fully covariant) relativistic phase space. 

Next, in section~\ref{Sec:ActionAngle} we focus our attention on $\Gamma_{bound}$ and introduce generalized action-angle variables $(J_\alpha,Q^\alpha)$. Although such variables have been introduced and used previously in the literature, see e.g.~\cite{wS02,tHeF08,rFwH09}, we are not aware of rigorous results regarding their global properties on $\Gamma_{bound}$. Two difficulties arise; the first one is related to the fact that the invariant sets are not compact, such that one cannot immediately apply the Liouville-Arnold theorem to construct action-angle variables. However, by showing that the Hamiltonian vector fields associated with the integrals of motion are complete on each invariant set, one can generalize the construction of the Liouville-Arnold theorem~\cite{eFgGgS03} and construct $(J_\alpha,Q^\alpha)$ in an open neighborhood of any invariant set. Whereas the action variables $J_1$, $J_2$, $J_3$ are topological invariants associated with each invariant set, there is an ambiguity in the choice for $J_0$, originating from the presence of its non-compact factor. In this article, we choose $J_0$ equal to the integral of motion corresponding to the rest mass of the particle, and we explain the advantages of our choice. By construction, on each invariant set the variables $J_\alpha$ are constant while the variables $Q^\alpha$ are globally well-defined coordinates on this set, $Q^1$, $Q^2$, and $Q^3$ providing angles parametrizing the torus $T^3$ and $Q^0$ parametrizing the non-compact factor $\Real$. The second difficulty is related to the question of whether or not the action variables $J_\alpha$ uniquely characterize each invariant set; that is, whether the variables $(J_\alpha,Q^\alpha)$ provide a global chart of $\Gamma_{bound}$. Although this can be established in some limiting cases, like the Schwarzschild limit or when restricting to orbits with sufficiently large radii, we do not address this question in the present article. To circumvent this problem, we replace the $J$-variables with the variables $P_\alpha$ describing the constants of motion, and work with the globally-defined (but non-canonical) variables $(P_\alpha,Q^\alpha)$ instead of $(J_\alpha,Q^\alpha)$. The main results of section~\ref{Sec:ActionAngle} are summarized in Theorems~\ref{Thm:Global1} and~\ref{Thm:Global2}. Finally, in this section we also express the variables $(J_\alpha,Q^\alpha)$ explicitly in terms of Legendre's elliptic integrals, we compute the fundamental frequencies $\omega^\alpha$ characterizing the motion and discuss their meaning, and we show that the free-particle (Liouville) flow on $\Gamma_{bound}$ is trivialized.

Section~\ref{Sec:Mixing} is devoted to the main results of this article and states the strong Jeans and mixing theorems. To this purpose, we start with a reduction to a problem on a six-dimensional phase space foliated by invariant tori $T^3$, for which the motion is characterized by the standard winding around the tori with associated frequencies $\boldsymbol{\omega} = (\omega^1,\omega^2,\omega^3)$. This reduction is performed by fixing the particles' rest mass, by considering constant Boyer-Lindquist time hypersurfaces and by identifying points lying on the same Killing orbit generated by the asymptotically timelike unit Killing vector field.\footnote{An alternative approach which does not rely on the introduction of a specific foliation has been discussed in~\cite{pRoS18}.} The reduced problem allows one to easily prove that the Cauchy problem associated with the Vlasov equation on $\Gamma_{bound}$ is well-posed and that its propagator is described by a strongly continuous unitary group. Next, we introduce the \emph{$A$-nondegeneracy condition} which essentially states that the Jacobian of $\boldsymbol{\omega}$ is almost everywhere invertible and show that this implies that the frequencies are non-resonant for almost all values of the constants of motion.\footnote{Resonant orbits and their implications for perturbations of Kerr black holes have been analyzed in Refs.~\cite{jBmGtH13,jBmGtH15}.} Based on these observations, we first formulate the strong Jeans theorem for our setting (Theorem~\ref{Thm:StrongJeans}), stating that a stationary DF is independent of all angle variables if the $A$-nondegeneracy condition holds. A similar theorem (Theorem~\ref{Thm:StrongJeansBis}) is proven stating that under an analogous \emph{$B$-nondegeneracy condition}, invariance of the DF under the Carter flow implies that it must be independent of the angle variables. The mixing theorem is stated in Theorem~\ref{Thm:Mixing}, and we formulate it for any dual pair $(F,G)$, where $F$ denotes the initial distribution function and $G$ the test function. Our theorem only requieres very mild regularity conditions on $F$ and $G$; in particular it works for any $F\in L^p$, $1 < p < \infty$, and $G$ lying in the dual space $L^q$ with $1/p + 1/q = 1$.

Next, in section~\ref{Sec:DeterminantConditions} we provide a more detailed analysis in the \emph{Keplerian limit} of orbits lying far from the black hole. To this purpose we first show that  the relevant quantities are analytic in the parameter $\mu$, which represents the inverse square root of the \emph{semi-latus rectum} of the orbit, in a vicinity of $\mu=0$. Next, we show that for small enough $|\mu|$, the energy, Carter constant and the action variables can be expressed in the form of convergent power series in $\mu$. Further, we prove that for small enough $|\mu|$, the mapping from the constants of motion to the action variables can be inverted, and using this result, we express the energy and Carter constant as a power series of the action variables, up to the desired accuracy in $\mu$. We perform this expansion including terms of the order of $\mu^5$ in the energy and terms of the order of $\mu^2$ in the Carter constant. This allows one to compute the fundamental frequencies $\boldsymbol{\omega}$ including terms of the order $\mu^6$ and to prove that both the $A$- and $B$-degeneracy conditions are satisfied for small enough values of $\mu > 0$ (see Theorem~\ref{Thm:Determinant}), provided the rotation parameter of the black hole is nonzero. This shows that mixing is taking place in the far exterior region of a rotating Kerr black hole. The validity of the non-degeneracy conditions for generic bound orbits then follows from the analytic dependency of the fundamental frequency on the constants of motion.

Finally, in section~\ref{Sec:Conclusions} we provide a summary and discussion of our results. Several technical points, such as a thorough analysis of the effective potentials describing the motion in the polar and radial directions, the parametrization of orbits and the explicit evaluation of the generalized action-angle variables $(J_\alpha,Q^\alpha)$ are discussed in appendices~\ref{App:Polar}--\ref{App:EllipticIntegrals}.

\subsection{Notation and conventions}

Throughout this article, we use the same notation as in Ref.~\cite{pRoS16}. In particular, $T^* M$ denotes the cotangent bundle associated with the spacetime manifold $(M,g)$ (which is assumed to be a smooth and time-oriented Lorentz manifold), the relativistic one-particle phase space corresponding to a simple gas of massive particles is the subspace
\begin{equation}
\Gamma := \{ (x,p) \in T^* M : \hbox{$p^\sharp$ is future-directed timelike} \},
\end{equation}
where $p^\sharp := g_x^{-1}(p,\cdot)$ denotes the vector associated with the momentum covector $p\in T_x^* M$ at $x\in M$, and the Liouville equation, which describes the evolution of a collisionless DF $f: \Gamma\to \Real$ is
\begin{equation}
X_H [f] := g^{\mu\nu}(x) p_\nu\frac{\partial f}{\partial x^\mu}
 - \frac{1}{2}p_\alpha p_\beta\frac{\partial g^{\alpha\beta}(x)}{\partial x^\mu}
 \frac{\partial f}{\partial p_\mu} = 0,
\label{Eq:Liouville}
\end{equation}
in adapted local coordinates $(x^\mu,p_\mu)$ on $T^* M$. Here, $X_H$ refers to the Liouville vector field which can be defined invariantly as the Hamiltonian vector field associated with the free-particle Hamiltonian $H(x,p) := \frac{1}{2} g_x^{-1}(p,p)$ and the natural symplectic form $\Omega_s = dp_\mu\wedge dx^\mu$ on $T^* M$. Recall that for an arbitrary smooth function $F: T^*M\to \Real$, the associated Hamiltonian vector field $X_F$  is the unique vector field on $T^* M$ such that $dF = \Omega_s(\cdot,X_F)$, and is given explicitly by
\begin{equation}
X_F = \frac{\partial F}{\partial p_\mu}\frac{\partial}{\partial x^\mu}
 - \frac{\partial F}{\partial x^\mu}\frac{\partial}{\partial p_\mu}.
\end{equation}
The Poisson-bracket between two smooth functions $F,G: T^*M\to \Real$ is defined as
\begin{equation}
\{ F, G \} := \Omega_s(X_F,X_G) 
 = \frac{\partial F}{\partial p_\mu}\frac{\partial G}{\partial x^\mu}
 - \frac{\partial F}{\partial x^\mu}\frac{\partial G}{\partial p_\mu}
 = dG(X_F).
\end{equation}
Note that the symplectic form can be expressed as the exterior differential of the Poincar\'e one-form
\begin{equation}
\Theta = p_\mu dx^\mu,
\label{Eq:PoincareForm}
\end{equation}
see, for instance, Eq. (2) in Ref.~\cite{pRoS16} for a coordinate-invariant definition of $\Theta$.

Finally, recall that $\Gamma$ is equipped with the volume form
\begin{equation}
\eta_\Gamma := -\frac{1}{4!}\Omega_s\wedge\Omega_s\wedge\Omega_s\wedge\Omega_s 
 = -dx^0\wedge dx^1\wedge dx^2\wedge dx^3\wedge dp_0\wedge dp_1\wedge dp_2\wedge dp_3,
\end{equation}
which is invariant with respect to the Liouville flow: $\pounds_{X_H}\eta_\Gamma = 0$. For a recent review on the geometric structures of the cotangent bundle which are relevant for relativistic kinetic theory, see~\cite{rAcGoS22}.

\section{Geodesic motion on a Kerr black hole as an integrable Hamiltonian system}
\label{Sec:KerrGeodesics}

\begin{figure}[ht]
\centerline{\resizebox{9.5cm}{!}{\includegraphics{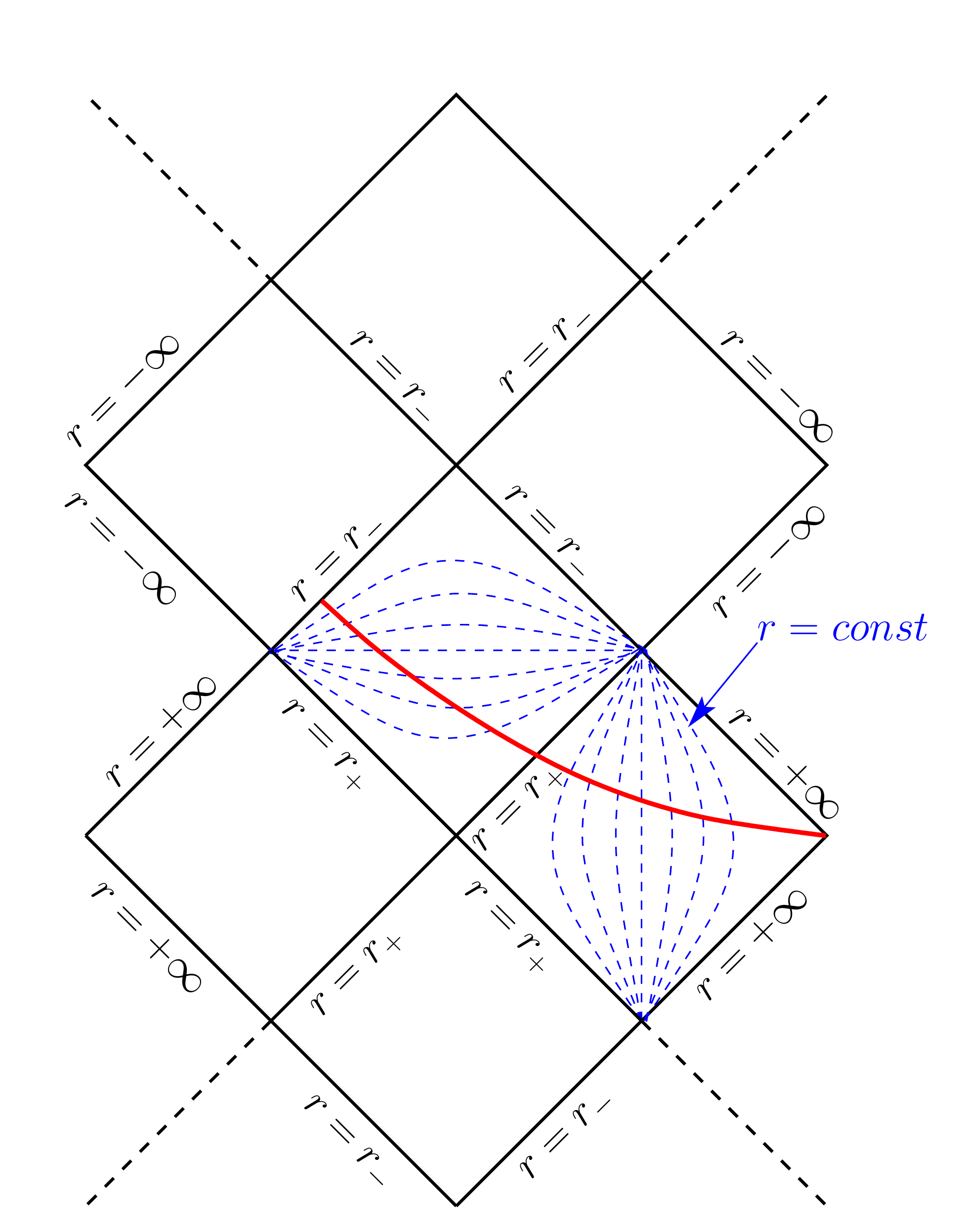}}}
\caption{Penrose diagram of the symmetry axis of the Kerr spacetime. The spacetime $(M,g)$ considered in this article corresponds to the two causal diamonds showing the dashed blue $r=const$ lines. The continuous red line illustrates the qualitative behavior of a $t=const$ hypersurface in the region $r > r_-$.}
\label{Fig:KerrPenroseDiagram}
\end{figure}

This section is devoted to a review of some of the most important results regarding the geodesic motion on a Kerr spacetime which are relevant for this article. We consider the spacetime $(M,g)$ describing a Kerr black hole of mass $M_H$ and rotation parameter $a_H$ satisfying $a_H^2 < M_H^2$. Using horizon-penetrating coordinates $(t,r,\vartheta,\varphi)$ which are related to the standard Kerr coordinates $(\tilde{V},r,\theta,\tilde{\phi})$ (see, for instance~\cite{MTW-Book}) through the relations $\tilde{V} = t + r$, $\theta = \vartheta$, $\tilde{\phi} = \varphi$, the Kerr metric is given by
\begin{eqnarray}
g &=& -dt^2 +dr^2 -2a_H\sin^2\vartheta dr d\varphi 
 + \left( r^2 + a_H^2 \right)\sin^2 \vartheta d\varphi ^2
 + \rho^2 d\vartheta ^2 
 + \frac{2M_H r}{\rho^2}  \left(dt + dr - a_H\sin ^2 \vartheta d\varphi \right)^2,
\label{Eq:KerrEDF}
\end{eqnarray}
with the function $\rho^2 = \rho^2(r,\vartheta) := r^2 + a_H^2\cos^2 \vartheta$. The representation of the metric in these coordinates is regular for all $r > 0$ and $0 < \vartheta < \pi$,\footnote{The coordinate singularities at the poles $\vartheta = 0,\pi$ could be removed by introducing Kerr-Schild coordinates $(x,y,z)$ defined by $x + i y = (r + i a_H)\sin\vartheta e^{i\varphi}$, $z = r\cos\vartheta$, see Ref.~\cite{MTW-Book}. In this article, we shall not consider polar orbits, i.e. bound orbits which cross the symmetry axis, such that we do not need to introduce Kerr-Schild coordinates.} in particular it is regular at the inner and outer horizons $r_\pm := M_H \pm \sqrt{M_H^2 - a_H^2}$. For the following, we restrict ourselves to the region of the (maximally extended) Kerr spacetime corresponding to $r > r_-$ which contains the future event horizon $r = r_+$, see Fig.~\ref{Fig:KerrPenroseDiagram}. For brevity, we shall denote this spacetime region by $(M,g)$. 

Note that the $t = const.$ hypersurfaces have the normal vector field
\begin{equation}
N := -g^{\mu\nu}(\nabla_\mu t)\frac{\partial}{\partial x^\nu}
 = \left( 1 + \frac{2M_H r}{\rho^2} \right)\frac{\partial}{\partial t}
 - \frac{2M_H r}{\rho^2}\frac{\partial}{\partial r},
\end{equation}
which satisfies $g(N,N) = -(1 + 2M r_H/\rho^2) < 0$; hence $N$ is timelike and the hypersurfaces $t = const.$ are spacelike, and $N$ provides a time orientation on $(M,g)$. Instead of $N$, it will result more convenient to use the vector fields\footnote{With respect to Boyer-Lindquist coordinates the vector field $Y$ is just $\Delta\frac{\partial}{\partial r}$.}
\begin{equation}
X := (r^2 + a_H^2)\frac{\partial}{\partial t} + a_H\frac{\partial}{\partial\varphi}\quad
\hbox{and}\quad
Y := 2M_Hr\frac{\partial}{\partial t} + a_H\frac{\partial}{\partial \varphi} 
 + \Delta\frac{\partial}{\partial r}
\label{Eq:XYDef}
\end{equation}
with the function $\Delta = \Delta(r) := r^2 -2M_H r + a_H^2$. They satisfy $-g(X,X) = g(Y,Y) = \rho^2\Delta$, $g(X,Y) = 0$ and $dt(X) > 0$, $dt(Y) > 0$; hence $X$ is future-directed timelike in the region $r > r_+$ and $Y$ is future-directed timelike in the region $r_- < r < r_+$.

The free-particle Hamiltonian for the metric~(\ref{Eq:KerrEDF}) is
\begin{eqnarray}
H(x,p) &=& \frac{1}{2\rho^2}\left[ 
 -\left( \rho^2 + 2M_H r \right)p_t^2 
 + 4M_H r p_t p_r  + 2a_H p_r p_\varphi + \Delta p_r^2 + p_\vartheta^2 
 + \frac{p_\varphi^2}{\sin^2\vartheta} \right].
\label{Eq:HKerr}
\end{eqnarray}
Since the underlying spacetime is stationary and axisymmetric, the following quantities are conserved along the particle trajectories:
\begin{eqnarray}
m = \sqrt{-2H} && \hbox{(rest mass)},\\
E = -p_t && \hbox{(energy)},\\
L_z = p_\varphi && \hbox{(azimuthal angular momentum)}.
\end{eqnarray}
Remarkably, there exists a fourth conserved quantity, discovered by Carter by separating the Hamilton-Jacobi equation~\cite{bC68}, which is defined by\footnote{In Kerr-Schild coordinates one has $L_z = x p_y - y p_x$ and
$$
{\cal K} = (r^2 + a_H^2)( p_x^2 + p_y^2 + p_z^2) - (x p_x + y p_y + z p_z)^2 - 2a_H E L_z
 + \frac{a_H^2}{r^2}\left[ (r^2-z^2) E^2 - r^2 p_z^2 + m^2 z^2 \right],
$$
which shows that ${\cal K}$ is everywhere regular for $r > 0$ including at the axis $\vartheta = 0,\pi$ where $x=y=0$.
}
\begin{equation}
{\cal K} := p_\vartheta^2 + \left(\frac{L_z}{\sin\vartheta} - E a_H \sin\vartheta \right)^2 
+ m^2 a_H^2\cos^2\vartheta\qquad \hbox{(Carter constant)}.
\end{equation}
Note that this constant is manifestly positive, and moreover, in the Schwarzschild limit $a_H = 0$ it reduces to the square of the total angular momentum $L^2$. For these reasons, we shall use the notation $L^2$ instead of ${\cal K}$ even in the rotating case.

For the following, we introduce the smooth functions $F_0,F_1,F_2,F_3: T^*M\to \Real$ on the cotangent bundle which define these integrals of motion:
\begin{eqnarray}
F_0(x,p) &:=& -H(x,p),\quad
F_1(x,p) := -p_t,\quad
F_2(x,p) := p_\varphi,
\nonumber\\
F_3(x,p) &:=& p_\vartheta^2 
 + \left( \frac{p_\varphi}{\sin\vartheta} + a_H\sin \vartheta p_t \right)^2
 - 2a_H^2\cos^2\vartheta H(x,p).
\label{Eq:FalphaDef}
\end{eqnarray}
The presence of these four functions imply that the geodesic motion in the Kerr spacetime yields an integrable Hamiltonian system. The precise formulation of this statement is contained in the following two proposition whose proof will be given below.

\begin{proposition}
\label{Prop:Commutators}
The functions $F_\alpha$ defined in Eq.~(\ref{Eq:FalphaDef}) Poisson-commute with each other:
$$
\{ F_\alpha, F_\beta \} = 0, \qquad \alpha,\beta=0,1,2,3.
$$
\end{proposition}

\begin{proposition}
\label{Prop:LI}
The differentials $dF_0,dF_1,dF_2,dF_3$ are linearly independent from each other at each $x\in \Gamma_0$, where $\Gamma_0\subset\Gamma$ is a dense subset of the phase space $\Gamma$ which is invariant with respect to the flows generated by the Hamiltonian vector fields $X_\alpha := X_{F_\alpha}$, $\alpha = 0,1,2,3$.
\end{proposition}

Next, one considers for each given value of $(m,E,L_z,L)$ the (possibly empty) subset
\begin{equation}
\Gamma_{m,E,L_z,L} 
 := \{ (x,p)\in \Gamma : F_0(x,p) = m^2/2, F_1(x,p) = E, F_2(x,p) = L_z, F_3(x,p) = L ^2 \}
\label{Eq:InvariantSets}
\end{equation}
of the one-particle phase space $\Gamma$, which by definition is invariant with respect to the Hamiltonian flows associated with $F_0$, $F_1$, $F_2$ and $F_3$. As a corollary of Proposition~\ref{Prop:LI} one has

\begin{corollary}
\label{Cor:InvSets}
Suppose $\Gamma_{m,E,L_z,L} \subset \Gamma_0$ is contained in the set $\Gamma_0$ appearing in the statement of the previous proposition. Then, $\Gamma_{m,E,L_z,L}$ is a four-dimensional smooth submanifold of $\Gamma$ whose tangent spaces are spanned by the vectors $\left. X_\alpha \right|_{(x,p)}$, $\alpha=0,1,2,3$, at each point $(x,p)\in \Gamma_{m,E,L_z,L}$. Furthermore, the restriction of the Poincar\'e one-form $\Theta$ to $\Gamma_{m,E,L_z,L}$ is closed.
\end{corollary}

\proof Since $\Omega_s(X_\alpha,X_\beta) = \{ F_\alpha, F_\beta \} = 0$ for all $\alpha,\beta = 0,1,2,3$ and the vectors $\left. X_\alpha \right|_{(x,p)}$ span the tangent spaces of $\Gamma_{m,E,L_z,L}$, it follows that the restriction of $\Omega_s = d\Theta$ on $\Gamma_{m,E,L_z,L}$ vanishes.
\qed

\begin{remark}
The Hamiltonian vector fields $X_\alpha = X_{F_\alpha}$ have the following interpretation: $-X_0$ is the Liouville vector field, $-X_1$ is the complete lift\footnote{See, for instance, Ref.~\cite{pRoS16} for a definition and a summary on the properties of the complete lift.} of the asymptotically timelike Killing vector field $\partial_t$ and $X_2$ the complete lift of the Killing vector field $\partial_\varphi$. The vector field $X_3$ generates the Carter flow, that is, the symmetry flow associated with the Carter constant, and it cannot be written as the complete lift of any spacetime vector field.
\end{remark}

For the following, we focus our attention on spatially bound orbits. This leads to the restriction of phase space $\Gamma$ on which the invariant submanifolds $\Gamma_{m,E,L_z,L}$ have topology $\Real \times T^3$. The next proposition characterizes the range for the parameters this subset corresponds to.

\begin{proposition}
\label{Prop:InvariantSets}
Let $\alpha := a_H/M_H\in [0,1)$ and denote for each $\beta\in (-1,1)$ by $L_{ms}(\alpha,\beta)$, $E_{min}(\alpha,\beta,L)$ and $E_{max}(\alpha,\beta,L)$ the quantities (whose precise form is unimportant for the moment) corresponding to the dimensionless quantities $\lambda_{ms}$, $\varepsilon_{min}$, $\varepsilon_{max}$ defined in Lemma~\ref{Lem:EffectivePotential} in appendix~\ref{App:Radial}. Finally, denote by $\Omega$ the open set of four-tuples $(m,E,L_z,L)$ satisfying
\begin{equation}
m > 0,\quad
L > L_{ms}(\alpha,\beta),\quad
E_{min}(\alpha,\beta,L) < E < \min\{ m, E_{max}(\alpha,\beta,L) \},\quad
L_z = \beta L + a_H E\neq 0,
\label{Eq:LELzConditions}
\end{equation}
for some $\beta\in (-1,1)$.

Then, $\Gamma_{m,E,L_z, L}$ has a unique connected component which lies entirely in the exterior region $r > r_+$. This component, which we denote by $\Gamma^{(ext)}_{m,E,L_z, L}$ in the following, has topology $\Real \times T^3$.
\end{proposition}

The remaining part of this section is devoted to the proofs of Propositions~\ref{Prop:Commutators}, \ref{Prop:LI} and \ref{Prop:InvariantSets}. Some of the intermediate results in these proofs will be used in the next section as well.

\subsection{Proof of Proposition~\ref{Prop:Commutators}}

The only commutator whose vanishing is not immediately evident is $\{ F_0, F_3 \}$. To compute it, we rewrite $F_3$ in the form
\begin{equation}
F_3 = p_\vartheta^2 + Q^2 - 2a_H^2\cos^2\vartheta H,\qquad
Q := \frac{p_\varphi}{\sin\vartheta} + a_H\sin \vartheta p_t.
\label{Eq:F3Q}
\end{equation}
Using the identity $\{ H, AB \} = \{ H, A \} B + A\{ H,B \}$ we obtain first
\begin{equation}
\{ F_0, F_3 \} = -2\{ H, p_\vartheta \} p_\vartheta - 2\{ H, Q \} Q 
 + 2a_H^2\{ H,\cos^2\vartheta \} H.
\end{equation}
Next, an explicit computation reveals that
\begin{eqnarray}
\{ H, p_\vartheta \} &=& -\frac{\partial H}{\partial\vartheta} 
 = \frac{\cot\vartheta}{\rho^2}\left[ \frac{p_\varphi^2}{\sin^2\vartheta}
  - a_H^2\sin^2\vartheta(2H + p_t^2) \right],\\
\{H, Q \} &=& \frac{\partial H}{\partial p_\vartheta}\frac{\partial Q}{\partial\vartheta}
 = -\cot\vartheta\frac{p_\vartheta}{\rho^2}
\left[ \frac{p_\varphi}{\sin\vartheta} - a_H\sin\vartheta p_t \right],\\
\{ H,\cos^2\vartheta \} &=& 
 \frac{\partial H}{\partial p_\vartheta}\frac{\partial \cos^2\vartheta}{\partial\vartheta}
 = -2\cos\vartheta\sin\vartheta\frac{p_\vartheta}{\rho^2},
\end{eqnarray}
from which one concludes easily that $\{ F_0, F_3 \} = 0$.
\qed

\subsection{Proof of Proposition~\ref{Prop:LI}}

The proof of Proposition~\ref{Prop:LI} and the determination of the dense invariant subset $\Gamma_0$ is more involved than the proof of the previous proposition. In order to proceed, we consider for each given value of $(m,E,L_z,L)$ the (possibly empty) subset $\Gamma_{m,E,L_z, L}\subset \Gamma$ defined in Eq.~(\ref{Eq:InvariantSets}), which by definition is invariant with respect to the Hamiltonian flows generated by $F_0$, $F_1$, $F_2$ and $F_3$. Clearly, each point $(x,p)\in \Gamma$ is contained in precisely one of these invariant sets. Furthermore, $(x,p)\in  \Gamma_{m,E,L_z,L}$ if and only if $p$ is future-directed and if the conjugate pairs $(t,p_t)$, $(\varphi, p_\varphi)$, $(\vartheta,p_\vartheta)$, $(r,p_r)$ fulfill the following restrictions:
\begin{eqnarray}
(t,p_t) &:& p_t = -E,
\label{Eq:tPlaneRestriction}\\
(\varphi,p_\varphi) &:& p_\varphi = L_z,
\label{Eq:phiPlaneRestriction}\\
(\vartheta,p_\vartheta) &:& p_\vartheta^2 + K(\vartheta) = L ^2, \qquad 
K(\vartheta) := \left(\frac{L_z}{\sin\vartheta} - a_H\sin \vartheta E \right) ^2 
 + a_H^2 m^2 \cos^2 \vartheta,
\label{Eq:thetaPlaneRestriction}\\
(r,p_r) &:& \Delta p_r^2 - 2(2M_H E r - a_H L_z) p_r - (r^2 + a_H^2 + 2M_H r) E^2 
 + 2a_H E L_z + L ^2 + m^2 r^2 = 0,
\label{Eq:rPlaneRestriction}
\end{eqnarray}
where the last of these restrictions is an immediate consequence of the following identity:
\begin{equation}
2r^2 H(x,p) = \Delta p_r^2  + 2(a_H p_\varphi + 2M_H r p_t) p_r
 -(r^2 + a_H^2 + 2M_H r) p_t^2 - 2a_H p_t p_\varphi + F_3(x,p).
\label{Eq:HamSeparability}
\end{equation}

As long as $\Delta\neq 0$ Eq.~(\ref{Eq:rPlaneRestriction}) is equivalent to
\begin{equation}
\left( \Delta p_r - 2M_H E r + a_H L_z \right)^2 = R(r),\qquad
R(r) := \left[ E(r^2 + a_H^2 ) - a_H L _z  \right]^2 - \Delta\left(L^2 + m^2 r^2 \right).
\label{Eq:rPlaneRestrictionBis}
\end{equation}
Note that Hamilton's equations of motion imply that
\begin{equation}
\dot{r} = \frac{\partial H}{\partial p_r} 
 = \frac{1}{\rho^2}\left( \Delta p_r - 2M_H E r + a_H L_z \right),
\end{equation}
and hence the expression inside the parenthesis on the left-hand side of Eq.~(\ref{Eq:rPlaneRestrictionBis}) is just $\rho^2$ times the radial velocity $\dot{r}$. When $\Delta = 0$, Eq.~(\ref{Eq:rPlaneRestriction}) reduces to
\begin{equation}
2(2M_H E r - a_H L_z)(p_r + E) = L^2 + m^2 r^2.
\label{Eq:rPlaneRestrictionBisHor}
\end{equation}

The next lemma characterizes the set of points in phase space $\Gamma$ for which the differentials $dF_0$, $dF_1$, $dF_2$ and $dF_3$ fail to be linearly independent from each other:

\begin{lemma}
\label{Lem:LI}
Let $(x,p)\in \Gamma_{m,E,L_z,L}$ be such that $0 < \vartheta < \pi$. Then, $dF_0$, $dF_1$, $dF_2$, $dF_3$ are linearly independent unless one (or both) of the following cases occur:
\begin{enumerate}
\item[(a)] $p_\vartheta = 0$ and $K'(\vartheta) = 0$,
\item[(b)] $r > r_+$ and $R(r) = R'(r) = 0$.
\end{enumerate}
\end{lemma}

\begin{remark}
Case (a) corresponds to particle trajectories which are confined to the equatorial plane or to certain cones of constant $\vartheta$ (see appendix~\ref{App:Polar}), while case (b) corresponds to spherical trajectories.
\end{remark}

\proofof{Lemma~\ref{Lem:LI}} We note first that $dF_1 = -dp_t\neq 0$ and $dF_2 = dp_\varphi\neq 0$ are linearly independent from each other. Next, we consider instead of $dF_0 = -dH$ and $dF_3$ the one-forms
\begin{equation}
\Lambda_0 := 2r^2 dH - dF_3,\qquad
\Lambda_3 := dF_3 + 2a_H^2\cos^2\vartheta dH.
\end{equation}
The linear transformation that maps $(dH,dF_3)$ to $(\Lambda_0,\Lambda_3)$ has determinant equal to $2\rho^2 > 0$, and thus it is invertible. Consequently, the statement of the Lemma is equivalent to the verification that the one-forms $dF_1,dF_2,\Lambda_0,\Lambda_3$ are linearly independent from each other.

An explicit calculation taking into account the definition of $Q$ defined in Eq.~(\ref{Eq:F3Q}) reveals that\footnote{Using Kerr-Schild coordinates one finds the following expression for $\Lambda_3$ on the symmetry axis:
$$
\left. \Lambda_3 \right|_{(x,y) = (0,0)} = 2z\left[ -p_z(p_x dx + p_y dy) + (p_x^2 + p_y^2) dz \right] + 2(r^2 + a_H^2)(p_x dp_x + p_y dp_y) + 2 a_H p_t dp_\varphi,
$$
which shows that $\Lambda_3$ is linearly independent from $dp_t$ and $dp_\varphi$ unless $(p_x,p_y) = (0,0)$, that is, the motion is confined to the axis.
}
\begin{equation}
\Lambda_3 = 2p_\vartheta dp_\vartheta 
 + 2Q\left( \frac{dp_\varphi}{\sin\vartheta} + a_H\sin\vartheta dp_t \right)
 + K'(\vartheta) d\vartheta,
\end{equation}
which shows that $\Lambda_3$ is linearly independent from $dF_1$ and $dF_2$ unless the conditions in case (a) are met. In order to analyze $\Lambda_0$, it is convenient to introduce the quadratic form
\begin{equation}
{\cal Q}(r,p_r) :=  \Delta p_r^2 - 2(2M_H E r - a_H L_z) p_r - (r^2 + a_H^2 + 2M_H r) E^2 
 + 2a_H E L_z + L ^2 + m^2 r^2,
\end{equation}
such that ${\cal Q}(r,p_r) = 0$ if $(x,p)\in \Gamma_{m,E,L_z,L}$, see Eq.~(\ref{Eq:rPlaneRestriction}). Using the identity~(\ref{Eq:HamSeparability}) one finds
\begin{equation}
\Lambda_0 = \frac{\partial{\cal Q}}{\partial p_r}(r,p_r) dp_r
 + \frac{\partial{\cal Q}}{\partial r}(r,p_r) dr
 + \left( \ldots \right) dp_t + \left( \ldots \right) dp_\varphi,
\end{equation}
which shows that $\Lambda_0$ is linearly independent form $dF_1$, $dF_2$ and $\Lambda_3$ unless
\begin{equation}
\frac{\partial{\cal Q}}{\partial p_r}(r,p_r) 
 = 2\left(\Delta p_r - 2M_H E r + a_H L_z \right) = 0\hbox{  and  }
\frac{\partial{\cal Q}}{\partial r}(r,p_r) = 0.
\label{Eq:LI}
\end{equation}
When $\Delta = 0$ the first of these equations and Eq.~(\ref{Eq:rPlaneRestrictionBisHor}) imply that $L^2 + m^2 r^2 = 0$ which is a contradiction since $m > 0$ and $r > r_-$. When $\Delta\neq 0$ the first equation and Eq.~(\ref{Eq:rPlaneRestrictionBis}) imply $R(r) = 0$; however, this situation cannot occur when $r_- < r < r_+$ since in this case $\Delta < 0$ and thus $R(r) > 0$. When $r > r_+$, the quadratic form ${\cal Q}$ can also be written as
\begin{equation}
{\cal Q}(r,p_r) = \frac{1}{\Delta}\left[ \left( \Delta p_r - 2M_H E r + a_H L_z \right)^2 - R(r) \right],
\end{equation}
and differentiating both sides with respect to $r$ shows that when $R(r) = 0$, the second condition in Eq.~(\ref{Eq:LI}) is equivalent to $R'(r) = 0$.
\qed

To conclude the proof of the Proposition, we define $\Gamma_0$ to be the set of points $(x,p)\in\Gamma$ for which $dF_0,dF_1,dF_2,dF_3$ are linearly independent from each other. It follows from Lemma~\ref{Lem:LI} that the complement $\Gamma\setminus\Gamma_0$ is a zero measure set in $(\Gamma,\eta_\Gamma)$ since the sets $\{ p_\vartheta = 0 \}$ and $\{ R(r) = 0 \} = \{ \Delta p_r = -2M_H r p_t + a_H p_\varphi \}$ are already zero measure sets. Finally, it is not difficult to verify that $\Gamma\setminus\Gamma_0$ is invariant with respect to the flows generated by $X_{F_\alpha}$.
\qed

\subsection{Proof of Proposition~\ref{Prop:InvariantSets}}

As discussed previously, $(x,p)\in \Gamma_{m,E, L_z, L}$ if and only if $p$ is future-directed and if the pairs $(t,p_t)$, $(\varphi, p_\varphi)$, $(\vartheta,p_\vartheta)$, $(r,p_r)$ satisfy Eqs.~(\ref{Eq:tPlaneRestriction}--\ref{Eq:rPlaneRestriction}). From these conditions it is clear that the time coordinate $t$ and the azimuthal angle $\varphi$ are free, giving rise to one $\Real\times S^1$ factor.

Next, one uses the qualitative features of the function $K: (0,\pi)\to \Real$ which are summarized in appendix~\ref{App:Polar}. For the present case in which $|L_z - a_H E| < L$ and $0 < E_{ms} < E < m$ it turns out that $K$ decreases monotonously from $\infty$ to $(L_z-a_H E)^2$ and then increases monotonously again to $\infty$ as $\vartheta$ increases from $0$ to $\pi/2$ to $\pi$. Consequently, the projections of the sets $\Gamma_{m,E, L_z, L}$ onto the $(\vartheta,p_\vartheta)$-plane are closed curves which are topologically equivalent to $S^1$.
 
Finally, we analyze the projection of the set $\Gamma_{m,E, L_z, L}$ onto the $(r,p_r)$-plane. When $r\neq r_+$ (that is, $\Delta\neq 0$), this set is described by Eq.~(\ref{Eq:rPlaneRestrictionBis}), where the function $R$ is a fourth order polynomial in the variable $r$, which depends on the six parameters $a_H$, $M_H$, $m$, $L_z$, $L$ and $E$. In the following, we determine the $r$-intervals on which this polynomial is non-negative. To perform this analysis, we first note that for $r_- < r < r_+$ the function $R(r)$ is manifestly positive and hence Eq.~(\ref{Eq:rPlaneRestrictionBis}) has always two solutions given by
\begin{equation}
p_r = p_{r\pm}(r) = \frac{2M_H E r - a_H L_z \pm \sqrt{R(r)}}{\Delta}
 = -\frac{(r^2 + a_H^2 + 2M_H r)E^2 - 2a_H E L_z - (L^2 + m^2 r^2)}{2M_H E r - a_H L_z \mp \sqrt{R(r)}},
\label{Eq:prm}
\end{equation}
where the second representation is useful to understand the limits when $r\to r_\pm$.
The next lemma shows that only the solution corresponding to $p_r = p_{r-}(r)$ yields a future-directed momentum $p$ and is relevant for the purpose of this work.

\begin{lemma}
\label{Lem:Future1}
Let $(x,p)\in \Gamma_{m,E, L_z, L}$ such that $r_- < r < r_+$. Then $p_r = p_{r-}(r)$.
\end{lemma}

\proof Consider the vector field $Y$ defined in Eq.~(\ref{Eq:XYDef}) which is future-directed timelike in the region $\Delta < 0$. Since $(x,p)\in \Gamma_{m,E, L_z, L}$, $p$ must be future-directed and thus
$$
0 < -p(Y) = 2M_H E r - a_H L_z - \Delta p_r = \mp \sqrt{R},
$$
where we have used the fact that $p_r = p_{r\pm}(r)$ in the last step. This proves that only the lower sign is possible as claimed.
\qed
 
\begin{remark}
As $r\to r_\pm$, the relevant function $p_{r-}(r)$ has a well-defined limit provided $2M_H E r_\pm - a_H L_z > 0$.
\end{remark}

Next, we restrict our attention to the region $r > r_+$. For this, it is convenient to replace $ L_z$ with the new parameter
\begin{equation}
\hat{L}_z := L_z - a_H E,
\label{Eq:Lhatz}
\end{equation}
whose square corresponds to the minimal value of the function $K(\vartheta)$ when $a_H^2(E^2-m^2)\leq L_z^2$, see appendix~\ref{App:Polar}. With this change we can rewrite the function $R$  in the form
\begin{equation}
R(r) = r^4[E - W_+(r)] [ E - W_-(r) ],
\label{Eq:RWpmRel}
\end{equation}
where
\begin{equation}
W_\pm(r) := \frac{a_H\hat{L}_z}{r^2} \pm \frac{\sqrt{\Delta\left(m^2 r^2 + L ^2 \right)}}{r^2}.
\label{Eq:Wpm}
\end{equation} 
Since $W_+(r) > W_-(r)$ for $r > r_+$ it follows that $R(r)\geq 0$ if and only if $r$ lies in either one of the disjoint sets
$$
\{ r > r_+ : W_+(r) \leq E \} \hbox{ or } \{ r > r_+ : W_-(r) \geq E \}.
$$
However, as shown in the following Lemma, only the first of these two sets corresponds to  future-directed trajectories and hence belongs to the invariant set $\Gamma_{m,E, L_z, L}$:

\begin{lemma}
\label{Lem:Future2}
Let $(x,p)\in \Gamma_{m,E, L_z, L}$ such that $r > r_+$. Then, $W_+(r)\leq E$.
\end{lemma}

\proof If $(x,p)\in \Gamma_{m,E, L_z, L}$, then $p$ is, by definition, future-directed. On the other hand, the vector field $X$ defined in Eq.~(\ref{Eq:XYDef}) is also timelike future-directed in the region $\Delta > 0$. Therefore, we must have
$$
0 < -p(X) = (r^2 + a_H^2) E - a_H L_z = r^2 E - a_H\hat{L}_z.
$$
However, if $E\leq W_-(r)$ then
$$
r^2 E - a_H\hat{L}_z \leq  -\sqrt{\Delta(m^2 r^2 + L^2)} < 0,
$$
so in this case $p$ is past-directed and we obtain a contradiction. On the other hand, if $E\geq W_+(r)$ then
$$
r^2 E - a_H\hat{L}_z \geq +\sqrt{\Delta(m^2 r^2 + L^2)} > 0,
$$
and $p$ is future-directed.
\qed

\begin{remark}
It follows from Lemmas~\ref{Lem:Future1} and \ref{Lem:Future2} that the projection of $\Gamma_{m,E, L_z, L}$ onto the $(r,p_r)$-plane consists of the points for which
$$
r_- < r < r_+ \hbox{ and } p_r = p_{r-}(r) \hbox{ or }
r\geq r_+ \hbox{ and } W_+(r)\leq E \hbox{ and } p_r = p_{r\pm}(r),
$$
with $p_{r\pm}(r)$ given by Eq.~(\ref{Eq:prm}). In the limit $r\to r_+$ one obtains the condition $E\geq W_+(r_+) = a_H\hat{L}_z/r_+^2$ which implies that $2M_H E r_+ - a_H L_z = (2M_H r_+ - a_H^2)E - a_H\hat{L}_z = r_+^2 E - a_H\hat{L}_z\geq 0$; hence for values of $E$ such that $E > a_H\hat{L}_z/r_+^2$ the curve $p_r = p_{r-}(r)$ intersects the event horizon. This curve corresponds to infalling particles that are absorbed by the black hole.
\end{remark}

\begin{remark}
The properties of the projection of $\Gamma_{m,E, L_z, L}$ onto the $(r,p_r)$-plane in the region $r > r_+$ depend on the shape of the effective potential $W_+$ for a given energy level $E$. A detailed analysis of the qualitative features of the function $W_+$ is given in appendix~\ref{App:Radial}. We summarize the relevant results in the following Lemma.
\end{remark}

\begin{figure}[ht]
\centerline{\resizebox{11.5cm}{!}{\includegraphics{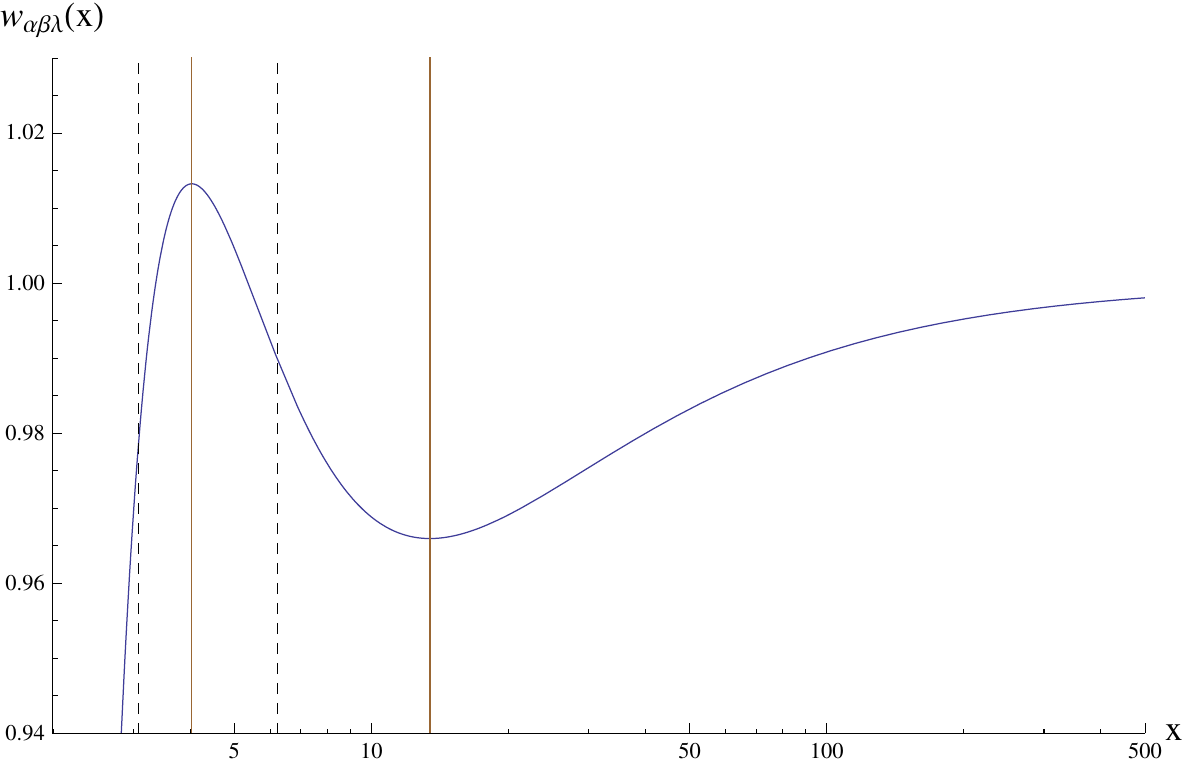}}}
\caption{The effective potential $w_{\alpha,\beta,\lambda}(x) = W_+(r)/m$ as a function of $x = r/M_H$ for the parameter values $\alpha = 0.12$, $\beta = -0.61$ and $\lambda = 4.25$. The vertical solid lines show the locations of the maximum and minimum, $r_{max}/M_H\approx 4.034$ and $r_{min}/M_H\approx 13.45$. The dashed vertical lines indicate the locations of $r_{ph}/M_H\approx 3.077$ and $r_{ms}/M_H\approx 6.223$.}
\label{Fig:KerrPotential}
\end{figure}

\begin{lemma}
Suppose that $|a_H| < M_H$ and $|\hat{L}_z| < L$ and let $L_{ms}:= M_H m\lambda_{ms}(\alpha,\beta)$, $r_{ms} := M_H x_{ms}(\alpha,\beta)$, $r_{ph} := M_H x_{ph}(\alpha,\beta)$ be defined as in appendix~\ref{App:Radial} (in particular, see Lemmas~\ref{Lem:ph} and~\ref{Lem:ms} for the definitions of $x_{ph}$ and $x_{ms}$). Then, the function $W_+ : [r_+,\infty)\to \Real$ defined by Eq.~(\ref{Eq:Wpm}) satisfies the following properties:
\begin{itemize}
\item When $L \leq L_{ms}$ the function $W_+$ is monotonously increasing.
\item For each $L > L_{ms}$ the function $W_+$ has a unique local maximum (the centrifugal barrier) located at $r_{max}\in (r_{ph},r_{ms})$ and it has a potential well with a local minimum located at $r_{min}\in (r_{ms},\infty)$ which is enclosed between the centrifugal barrier and its asymptotic value $m$ (see Fig.~\ref{Fig:KerrPotential}).
\item As $L$ increases from $L_{ms}$ to $\infty$, $r_{max}$ decreases monotonously from $r_{ms}$ to $r_{ph}$ and $W_+(r_{max})$ increases monotonously from $E_{ms}$ to $\infty$ while $r_{min}$ increases monotonously from $r_{ms}$ to $\infty$ and $W_+(r_{min})$ increases monotonously from $E_{ms}$ to $m$.
\end{itemize}
\end{lemma}
It follows from this Lemma that for the given ranges for $L$ and $E$ in the hypothesis of the proposition, there are closed trajectories giving rise to the remaining $S^1$-factor in the topology of $\Gamma^{(ext)}_{m,E, L_z, L}$. This concludes the proof of the proposition.
\qed

\section{Generalized action-angle variables on $\Gamma_{bound}$}
\label{Sec:ActionAngle}

This section is devoted to the construction of generalized action-angle variables which provide new symplectic coordinates $(J_\alpha,Q^\alpha)$ on the subset of the relativistic one-particle phase space $\Gamma$ corresponding to bound orbits in the Kerr exterior region. The construction largely follows standard methods from classical mechanics, see for instance~\cite{Arnold-Book,Zehnder-Book}. Nevertheless, one should point out that in our case the invariant sets $\Gamma^{(ext)}_{m,E,L_z, L}$ are not compact, such that one cannot directly apply the famous Liouville-Arnold theorem. However, by showing that the Hamilton vector fields $X_\alpha$ associated with the integrals of motion $F_\alpha$ defined in Eq.~(\ref{Eq:FalphaDef}) are complete, it is not difficult to generalize the proof to the present case, obtaining one extended coordinate $Q^0\in \Real$ in addition to the $2\pi$-periodic angle variables $Q^1,Q^2,Q^3$. For a generalization of the Liouville-Arnold theorem to non-compact invariant manifolds of completely (or partially) integrable systems, see~\cite{eFgGgS03}.

For the following, we recall the set $\Omega$ defined by the conditions~(\ref{Eq:LELzConditions}) in Proposition~\ref{Prop:InvariantSets}, and we introduce the following subset of $\Gamma$:
\begin{equation}
\Gamma_{bound} := \bigcup\limits_{(m,E,L,L_z)\in \Omega} \Gamma^{(ext)}_{m,E,L_z, L},
\label{Eq:GammaBound}
\end{equation}
which (with the exception of the special orbits characterized in Lemma~\ref{Lem:LI}) describes the phase space of bound orbits in the exterior Kerr spacetime. In the next subsection, we provide a formal definition for the construction of the generalized action-angle variables $(J_\alpha,Q^\alpha)$. In subsection~\ref{SubSec:Action} we determine the action variables and analyze the local invertibility of the transformation $\II: (P_\alpha)\mapsto (I_\alpha)$ which maps the conserved quantities $(P_\alpha) := (m,E,L_z,L)$ to the action variables. Next, in subsection~\ref{SubSec:Completeness} we show that the Hamiltonian vector fields $X_\alpha$ associated with $F_\alpha$ are complete in $\Gamma_{bound}$. This allows one to show in subsection~\ref{SubSec:QVars} that the (multi-valued) $Q^\alpha$-variables are globally well-defined, and leads to the main theorems~\ref{Thm:Global1},\ref{Thm:Global2} of this section. Finally, in subsection~\ref{SubSec:ExplicitAA} we provide explicit expressions for the variables $(J_\alpha,Q^\alpha)$ in terms of Legendre's elliptic integrals.

\subsection{Formal definition of the action-angle variables}

Recall that each set $\Gamma^{(ext)}_{m,E,L_z,L}$ in the union~(\ref{Eq:GammaBound}) is invariant under the Liouville flow and topologically of the form $\Real\times T^3$. We have also seen that the torus $T^3$ is naturally generated by the three $S^1$-factors corresponding to the motion in the azimuthal, polar and radial directions, respectively. Each of these $S^1$ factors gives rise to the variable
\begin{equation}
I_k(m,E,L_z,L) := \frac{1}{2\pi} \oint\limits_{\gamma_k} \Theta,\qquad k = 1,2,3,
\label{Eq:Ik}
\end{equation}
where $\gamma_k$ denotes a closed curve confined to $\Gamma^{(ext)}_{m,E, L_z, L}$ which circumscribes once the $k$'th factor $S^1$ and can be contracted to a single point with respect to the other $S^1$ factors. Since the restriction of the Poincar\'e one-form $\Theta$ to $\Gamma^{(ext)}_{m,E,L_z,L}$ is closed, the variables $I_k$ are invariant with respect to deformations of the curves $\gamma_k$ within $\Gamma^{(ext)}_{m,E,L_z,L}$, and thus they describe topological invariants of $\Gamma^{(ext)}_{m,E, L_z, L}$.

However, the choice for $I_0$ is more subtle. One could imagine defining $I_0$ in analogy with Eq.~(\ref{Eq:Ik}) to be proportional to
$$
\int\limits_{\gamma_0} \Theta,
$$
with $\gamma_0$ a curve in $\Gamma^{(ext)}_{m,E, L_z, L}$ which is contractible to a single point with respect to each $S^1$ factor. However, since $\Real$ is not compact, in general the result depends on the end points of $\gamma_0$ and fails to represent an invariant quantity.\footnote{In our case, this integral is equal to $-E(t_2-t_1)$ with $E = -p_t$ the conserved energy and $t_1$ and $t_2$ the times at the endpoints of the curve $\gamma_0$. 
One possibility of getting rid of the dependency of the end points is to restrict $\gamma_0$ to be an integral curve of the complete lift of the Killing vector field $\partial_t$ and divide the resulting integral by the time interval $(t_2-t_1)$, where $t$ is interpreted as the time-parameter along this integral curve. Up to a sign, this corresponds to the choice made in previous work~\cite{wS02,tHeF08,pRoS18}.} For this reason, we adopt a different choice for $I_0$, already proposed by us in~\cite{pRoS20}, which turns out the be convenient for our purposes, namely 
\begin{equation}
I_0(m,E,L_z,L) := m.
\label{Eq:I0}
\end{equation}
The quantities $(I_\alpha) := (I_0,I_1,I_2,I_3)$ defined by Eqs.~(\ref{Eq:Ik},\ref{Eq:I0}) yield a smooth map $\II: \Omega\to \Real^4$ which, as we show below, is (at least) locally invertible. The generalized action variables $J_\alpha: \Gamma_{bound}\to \Real$ are defined by the functions
\begin{equation}
J_\alpha(x,p) = I_\alpha\left(\sqrt{-2H(x,p)},F_1(x,p),F_2(x,p),\sqrt{F_3(x,p)}\right),\qquad
(x,p)\in \Gamma_{bound},\qquad \alpha = 0,1,2,3.
\end{equation}
Note that unlike $J_1$, $J_2$ and $J_3$ which have the units of an action (mass squared in geometrized units), $J_0$ has units of mass. Consequently, the conjugate variable $Q^0$ (defined below) has units of length whereas the angle variables $Q^1$, $Q^2$ and $Q^3$ are dimensionless. The main advantage for the choice in Eq.~(\ref{Eq:I0}) lies in the fact that the one-particle Hamiltonian~(\ref{Eq:HKerr}) and Liouville vector field assume the simple form
\begin{equation}
H = -\frac{1}{2} J_0^2,\qquad
X_H = -J_0\frac{\partial}{\partial Q^0}.
\end{equation}
An immediate consequence is that any DF satisfying the collisionsless Boltzmann equation whose support lies in the closure $\overline{\Gamma_{bound}}$ of $\Gamma_{bound}$ can be represented by a suitable function depending only on the variables $(Q^1,Q^2,Q^3,J_0,J_1,J_2,J_3)$, provided they are globally well-defined on $\Gamma_{bound}$.

The generalized angle variables $Q^\alpha$ can be formally defined as
\begin{equation}
Q^\alpha(x,p) = \left. \frac{\partial S}{\partial I_\alpha} \right|_x(\gamma_x; J_0(x,p),J_1(x,p),J_2(x,p),J_3(x,p)),\qquad
(x,p)\in \Gamma_{bound},
\label{Eq:AngleVariables}
\end{equation}
where the generating function $S$ is given by
\begin{equation}
S(\gamma_x; I_0,I_1,I_2,I_3) := \int\limits_{\gamma_x} \Theta,\qquad
(I_0,I_1,I_2,I_3)\in \II(\Omega).
\label{Eq:GeneratingFunction}
\end{equation}
Here, the line integral is performed along a curve $\gamma_x$ which is confined to the set $\Gamma^{(ext)}_{m,E, L_z, L}$ with $\II(m,E,L_z,L) = (I_0,I_1,I_2,I_3)$ and connects a given reference point $(x_0,p_0)$ on this set to a point in the intersection between $\Gamma^{(ext)}_{m,E, L_z, L}$ and the fibre over $x\in M$. We choose the reference point $(x_0,p_0)$ as the one parametrized by $(t,p_t) = (t_1,-E)$, $(\varphi,p_\varphi) = (\varphi_1, L_z)$, $(\vartheta,p_\vartheta) = (\vartheta_1,\sqrt{ L^2 -K(\vartheta_1)})$ and $(r,p_r) = (r_1,p_{r+}(r_1) = p_{r-}(r_1))$, where the functions $p_{r\pm}(r)$ are defined in Eq.~(\ref{Eq:prm}). Here, $\vartheta_1$ and $r_1$ refer to the left turning points in the $(\vartheta,p_\vartheta)$ and $(r,p_r)$-planes, respectively, and the values of $t_1$ and $\varphi_1$ will be determined shortly. We close this subsection by remarking that the generating function $S$ and $Q^\alpha$-variables are multi-valued, since the $S$ depends on the winding of the curve $\gamma_x$ around the torus $T^3$. Nevertheless, as we show below, the $Q^\alpha$-variables are globally well-defined on $\Gamma_{bound}$, with $Q^1$, $Q^2$, $Q^3$ being $2\pi$-periodic.

\subsection{Generalized action variables and local invertibility}
\label{SubSec:Action}

Next, we provide an explicit integral representation for the generalized action variables $J_\mu$ and ask whether they provide ``good" labels for the invariant sets. In order to do so, we start with the quantities $I_\alpha$ in Eqs.~(\ref{Eq:Ik},\ref{Eq:I0}) which define the action variables. Using the fact that $\Theta = p_t dt + p_\varphi d\varphi + p_r dr + p_\vartheta d\vartheta$ and Eqs.~(\ref{Eq:thetaPlaneRestriction},\ref{Eq:prm}) we conclude that
\begin{eqnarray}
I_0(m,E,L_z,L) &=& m,
\label{Eq:DefI0}\\
I_1(m,E,L_z, L) &=& L_z,
\label{Eq:DefI1}\\
I_2(m,E, L_z, L) &=& \frac{1}{2\pi}\oint p_\vartheta d\vartheta
 = \frac{1}{\pi} \int \limits_{\vartheta_1}^{\vartheta_2} \sqrt{L ^2 - K(\vartheta)} d\vartheta,
\label{Eq:DefI2}\\ 
I_3(m,E, L_z, L) &=& \frac{1}{2\pi}\oint p_r dr
 = \frac{1}{\pi} \int \limits_{r_1}^{r_2} \frac{\sqrt{R(r)}}{\Delta(r)} dr,
\label{Eq:DefI3}
\end{eqnarray}
where here and in the following $\vartheta_1 < \vartheta_2$ and $r_1 < r_2$ refer to the turning points. Similarly, the generating function~(\ref{Eq:GeneratingFunction}) yields
\begin{equation}
S(\gamma_x; I_0,I_1,I_2,I_3) = -E(t - t_1) + L_z(\varphi - \varphi_1) 
 + \int\limits_{\gamma_r} p_r dr 
 + \int\limits_{\gamma_\vartheta} p_\vartheta d\vartheta,
\label{Eq:GeneratingFunctionBis}
\end{equation}
where the last two integrals should be interpreted as line integrals. More specifically, the first integral is a line integral along the projection $\gamma_r$ of the curve $\gamma_x$ onto the $(r,p_r)$-plane, and similarly, the second integral is a line integral along the projection $\gamma_\vartheta$ of $\gamma_x$ onto the $(\vartheta,p_\vartheta)$-plane. For definiteness, in the following, we restrict the curve $\gamma_x$ to be such that its projections $\gamma_r$ and $\gamma_\vartheta$ are oriented clockwise. In order to simplify the calculations, it is useful to represent $p_r$ in the form
\begin{equation}
p_r = \frac{2M_H E r - a_H L_z + V}{\Delta},\qquad
V^2 = R(r).
\label{Eq:prV}
\end{equation}
Then, Eq.~(\ref{Eq:GeneratingFunctionBis}) can be rewritten as
\begin{equation}
S(\gamma_x; I_0,I_1,I_2,I_3) = -E t_{BL} + L_z\varphi_{BL}
 + \int\limits_{\gamma_r} V \frac{dr}{\Delta} 
 + \int\limits_{\gamma_\vartheta} p_\vartheta d\vartheta,
\label{Eq:GeneratingFunctionBisBis}
\end{equation}
with $V^2 = R(r)$ and $p_\vartheta^2 = L^2 - K(\vartheta)$ and where
\begin{equation}
t_{BL} := t - 2M_H\int\limits^r \frac{r dr}{\Delta},\qquad
\varphi_{BL} := \varphi - a_H\int\limits^r \frac{dr}{\Delta},
\end{equation}
turn out to be the Boyer-Lindquist time and azimuthal coordinates. In deriving these equations, we have adjusted the free constants $t_1$ and $\varphi_1$ determining the reference point $(x_0,p_0)$ such as to absorb a $r_1$-dependent constant. In subsection~\ref{SubSec:ExplicitAA} the line integrals in Eqs.~(\ref{Eq:DefI2},\ref{Eq:DefI3},\ref{Eq:GeneratingFunctionBisBis}) will be represented more explicitly in terms of Legendre's elliptic integrals.

Before we proceed with the computation of the variables $Q^\mu$ we need the following result.

\begin{lemma}
\label{Lem:IIInv}
The map $\II : \Omega\to \Real^4, P\mapsto (I_0,I_1,I_2,I_3)(P),$ is locally invertible in a vicinity of each point $P = (m,E,L_z,L)\in \Omega$.
\end{lemma}

\proof According to the inverse function theorem, it is sufficient to show that the linearization $D\II(P)$ of the map $\II$ is invertible at each $P\in \Omega$. The associated Jacobi matrix has the form
\begin{eqnarray}
{\bf M} = \left( \begin{array}{cccc}
1 & 0 & 0 & 0\\ 
0 & 0 & 1 & 0 \\
I_{20}&I_{21}&I_{22}&I_{23}\\
I_{30}&I_{31}&I_{32}&I_{33}
\end{array} \right),
\qquad
I_{\alpha\beta} := \frac{\partial I_\alpha}{\partial P_\beta},
\label{Eq:MDef}
\end{eqnarray}
and its determinant is
\begin{equation}
\det {\bf M} = I_{23} I_{31} - I_{21} I_{33}
 = \frac{\partial I_2}{\partial L}\frac{\partial I_3}{\partial E} 
 -\frac{\partial I_2}{\partial E}\frac{\partial I_3}{\partial L}.
\end{equation}
Using the expressions~(\ref{Eq:dSthetadE},\ref{Eq:dSthetadL},\ref{Eq:dSrdE},\ref{Eq:dSrdL}) from appendix~\ref{App:AngleVars} and taking into account that $I_2 = S_\vartheta(\vartheta_1)/\pi$ and $I_3 = S_r(r_2)/\pi$, one finds
\begin{equation}
\det {\bf M} = \frac{L}{\pi ^2}\int \limits_{r_1}^{r_2} \int\limits_{\vartheta_1}^{\vartheta_2}   
\left[ \Delta\rho^2 E + 2M_H r( r^2 E - a_H\hat{L}_z ) \right]
\frac{d\vartheta}{\sqrt{L^2 - K(\vartheta)}}\frac{dr}{\Delta(r) \sqrt{R(r)}},
\end{equation}
where we recall that $\hat{L}_z = L_z - a_H E$. Since on $\Gamma_{bound}$ one has $E > 0$ and $ a_H\hat{L}_z + \sqrt{\Delta(m^2 r^2 + L^2)} = r^2 W_+(r) \leq r^2 E$, such that $r^2 E - a_H\hat{L}_z > 0$, if follows that $\det{\bf M}$ is positive.
\qed

\begin{remark}
Denoting by $\Omega_I := \II(\Omega)$ the image of $\,\II$, the question regarding the global invertibility of the map $\II: \Omega\to \Omega_I$ is, of course, a difficult question. In the Schwarzschild limit $a_H = 0$, the integral $I_2$ can be computed explicitly and yields $I_2(P) = L - |L_z|$, which does not depend on $E$. It is then sufficient to show that for fixed values of $(m,L_z,L)$ the map $E\mapsto I_3(m,E,L_z,L)$ is invertible which is the case since $\partial I_3/\partial E$ is positive. However, in the Kerr case, $I_2$ is not independent of $E$, and hence one needs to show that for fixed $(m,L_z)$ the map $(E,L)\mapsto (I_2,I_3)(m,E,L_z,L)$ is invertible. We will not pursue this question further in this work.
\end{remark}

\subsection{Completeness of the generators and global coordinates on the invariant submanifolds}
\label{SubSec:Completeness}

In this subsection we show:

\begin{lemma}
\label{Lem:CompleteGenerators}
Let $(m,E,L_z,L)\in \Omega$. Then, the Hamiltonian vector fields $X_\alpha = X_{F_\alpha}$ associated with the integrals of motion $F_\alpha$ defined in Eq.~(\ref{Eq:FalphaDef}) are complete on $\Gamma^{(ext)}_{m,E,L_z, L}$.
\end{lemma}

\proof Recall that the fields $X_\alpha$ are tangent to $\Gamma^{(ext)}_{m,E,L_z, L} = \Real\times T^3$ and that $-X_1$ and $X_2$ correspond to the complete lifts of the Killing vector fields, which are complete in the exterior region. By introducing angles $\chi$ and $\phi$ on the $S^1$-factors corresponding to the radial and polar motions and transporting them along the flows of $-X_1$ and $X_2$ one obtains a global coordinates system $(t,\varphi,\chi,\phi)$ on $\Gamma^{(ext)}_{m,E,L_z, L}$, such that
\begin{equation}
X_1 = -\frac{\partial}{\partial t},\qquad
X_2 = \frac{\partial}{\partial \varphi},
\end{equation}
and $X_0$ and $X_3$ have the form
\begin{equation}
X_\alpha = A_\alpha(\chi,\phi)\frac{\partial}{\partial t} + Y_\alpha,\qquad
Y_\alpha = Y_\alpha^\varphi(\chi,\phi)\frac{\partial}{\partial\varphi} 
+ Y_\alpha^\chi(\chi,\phi)\frac{\partial}{\partial\chi}
+ Y_\alpha^\phi(\chi,\phi)\frac{\partial}{\partial\phi},\qquad \alpha = 0,3,
\end{equation}
with components $A_\alpha$ and $Y_\alpha^\varphi$, $Y_\alpha^\chi$, $Y_\alpha^\phi$ which are independent of $(t,\varphi)$. The vector fields $Y_\alpha$ are tangent to the compact manifold $T^3$ and hence they are complete. Together with the boundedness of $A_\alpha$, this implies that $X_\alpha$ are also complete.
\qed

For what follows, it turns out to be useful to provide explicit definitions for the angles $\chi$ and $\phi$. Of course, there are many possibilities for introducing such angles. Our choice has the advantage of allowing one to write the variables $Q^\mu$ explicitly in terms of Legendre's elliptic integrals. We start with the definition of the angle $\chi$ which has period $\pi$ and parametrizes the invariant curve in the $(r,p_r)$-plane according to
\begin{eqnarray}
r &=& r_4 + \frac{r_1 - r_4}{1 - b^2\sin^2\chi},
\label{Eq:rchi}\\
V &=& \frac{M_H m b^2}{C}(r_1 - r_4)\frac{\sqrt{1 - k^2\sin^2\chi}}{(1-b^2\sin^2\chi)^2}\sin(2\chi),
\label{Eq:Vchi}
\end{eqnarray}
where here $r_3 < r_4 < r_1 < r_2$ denote the roots of the polynomial $R(r)$ defined in Eq.~(\ref{Eq:rPlaneRestrictionBis}) (with $r_1 < r_2$ the turning points, as before), $p_r$ is obtained from $V$ according to Eq.~(\ref{Eq:prV}), and
\begin{equation}
b := \sqrt{\frac{r_2 - r_1}{r_2 - r_4}},\qquad
k := \sqrt{\frac{r_4 - r_3}{r_1 - r_3}} b,\qquad
C := \sqrt{\frac{2M_H(r_1+r_2+r_3+r_4)}{(r_1 - r_3)(r_2 - r_4)}}.
\label{Eq:bkC}
\end{equation}
It follows from Eqs.~(\ref{Eq:rchi},\ref{Eq:Vchi}) that
\begin{equation}
\frac{dr}{V} = \frac{C}{M_H m}\frac{d\chi}{\sqrt{1 - k^2\sin^2\chi}}.
\label{Eq:drV}
\end{equation}
The invariant curve in the $(\vartheta,p_\vartheta)$-plane is parametrized in terms of the $2\pi$-periodic angle $\phi$, such that
\begin{eqnarray}
\cos\vartheta &=& -\cos\vartheta_1\sin\phi,
\label{Eq:thetaphi}\\
p_\vartheta &=& \frac{\sqrt{L^2 - \hat{L}_z^2}}{\sin\vartheta}\sqrt{1 - k_1^2\sin^2\phi}\cos\phi,
\label{Eq:pthetaphi}
\end{eqnarray}
where here $\vartheta_1$ denotes the left turning point of the polar motion and $k_1 := \zeta_1/\zeta_2$ denotes the ratio between the two positive roots of the polynomial $q(\zeta)$ defined in Eq.~(\ref{Eq:qDef}). It follows from Eqs.~(\ref{Eq:thetaphi},\ref{Eq:pthetaphi}) that
\begin{equation}
\frac{d\vartheta}{p_\vartheta} = \frac{\cos\vartheta_1}{\sqrt{L^2 - \hat{L}_z^2}}
\frac{d\phi}{\sqrt{1 - k_1^2\sin^2\phi}}.
\label{Eq:dvarthetapvartheta}
\end{equation}
The coordinates $(z^\mu) := (t,\varphi,\chi,\phi)$ provide global coordinates on each invariant submanifold $\Gamma^{(ext)}_{m,E,L_z, L}$ with $(m,E,L_z,L)\in \Omega$.

For completeness, we compute the components of the generators $-X_0$ and $X_3$ of the Liouville and Carter flows with respect to these coordinates. Using $dz^\mu(X_\alpha) = \{F_\alpha, z^\mu \}$ and the fact that $\{F_\alpha, F_\beta \} = 0$ one finds
\begin{equation}
dz^\mu(X_\alpha) = \frac{\partial z^\mu}{\partial x^\nu} \{ F_\alpha, x^\nu\}
 = \frac{\partial F_\alpha}{\partial p_\nu}\frac{\partial z^\mu}{\partial x^\nu}.
\end{equation}
Taking into account Eqs.~(\ref{Eq:drV},\ref{Eq:dvarthetapvartheta}) a straightforward calculation yields
\begin{eqnarray}
\rho^2 X_0 &=& -\left[ 2M_H r(p_r + E) + \rho^2 E \right] \frac{\partial}{\partial t}
 - \left( a_H p_r + \frac{L_z}{\sin^2\vartheta} \right) \frac{\partial}{\partial \varphi}
\nonumber\\
 &-& \frac{M_H m}{C}\sqrt{1 - k^2\sin^2\chi} \frac{\partial}{\partial\chi}
 - \frac{\sqrt{L^2 - \hat{L}_z^2}}{\cos\vartheta_1}
 \sqrt{1 - k_1^2\sin^2\phi} \frac{\partial}{\partial \phi},
\\
\rho^2 X_3 &=& 2a_H
\left[ -2a_H\cos^2\vartheta M_H r(p_r + E) + \rho^2\hat{L}_z \right]\frac{\partial}{\partial t}
 + 2\left[ -a_H^3\cos^2\vartheta(p_r + E) + \frac{r^2}{\sin^2\vartheta}(\hat{L}_z + a_H\cos^2\vartheta E) \right]\frac{\partial}{\partial \varphi}
\nonumber\\
 &-& 2\frac{M_H m}{C}a_H^2\cos^2\vartheta
 \sqrt{1 - k^2\sin^2\chi} \frac{\partial}{\partial\chi}
  + 2r^2\frac{\sqrt{L^2 - \hat{L}_z^2}}{\cos\vartheta_1}
  \sqrt{1 - k_1^2\sin^2\phi} \frac{\partial}{\partial \phi},
\end{eqnarray}
where here $r$, $\vartheta$ and $p_r$ are functions of $(\chi,\phi)$, which are determined by the relations in Eqs.~(\ref{Eq:rchi},\ref{Eq:Vchi},\ref{Eq:prV},\ref{Eq:thetaphi}).

In the next subsection, a coordinate transformation $(t,\varphi,\chi,\phi)\mapsto (Q^0,Q^1,Q^2,Q^3)$ is introduced which brings the vector fields $X_0$ and $X_3$ in simpler form. Before we proceed, we note the following consequence of Lemma~\ref{Lem:CompleteGenerators} which will turn out to be useful later. Denote by $\varphi_\alpha^{\lambda^\alpha}$ the flows associated with $X_\alpha$ and define for each $(\lambda^0,\lambda^1,\lambda^2,\lambda^3)\in \Real^4$ the map
\begin{equation}
\varphi^\lambda := \varphi_0^{\lambda^0}\circ\varphi_1^{\lambda^1}\circ\varphi_2^{\lambda^2}\circ\varphi_3^{\lambda^3}.
\label{Eq:varphiDef}
\end{equation}
Since the vector fields $X_\alpha$ are complete, this map is well-defined for all $\lambda\in \Real^4$ and it satisfies $\varphi^0 = \id$ and $\varphi^{\lambda + \mu} = \varphi^\lambda\circ\lambda^\mu$ for all $\lambda,\mu\in \Real^4$, due to the fact that the vector fields $X_\alpha$ commute with each other. Therefore, the maps $\varphi^\lambda$ define an action of the group $(\Real^4,+)$ on $\Gamma_{bound}$. Restricting to a particular invariant submanifold $\Gamma^{(ext)}_{m,E,L_z, L}$, one has:

\begin{lemma}
\label{Lem:GlobalFlow}
Let $(m,E,L_z,L)\in \Omega$ and consider the restriction of the map $\varphi^\lambda$ defined in Eq.~(\ref{Eq:varphiDef}) on $\Gamma^{(ext)}_{m,E,L_z, L}$. Denote by
$$
H_{Iso} := \{ \lambda\in\Real^4 : \varphi^\lambda(x,p) = (x,p) \hbox{ for all $(x,p)\in \Gamma^{(ext)}_{m,E,L_z, L}$}  \}
$$
the isotropy subgroup. Then, $\varphi^\lambda$ provides a diffeomorphism of $\Real^4/H_{Iso}$ onto $\Gamma^{(ext)}_{m,E,L_z, L}$.
\end{lemma}

\proof Let us abbreviate $\Gamma^{(0)} := \Gamma^{(ext)}_{m,E,L_z, L}$, and let $(x_0,p_0)\in \Gamma^{(0)}$ be fixed. Consider the map
\begin{equation}
G: \Real^4\to \Gamma^{(0)}, 
\lambda \mapsto 
G(\lambda) = \varphi^\lambda(x_0,p_0).
\label{Eq:GMap}
\end{equation}
Its linearization $DG(\lambda)$ maps the vector fields $e_\alpha$ of the standard basis of $\Real^4$ to the vectors $\left. X_\alpha \right|_{\varphi^\lambda(x_0,p_0)}$, which are linearly independent, and hence $G$ is a local diffeomorphism. Consequently, the image of $G$ is both open and closed in $\Gamma^{(0)}$, which implies that $G(\Real^4) = \Gamma^{(0)}$. Hence, when restricted to the quotient $\Real^4/H_{Iso}$, $G$ provides a diffeomorphism of $\Real^4/H_{Iso}$ to $\Gamma^{(0)}$.
\qed

\begin{remark}
Since $\Gamma^{(ext)}_{m,E,L_z, L} = \Real\times T^3$, the isotropy subgroup $H_{Iso}$ is a lattice group generated by three linearly independent vectors $E_\varphi,E_\chi,E_\phi\in \Real^4$. They will be computed explicitly in subsection~\ref{SubSec:ExplicitAA} below.
\end{remark}

\subsection{The generalized angle variables}
\label{SubSec:QVars}

In this subsection we analyze the properties of the $Q^\alpha$ variables which are formally defined by Eq.~(\ref{Eq:AngleVariables}). We first prove that they are locally well-defined and that $(J_\alpha,Q^\alpha)$ are local symplectic coordinates.

\begin{lemma}
\label{Lem:QJLocal}
For each $(x,p)\in \Gamma_{bound}$ there exists an open neighborhood $U$ of $(x,p)$ in $\Gamma_{bound}$ on which the variables $Q^\alpha$ are well-defined and smooth and the coordinates $(J_\alpha,Q^\alpha)$ are symplectic.
\end{lemma}

\proof Let $(x,p)\in \Gamma_{bound}$ and set $(P_\alpha) := (\sqrt{2F_0(x,p)},F_1(x,p),F_2(x,p),\sqrt{F_3(x,p)})\in \Omega$, see Eq.~(\ref{Eq:InvariantSets}). Next, choose an open neighborhood $\Omega'\subset \Omega$ of $(P_\alpha)$ on which the map $\II$ is injective and consider the corresponding subset
$$
\Gamma_{bound}' := \bigcup\limits_{(m,E,L,L_z)\in \Omega'} \Gamma^{(ext)}_{m,E,L_z, L}
$$ 
of $\Gamma_{bound}$. Next, rewrite the generating function $S$ in Eq.~(\ref{Eq:GeneratingFunctionBisBis}) in the form
$S(\gamma_x; I_\alpha) = -E t_{BL} + L_z\varphi_{BL} 
 + S_r(\gamma_r; I_\alpha)  + S_\vartheta(\gamma_\vartheta; I_\alpha)$,
with the generating functions $S_r$ and $S_\vartheta$ corresponding to the radial and polar motion, respectively, defined by
\begin{equation}
S_r(\gamma_r; I_\alpha) = \int\limits_{\gamma_r} V\frac{dr}{\Delta},\qquad
S_\vartheta(\gamma_\vartheta; I_\alpha) := \int\limits_{\gamma_\vartheta} p_\vartheta d\vartheta,
\label{Eq:SrStheta}
\end{equation}
where we recall that $V$ and $p_\vartheta$ satisfy $V^2 = R(r)$ and $p_\vartheta^2 = L^2 - K(\vartheta)$. Unless $(x,p)$ corresponds to a turning point for the radial or polar motions, we can choose an open neighborhood $U$ of $(x,p)$ in $\Gamma_{bound}'$ on which $S_r$ and $S_\vartheta$ are well-defined functions of $(r,I_\alpha)$ and $(\vartheta,I_\alpha)$, respectively. On such a neighborhood, the functions
\begin{equation}
Q^\alpha(x,p) := \left. \frac{\partial S}{\partial I_\alpha} \right|_x (\gamma_x, J_\alpha(x,p)),
\qquad (x,p)\in U,
\label{Eq:QalphaS}
\end{equation}
are well-defined and the map $(x^\mu,p_\mu)\mapsto (Q^\mu,J_\mu)$, $J_\mu := I_\mu(P(x,p))$, is smooth and symplectic since
\begin{equation}
\Omega_s = dp_\mu\wedge dx^\mu
 = d\left( \frac{\partial S}{\partial x^\mu} \right)\wedge dx^\mu
 = \frac{\partial^2 S}{\partial x^\mu\partial I_\alpha} dJ_\alpha\wedge dx^\mu
 + \frac{\partial^2 S}{\partial x^\mu\partial x^\nu} dx^\nu\wedge dx^\mu
 = dJ_\alpha\wedge d\left( \frac{\partial S}{\partial I_\alpha} \right)
 = dJ_\alpha\wedge dQ^\alpha.
\end{equation}
If the point $(x,p)$ corresponds to a turning point of the polar motion, say, we use $p_\vartheta$ instead of $\vartheta$ to parametrize the curve $\gamma_\vartheta$ which leads to an alternative generating function $S_\vartheta'(p_\vartheta,I_\alpha)$. Specifically, we rewrite
$$
S_\vartheta(\gamma_\vartheta; I_\alpha) = p_\vartheta \vartheta + S'_\vartheta(p_\vartheta,I_\alpha),\qquad
S'_\vartheta(p_\vartheta,I_\alpha) 
 = -\int\limits_{\gamma_{p_\vartheta}} \vartheta dp_\vartheta,
$$
and notice that for points away from the turning point
$$
\left. \frac{\partial S_\vartheta}{\partial I_\alpha} \right|_\vartheta 
 = \left( \vartheta + \frac{\partial S_\vartheta'}{\partial p_\vartheta} \right)
 \frac{\partial p_\vartheta}{\partial I_\alpha} 
 + \frac{\partial S_\vartheta'}{\partial I_\alpha}
 = \frac{\partial S_\vartheta'}{\partial I_\alpha},
$$
such that $Q^\alpha$ can be obtained by the same equation~(\ref{Eq:QalphaS}) replacing $S_\vartheta$ with $S_\vartheta'$ in the generating function. Again, the transformation $(x^\mu,p_\mu) \mapsto (Q^\mu,J_\mu)$ is symplectic. A similar procedure is used if $(x,p)$ is a turning point of the radial motion.
\qed

Although they are locally well-defined, the variables $Q^\alpha$ are not globally uniquely defined on $\Gamma_{bound}'$, due to their dependency on the winding of the curve $\gamma_x$ connecting the reference point $(x_0,p_0)$ to the end point $(x,p)$. If $\gamma_x$ and $\gamma_x'$ are two curves connecting $(x_0,p_0)$ and $(x,p)$, one has, according to  Eqs.~(\ref{Eq:Ik},\ref{Eq:GeneratingFunction}):
\begin{equation}
S(\gamma_x'; I_\alpha) = S(\gamma_x; I_\alpha) + 2\pi\sum\limits_{j=1}^3 n^j I_j,
\end{equation}
with $n = (n^1,n^2,n^3)\in \Integer^3$ the difference in the winding numbers between the two curves. As a consequence, the corresponding $Q^\alpha$-variables are related to each other by
\begin{equation}
(Q')^\alpha(x,p) = Q^\alpha(x,p) + 2\pi\delta^\alpha{}_j n^j,
\end{equation}
that is, the variables $Q^j$ with $j=1,2,3$ are $2\pi$-periodic, whereas $Q^0$ is independent of the winding. We conclude this subsection by proving that (taking into account the periodicity of the angle variables $Q^j$) the $Q^\alpha$ variables provide global coordinates on each invariant set $\Gamma^{(ext)}_{m,E,L_z, L}$.

\begin{lemma}
\label{Lem:QInvertibility}
Let $(m,E,L_z,L)\in \Omega$. Then, $\mathcal{Q}: \Gamma^{(ext)}_{m,E,L_z, L}\to \Real\times T^3, (x,p)\mapsto (Q^0(x,p),Q^1(x,p),Q^2(x,p),Q^3(x,p))$ is a diffeomorphism.
\end{lemma}

\proof Denote by $Z_\alpha := X_{J_\alpha}$ the Hamiltonian vector fields associated with $J_\alpha$. Due to the fact that $(J_\alpha,Q^\alpha)$ are symplectic coordinates, they satisfy
\begin{equation}
Z_\alpha[Q^\beta] = \{ J_\alpha, Q^\beta \} = \delta^\beta{}_\alpha.
\end{equation}
Furthermore, it follows from $J_\alpha = I_\alpha(m,E,L_z,L)$ and $F_0 = m^2/2$ $F_1 = E$, $F_2 = L_z$, $F_3 = L^2$ on $\Gamma^{(ext)}_{m,E,L_z, L}$ (see Eq.~(\ref{Eq:InvariantSets})) that
\begin{equation}
\Omega_s(\cdot,Z_\alpha) = dJ_\alpha = {\bf N}_\alpha{}^\beta dF_\beta
 = {\bf N}_\alpha{}^\beta \Omega_s(\cdot, X_\beta),
\label{Eq:ZXRelation}
\end{equation}
with ${\bf N} := {\bf M}\circ {\bf D}^{-1}$, where ${\bf M}$ is the matrix defined in Eq.~(\ref{Eq:MDef}) which is invertible as has been shown in the proof of Lemma~\ref{Lem:IIInv}, and ${\bf D} = \diag(m,1,1,2L)$. It follows from Eq.~(\ref{Eq:ZXRelation}) that $Z_\alpha = {\bf N}_\alpha{}^\beta X_\beta$, and as a consequence of Lemma~\ref{Lem:CompleteGenerators}  and the fact that ${\bf N}$ is constant on each $\Gamma^{(ext)}_{m,E,L_z, L}$, these fields $Z_\alpha$ are complete and satisfy $[Z_\alpha,Z_\beta] = \{ J_\alpha, J_\beta \} = 0$. Next, denote by $\psi_\alpha^{\lambda^\alpha}$ the flow associated with $Z_\alpha$ and define for each $\lambda = (\lambda^0,\lambda^1,\lambda^2,\lambda^3)\in\Real^4$ the map
\begin{equation}
\psi^\lambda := \psi_0^{\lambda^0}\circ\psi_1^{\lambda^1}\circ\psi_2^{\lambda^2}\circ\psi_3^{\lambda^3},
\label{Eq:psiDef}
\end{equation}
in analogy with Eq.~(\ref{Eq:varphiDef}). Denote by $(x_0,p_0)$ the reference point of $\Gamma^{(ext)}_{m,E,L_z, L}$, and introduce the map\footnote{Note that this map is related to the corresponding map $G$ defined in Eq.~(\ref{Eq:GMap}) through
\begin{equation}
G'(\lambda) = G({\bf N}^T\lambda),\qquad \lambda\in\Real^4.
\end{equation}
}
\begin{equation}
G': \Real^4\to \Gamma^{(ext)}_{m,E,L_z, L},
\lambda\mapsto G'(\lambda) := \psi^\lambda(x_0,p_0).
\end{equation}
As in the proof of Lemma~\ref{Lem:GlobalFlow} it follows that $G'$ defines a diffeomorphism of $\Real^4/H'_{Iso}$ onto $\Gamma^{(ext)}_{m,E,L_z, L}$, where $H'_{Iso}$ denotes the isotropy subgroup of $\psi^\lambda$. We claim that this map is the inverse of $\mathcal{Q}$. To prove this, it is sufficient to observe that $Q^\mu(x_0,p_0) = 0$ and
$$
\frac{\partial}{\partial\lambda^\beta} Q^\alpha( G'(\lambda) ) 
 = \left. Z_\beta[Q^\alpha] \right|_{G'(\lambda)} = \delta^\alpha{}_\beta,
$$
which implies that $Q^\alpha( G'(\lambda) ) = \lambda^\alpha$ for all $\lambda\in \Real^4$ and that $H_{Iso}' = \{ \lambda = (0,\ve{n}) : \ve{n}\in 2\pi\Integer^3 \}$.
\qed

Lemmas~\ref{Lem:IIInv}, \ref{Lem:QJLocal} and \ref{Lem:QInvertibility} imply:

\begin{theorem}
\label{Thm:Global1}
Let $\Omega'\subset \Omega$ be an open subset of $\Omega$ on which the map $\II$ is injective and consider the corresponding subset
\begin{equation}
\Gamma_{bound}' := \bigcup\limits_{(m,E,L,L_z)\in \Omega'} \Gamma^{(ext)}_{m,E,L_z, L}
\end{equation}
of $\Gamma_{bound}$. Then, the map $\Phi: \Gamma_{bound}'\to \Real\times T^3\times \II(\Omega')$ defined by
\begin{equation}
\Phi(x,p) := (Q^0(x,p),Q^1(x,p),Q^2(x,p),Q^3(x,p),J_0(x,p),J_1(x,p),J_2(x,p),J_3(x,p)),
\qquad (x,p)\in \Gamma_{bound}',
\end{equation}
is a diffeomorphism which satisfies $\Omega_s = dJ_\alpha\wedge dQ^\alpha$.
\end{theorem}

As mentioned previously, it is not a priori clear whether or not the map $\II$ defined in Lemma~\ref{Lem:IIInv} is globally invertible, such that it is a priori not clear either whether or not the map $\Phi$ can be extended to all $\Gamma_{bound}$. To bypass this problem, one can also work with the variables $(Q^\alpha,P_\alpha)$ instead of $(Q^\alpha,J_\alpha)$ which are globally well-defined on $\Gamma_{bound}$ as follows from Lemma~\ref{Lem:QInvertibility}. These variables are not symplectic anymore; however the symplectic form $\Omega_s$ still has a simple representation.

\begin{theorem}
\label{Thm:Global2}
The map $\Psi: \Gamma_{bound}\to \Real\times T^3\times \Omega$ defined by
\begin{equation}
\Psi(x,p) := (Q^0(x,p),Q^1(x,p),Q^2(x,p),Q^3(x,p),P_0(x,p),P_1(x,p),P_2(x,p),P_3(x,p)),
\qquad (x,p)\in \Gamma_{bound},
\end{equation}
is a diffeomorphism which satisfies $\Omega_s = {\bf M}_\alpha{}^\beta dP_\beta\wedge dQ^\alpha$, with ${\bf M}$ the Jacobi matrix defined in Eq.~(\ref{Eq:MDef}).
\end{theorem}

\proof The fact that $\Psi$ defines a diffeomorphism follows from Lemmas~\ref{Lem:QJLocal} and \ref{Lem:QInvertibility}. To prove the validity of the claimed expression for the symplectic form, we take a subset $\Gamma_{bound}'$ on which the previous theorem applies and note that
$$
dJ_\alpha = {\bf M}_\alpha{}^\beta dP_\beta,
$$
which implies that $\Omega_s = {\bf M}_\alpha{}^\beta dP_\beta\wedge dQ^\alpha$ on $\Gamma_{bound}'$. Since ${\bf M}_\alpha{}^\beta$ only depends on $P_\alpha$ and $\Gamma_{bound}$ can be covered by sets of the form $\Gamma_{bound}'$ as in the previous theorem, the claim follows.
\qed

\subsection{Explicit expressions for the generalized action-angle variables in terms of Legendre's elliptic integrals}
\label{SubSec:ExplicitAA}

As mentioned above, the variables $J_\alpha$ and $Q^\alpha$ and  can be computed explicitly. We provide more details of the calculations in appendix~\ref{App:AngleVars}; the explicit representation is based on the roots of the polynomial $R(r)$ defined in Eq.~(\ref{Eq:rPlaneRestrictionBis}), those of the polynomial $q(\zeta)$ in Eq.~(\ref{Eq:qDef}) and the angles $\chi$ and $\phi$ defined through Eqs.~(\ref{Eq:rchi},\ref{Eq:Vchi},\ref{Eq:thetaphi},\ref{Eq:pthetaphi}). In terms of the functions $\GG^\alpha(\phi)$ and $\HH^\alpha(\chi)$ defined in Eqs.~(\ref{Eq:GG0Def}--\ref{Eq:GG3Def},\ref{Eq:HH0Def}--\ref{Eq:HH3Def}) and the abbreviations $\GG^\alpha := \GG^\alpha(\pi/2)$ and $\HH^\alpha := \HH^\alpha(\pi/2)$, the action variables can be written as
\begin{eqnarray}
J_0 &=& m,\\
J_1 &=& L_z,\\
J_2 &=& \frac{1}{\pi}\left[ M_H m\GG^0 + M_H\GG^1 + L_z\GG^2 + L\GG^3 \right],
\label{Eq:I2}\\ 
J_3 &=& \frac{1}{\pi}\left[ M_H m\HH^0 + M_H\HH^1 + L_z\HH^2 + L\HH^3 \right],
\label{Eq:I3}
\end{eqnarray}
where it is understood that $(P_\alpha) = (m,E,L_z,L)$ are determined from the values of $F_\alpha(x,p)$ according to the relations in Eq.~(\ref{Eq:InvariantSets}). The angle variables $Q^\alpha$ defined in Eq.~(\ref{Eq:AngleVariables}) can be computed by applying the chain rule:
\begin{equation}
Q^\alpha = \frac{\partial S}{\partial I_\alpha}
 = \frac{\partial S}{\partial P_\beta} \frac{\partial P_\beta}{\partial I_\alpha}
 = ({\bf M}^{-T})^\alpha{}_\beta \tilde{Q}^\beta,\qquad
\tilde{Q}^\beta := \frac{\partial S}{\partial P_\beta},
\label{Eq:Q}
\end{equation}
with ${\bf M}$ the Jacobi matrix of the map $\II: \Omega\to \Real^4$ and ${\bf M}^{-T}$ its inverse transposed. The partial derivatives of the generating function $S$ with respect to $P_\beta$ give
\begin{eqnarray}
\tilde{Q}^0 &=& \frac{\partial S}{\partial m} = M_H \left[ \GG^0(\phi) + \HH^0(\chi) \right],\\
\tilde{Q}^1 &=& \frac{\partial S}{\partial E} 
 = -t_{BL} + M_H\left[ \GG^1(\phi) + \HH^1(\chi) \right],\\
\tilde{Q}^2 &=& \frac{\partial S}{\partial L_z} = \varphi_{BL} + \GG^2(\phi) + \HH^2(\chi),\\
\tilde{Q}^3 &=& \frac{\partial S}{\partial L} = \GG^3(\phi) + \HH^3(\chi),
\end{eqnarray}
and it can be verified that these quantities satisfy $X_\alpha\left[ \tilde{Q}^\beta \right] = 0$ for $\alpha\neq \beta$ and $X_0[\tilde{Q}^0] = m$, $X_1[\tilde{Q}^1] = 1$, $X_2[\tilde{Q}^2] = 1$, $X_3[\tilde{Q}^3] = 2L$, with $X_\alpha$ the Hamiltonian vector field associated with $F_\alpha$. Hence, $\tilde{Q}^0/m$, $\tilde{Q}^1$, $\tilde{Q}^2$ and $\tilde{Q}^3/(2L)$ are transported along the Hamiltonian flows associated with $X_\alpha$ which implies that they are locally well-defined and multi-valued functions on each invariant set $\Gamma^{(ext)}_{m,E,L_z, L}$. However, they do not have the correct period: under full revolutions $\varphi\mapsto \varphi + 2\pi$, $\chi\mapsto \chi + \pi$, and $\phi\mapsto \phi + 2\pi$ about the $S^1$ factors, these variables change according to $\tilde{Q}^\alpha\mapsto\tilde{Q}^\alpha + E_a^\alpha$, with $E_\varphi,E_\chi,E_\phi/(2L)\in\Real^4$ the generators of the isotropy group $H_{Iso}$ (see the remark below Lemma~\ref{Lem:GlobalFlow}), given by $E_\varphi = (0,0,2\pi,0)$, $E_\chi = 2(M_H\HH^0,M_H\HH^1,\HH^2,\HH^3)$, and $E_\phi = 2(M_H\GG^0,M_H\GG^1,\GG^2,\GG^3)$.

The matrix ${\bf M}$ and its inverse transposed read
\begin{equation}
{\bf M} = \frac{1}{\pi}\left( \begin{array}{cccc}
\pi & 0 & 0 & 0\\ 
0 & 0 & \pi & 0 \\
M_H\GG^0 & M_H\GG^1 & \GG^2 & \GG^3\\
M_H\HH^0 & M_H\HH^1 & \HH^2 & \HH^3
\end{array} \right),\qquad
{\bf M}^{-T} = \left( \begin{array}{cccc}
1 & \omega^0 & 0 & M_H\eta^0 \\ 
0 & \omega^1 & 1 & M_H\eta^1 \\ 
0 & \omega^2 & 0 & M_H\eta^2 \\ 
0 & \omega^3 & 0 & M_H\eta^3 \\ 
\end{array} \right),
\label{Eq:MMmT}
\end{equation}
with
\begin{equation}
\omega^0 = -\frac{\GG^0\HH^3 - \GG^3\HH^0}{\GG^1\HH^3 - \GG^3\HH^1},\quad
\omega^1 = -\frac{1}{M_H} \frac{\GG^2\HH^3 - \GG^3\HH^2}{\GG^1\HH^3 - \GG^3\HH^1},\quad
\omega^2 = \frac{\pi}{M_H}\frac{\HH^3}{\GG^1\HH^3 - \GG^3\HH^1},\quad
\omega^3 = -\frac{\pi}{M_H}\frac{\GG^3}{\GG^1\HH^3 - \GG^3\HH^1},
\label{Eq:omegaDef}
\end{equation}
and
\begin{equation}
\eta^0 = -\frac{\GG^1\HH^0 - \GG^0\HH^1}{\GG^1\HH^3 - \GG^3\HH^1},\quad
\eta^1 = -\frac{1}{M_H} \frac{\GG^1\HH^2 - \GG^2\HH^1}{\GG^1\HH^3 - \GG^3\HH^1},\quad
\eta^2 = -\frac{\pi}{M_H}\frac{\HH^1}{\GG^1\HH^3 - \GG^3\HH^1},\quad
\eta^3 = \frac{\pi}{M_H}\frac{\GG^1}{\GG^1\HH^3 - \GG^3\HH^1}.
\label{Eq:etaDef}
\end{equation}
Note that according to the proof of Lemma~\ref{Lem:IIInv} one has $0 < \det({\bf M}) = -M_H(\GG^1\HH^3 - \GG^3\HH^1)/\pi^2$, such that these quantities are well-defined. Furthermore, note that ${\bf M}^{-T} E_\varphi = 2\pi(0,1,0,0)$, ${\bf M}^{-T} E_\chi = 2\pi(0,0,1,0)$ and ${\bf M}^{-T} E_\phi = 2\pi(0,0,0,1)$, such that the variables $Q^1$, $Q^2$ and $Q^3$ have the correct period, as is expected from their definition. Explicitly, one finds $Q^\alpha = q^\alpha - \omega^\alpha t_{BL}$ with
\begin{eqnarray}
q^0 &=& M_H\left\{
  \frac{\HH^1\GG^{03}(\phi) + \HH^3\GG^{10}(\phi) + \HH^0\GG^{31}(\phi)
 }{\GG^1\HH^3 - \GG^3\HH^1} 
 - \frac{\GG^1\HH^{03}(\chi) + \GG^3\HH^{10}(\chi) + \GG^0\HH^{31}(\chi)
 }{\GG^1\HH^3 - \GG^3\HH^1} 
 \right\},
\label{Eq:q0}\\
q^1 &=& \frac{\HH^1\GG^{23}(\phi) + \HH^2\GG^{31}(\phi) + \HH^3\GG^{12}(\phi)
 }{\GG^1\HH^3 - \GG^3\HH^1} 
 - \frac{\GG^1\HH^{23}(\chi) + \GG^2\HH^{31}(\chi) + \GG^3\HH^{12}(\chi)
 }{\GG^1\HH^3 - \GG^3\HH^1} + \varphi_{BL},
\label{Eq:q1}\\   
q^2 &=& \pi\frac{\HH^3\GG^1(\phi) - \HH^1\GG^3(\phi) - \HH^{13}(\chi)}{\GG^1\HH^3 - \GG^3\HH^1} ,
\label{Eq:q2}\\   
q^3 &=& \pi\frac{\GG^1\HH^3(\chi) - \GG^3\HH^1(\chi) + \GG^{13}(\phi)}{\GG^1\HH^3 - \GG^3\HH^1},
 \label{Eq:q3}
\end{eqnarray}
where we have introduced the functions $\GG^{\alpha\beta}(\phi) := \GG^\alpha\GG^\beta(\phi) - \GG^\beta\GG^\alpha(\phi)$ and $\HH^{\alpha\beta}(\chi) := \HH^\alpha\HH^\beta(\chi) - \HH^\beta\HH^\alpha(\chi)$ which are invariant with respect to the transformations $\phi\mapsto \phi + \pi/2$ and $\chi\mapsto \chi + \pi/2$. Like $(Q^1,Q^2,Q^3)$, the variables $(q^1,q^2,q^3)$ are angle coordinates associated with the azimuthal, polar and radial motion, respectively, and $\omega^1$, $\omega^2$ and $\omega^3$ are the corresponding frequencies describing their changes with respect to the Boyer-Lindquist time coordinate $t_{BL}$.\footnote{A coordinate-independent definition of these quantities is provided by noting that $\omega^\alpha = X_1[Q^\alpha]$ with $X_1$ the Hamiltonian vector field associated with the integral of motion $F_1$, which corresponds to minus the complete lift of the Killing vector field $\partial_t$, see the remark below Corollary~\ref{Cor:InvSets}.} Finally, note that $q^0$, $q^2$ and $q^3$ only depend on the conserved quantities $P_\alpha$ and the angles $\chi$ and $\phi$, while $q^1$ depends, in addition, on the azimuthal angle $\varphi_{BL}$. 

In the limit $\beta^2\to 1$ of equatorial orbits, it follows that $\GG^0 = 0$, $\GG^1 = -\alpha\GG^2$ and $\GG^3 = -\beta\GG^2$ which implies that $J_2 = 0$ and $\GG^{\alpha\beta}(\phi) = 0$, such that the above expressions for $q^0$, $q^1$ and $q^3$ simplify and become independent of $\phi$.

In the non-rotating limit $a_H = 0$ it follows that $\GG^0 = \GG^1 = \HH^2 = 0$ and $\GG^2(\phi) = -\beta\Pi_*(\phi,1-\beta^2,0)$, $\GG^3(\phi) = \phi + \pi/2$ and the expressions in Eq.~(\ref{Eq:omegaDef}) simplify considerably:
\begin{equation}
\omega^0 = -\frac{\HH^0}{\HH^1},\qquad
\pm\omega^1 = \omega^2 = -\frac{1}{M_H}\frac{\HH^3}{\HH^1},\qquad
\omega^3 = \frac{\pi}{M_H}\frac{1}{\HH^1},
\end{equation}
the $\pm$ sign corresponding to the sign of $L_z$. The fact that $\omega^1$ and $\omega^2$ are equal in magnitude reflects the fact that the motion is confined to a plane. Further, one finds $J_2 = L - |L_z|$ and
\begin{equation}
q^0 = -M_H\frac{\HH^{01}(\chi)}{\HH^1},\quad
q^1 = -\frac{1}{\pi}\GG^{23}(\phi) 
 +  \sign(\beta)\frac{\HH^{13}(\chi)}{\HH^1} + \varphi_{BL},
\quad
q^2 = \frac{\HH^{13}(\chi)}{\HH^1} + \phi + \pi/2,\quad
q^3 = \pi\frac{\HH^1(\chi)}{\HH^1}.
\end{equation}

Before closing this section, for completeness, we provide the explicit expressions for the Hamiltonian vector fields $X_\alpha$ in terms of the generalized action-angle variables $(P_\alpha,Q^\alpha)$. Using Theorem~\ref{Thm:Global2} and the relations $F_0 = m^2/2 = P_0^2/2$, $F_1 = E = P_1$, $F_2 = L_z = P_2$, $F_3 = L^3 = P_3^2$ from Eq.~(\ref{Eq:InvariantSets}) one finds
\begin{equation}
X_0 = P_0\frac{\partial}{\partial Q^0},\qquad
X_1 = \omega^\alpha\frac{\partial}{\partial Q^\alpha},\qquad
X_2 = \frac{\partial}{\partial Q^1},\qquad
X_3 = 2M_H P_3\eta^\alpha\frac{\partial}{\partial Q^\alpha}.
\label{Eq:Xalpha}
\end{equation}

\section{Mixing and strong Jeans theorem}
\label{Sec:Mixing}

In this section we apply the results from the previous section to analyze the late time dynamics of the solutions $f$ of the Liouville equation~(\ref{Eq:Liouville}), assuming that $f$ is supported in $\overline{\Gamma_{bound}}$, the subset of relativistic phase space consisting of bound timelike geodesic orbits in the Kerr exterior. According to the results from the previous section, the set $\Gamma_{bound}$ is diffeomorphic to $\Real\times T^3\times\Omega$ and in terms of the coordinates $(Q^\alpha,P_\alpha)$ the Liouville vector field is simply $-P_0\frac{\partial}{\partial Q^0}$, which implies that $f$ is a function depending only of the angle variables $Q^1,Q^2,Q^3$ and the conserved quantities $P_\alpha$.

For what follows, we consider a kinetic gas consisting of identical massive particle of fixed rest mass $m > 0$. We introduce a foliation $S_T$ of the Kerr exterior by three-dimensional hypersurfaces of constant Boyer-Lindquist time $t_{BL} = T$ and the associated six-dimensional subsets
\begin{equation}
\Sigma_T := \{ (x,p)\in \Gamma_{bound} : x\in S_T,  g_x(p,p) = -m^2 \}
\end{equation}
of $\Gamma_{bound}$. Owing to the fact that the Kerr exterior is stationary, the flow of the Killing vector field $\partial_t$ provides an isometry between the different sets $S_T$. Likewise, the flow $\varphi^t$ associated with the complete lift of $\partial_t$ provides a symplectic diffeomorphism between the different sets $\Sigma_T$, such that each of this set can be naturally identified with $\Sigma_0$, say.

In the next subsection we show how to reduce the considerations of the previous section to $\Sigma_0$, and we reformulate the Liouville equation~(\ref{Eq:Liouville}) as a Cauchy problem on $\Sigma_0$. Next, in subsection~\ref{SubSec:Jeans} we discuss the strong Jeans theorem and in subsection~\ref{SubSec:Mixing} we formulate sufficient conditions for phase space mixing to hold.

\subsection{Reduction to six-dimensional phase space and Cauchy problem}
\label{SubSec:Reduction}

We denote by $\hat{\Omega}$ the set of $3$-tuples $\ve{P} := (E,L_z,L)$ for which $(m,E,L_z,L)\in \Omega$. Each $\Sigma_T$ is foliated by the sets
\begin{equation}
\hat{\Gamma}_{E,L_z,L} 
 := \{ (x,p)\in \Sigma_T : F_1(x,p) = E, F_2(x,p) = L_z, F_3(x,p) = L^2 \},
\qquad (E,L_z,L)\in \hat{\Omega},
\end{equation}
which are topologically equal to $T^3$, and it follows from the results of the previous section that the variables $(\ve{P},\ve{q}) = (P_a,q^a)$, $a=1,2,3$, on $\Sigma_T$ are adapted to this foliation, where on each set $\hat{\Gamma}_{E,L_z,L}$, the $P_a$'s are constant and the quantities $q^1,q^2,q^3$ defined by Eqs.~(\ref{Eq:q1},\ref{Eq:q2},\ref{Eq:q3}) are angles. When restricted to $\Sigma_T$ the map $\Psi$ from Theorem~\ref{Thm:Global2} induces a diffeomorphism $\hat{\Psi}_T: \Sigma_T\to T^3\times\hat{\Omega}$ which is explicitly given by
\begin{equation}
\hat{\Psi}_T(x,p) = (Q^1(x,p),Q^2(x,p),Q^3(x,p),P_1(x,p),P_2(x,p),P_3(x,p)),
\qquad (x,p)\in \Sigma_T,
\end{equation}
with $Q^a = q^a - T\omega^a$, $a=1,2,3$. In terms of adapted local coordinates $(x^\mu,p_\mu)$ such that $x^0 = T$ the symplectic form $\Omega_s$ and volume form $\eta_\Gamma$ induce the forms
\begin{equation}
\hat{\Omega}_s = dp_i\wedge dx^i = {\bf\hat{M}}_a{}^b dP_b\wedge dq^a,
\end{equation}
and
\begin{equation}
\hat{\eta} = dx^1\wedge dx^2\wedge dx^3\wedge dp_1\wedge dp_2\wedge dp_3 
 = \det({\bf\hat{M}}) dq^1\wedge dq^2\wedge dq^3\wedge dP_1\wedge dP_2\wedge dP_3,
\end{equation}
on $\Sigma_T$ respectively, where ${\bf\hat{M}} = ({\bf M}_a{}^b)$ denotes the matrix consisting of the spatial components of the Jacobi matrix defined in Eq.~(\ref{Eq:MDef}). Note that both $\hat{\Omega}_s$ and $\hat{\eta}$ are invariant with respect to the complete lift of the Killing vector field $\partial_t$. For the following, we define $\hat{\Gamma} := T^3\times\hat{\Omega}$.

Now consider a solution $f$ of the Liouville equation~(\ref{Eq:Liouville}) with initial datum $f_0$ supported on the closure of the set $\Sigma_0$. By means of the flow $\varphi^t$ associated with the complete lift $-X_1$ of the Killing vector field $\partial_t$, we can describe the time evolution of the DF as a map $f_0\mapsto f_t$ on $L^1(\Sigma_0,\hat{\eta})$, where
\begin{equation}
f_t(x,p) := f(\varphi^t(x,p)),\qquad t\in\Real,\quad (x,p)\in \Sigma_0.
\end{equation}
On the other hand, it follows from $X_0[f] = 0$, $\frac{d}{dt} f_t = -X_1[f]$ and Eq.~(\ref{Eq:Xalpha}) that
\begin{equation}
f_t(x,p) = [U(t)F](\hat{\Psi}_0(x,p)),\qquad t\in \Real, \quad (x,p)\in \Sigma_0,
\label{Eq:ftSolution}
\end{equation}
with $F := f_0\circ\hat{\Psi}_0^{-1}$ the action-angle representation of the initial datum and where the operator $U(t)$ is defined by
\begin{equation}
(U(t)F)(\ve{q},\ve{P}) = F(\ve{q} - t\boldsymbol{\omega}(\ve{P}),\ve{P}),
\qquad t\in \Real,\quad (\ve{q},\ve{P}) \in \hat{\Gamma},
\label{Eq:UtDef}
\end{equation}
with $\boldsymbol{\omega} := (\omega^1,\omega^2,\omega^3)$ the frequencies defined in Eq.~(\ref{Eq:omegaDef}). Likewise, we define $\boldsymbol{\eta} := (\eta^1,\eta^2,\eta^3)$ with $\eta^a$ given in Eq.~(\ref{Eq:etaDef}).

For the following, we discuss several properties of the flow map~(\ref{Eq:UtDef}) on the function spaces $L^p(\hat{\Gamma},\hat{\eta})$ with measure $\hat{\eta} = \det(\hat{\bf M})d^3 q d^3 P$ and $1\leq p < \infty$. Unless when stated explicitly otherwise, we shall only employ the following properties:
\begin{enumerate}
\item[(i)] The maps $\boldsymbol{\omega},\boldsymbol{\eta}: \hat{\Omega}\to \Real^3$ are $C^1$.
\item[(ii)] The map $\hat{\bf M}: \hat{\Omega}\to \mbox{Mat}(3\times 3,\Real)$ is continuous and $\hat{\bf M}(\ve{P})$ is invertible for all $\ve{P}\in \hat{\Omega}$.
\end{enumerate}
Of course, these conditions are satisfied in our model describing bound Kerr geodesics in the exterior spacetime. In fact, it follows from the explicit representations in Eqs.~(\ref{Eq:omegaDef},\ref{Eq:etaDef}) which allow one to express $\boldsymbol{\omega},\boldsymbol{\eta}$ in terms of analytic functions of the simple roots of the polynomials $R(r)$ and $q(\zeta)$ (see Eqs.~(\ref{Eq:rPlaneRestrictionBis},\ref{Eq:qDef})), that these quantities are analytic in $\ve{P}$. However, the results in the following subsections can be generalized to any model of the same form as~(\ref{Eq:UtDef}) for which $\hat{\Gamma}$ has the form $\hat{\Gamma} = T^d\times\hat{\Omega}$ with $d\in \Natural$ and $\hat{\Omega}$ an open subset of some $\Real^d$ and for which the conditions (i) and (ii) with the dimension $3$ replaced with $d$ are satisfied.

\subsection{Well-posedness of the Cauchy problem and resonant frequencies}

Once it has been brought into the form~(\ref{Eq:UtDef}), the well-posedness of the Cauchy problem for the Liouville equation~(\ref{Eq:Liouville}) follows from the following standard result whose proof is included for the sake of completeness of the presentation.

\begin{lemma}
The map $U(t)$ defined by Eq.~(\ref{Eq:UtDef}) gives rise to a strongly continuous unitary group on $L^p(\hat{\Gamma},\hat{\eta})$.
\end{lemma}

\proof The group properties $U(0) = \id$ and $U(t)\circ U(s) = U(t+s)$ for all $t,s\in\Real$ are obvious. Next, for $F\in L^p(\hat{\Gamma},\hat{\eta})$ and $t\in \Real$ it follows that
$$
\int\limits_{\hat{\Gamma}} \left| [U(t) F](\ve{q},\ve{P}) \right|^p \hat{\eta}
 = \int\limits_{\hat{\Omega}} \int\limits_{T^3} |F(\ve{q} - t\boldsymbol{\omega}(\ve{P}),\ve{P})|^p d^3 q
 \det({\bf\hat{M}}) d^3 P
 = \int\limits_{\hat{\Gamma}} |F(\ve{q},\ve{P})|^p \hat{\eta},
$$
and hence, $U(t): L^p(\hat{\Gamma},\hat{\eta})\to L^p(\hat{\Gamma},\hat{\eta})$ is a well-defined linear map which preserves the norm. Since $U(t)^{-1} = U(-t)$ it follows that it is unitary. Finally, to show strong continuity, first take $F$ to lie in the space $C_0(\hat{\Gamma})$ of continuous functions with compact support. Then, $U(t) F\to F$ as $t\to 0$ follows from Lebesgue's dominated convergence theorem. By the density of $C_0(\hat{\Gamma})$ in $L^p(\hat{\Gamma},\hat{\eta})$ the same property holds for arbitrary $F\in L^p(\hat{\Gamma},\hat{\eta})$.
\qed

\begin{remark}
It follows from Eq.~(\ref{Eq:Xalpha}) that the DF is axisymmetric if and only if $F$ is independent of $q^1$. Furthermore, it is stationary if and only if $F$ is invariant with respect to the flow of $\omega^a\frac{\partial}{\partial q^a}$. Finally, $F$ is invariant under the Carter flow if and only if is invariant with respect to the flow associated with $\eta^a\frac{\partial}{\partial q^a}$. In the next subsection we show that the last two symmetry requirements have rather strong implications.
\end{remark}

For what follows, we consider the two continuous maps ${\bf A},{\bf B}: \hat{\Omega}\to \mbox{Mat}(3\times 3,\Real)$ defined by
\begin{equation}
{\bf A}^{ab} := \frac{\partial \omega^a}{\partial P_c}({\bf\hat{M}}^{-1})_c{}^b,\qquad
{\bf B}^{ab} := \frac{\partial \eta^a}{\partial P_c}({\bf\hat{M}}^{-1})_c{}^b,
\label{Eq:ABDef}
\end{equation}
where we recall that $\omega^a$ and $\eta^a$, $a=1,2,3$, refer to the quantities defined in Eqs.~(\ref{Eq:omegaDef},\ref{Eq:etaDef}) and ${\bf\hat{M}} = ({\bf M}_a{}^b)$ denotes the spatial components of the Jacobi matrix in Eq.~(\ref{Eq:MDef}).

\begin{lemma}
${\bf A}$ and ${\bf B}$ map $\hat{\Omega}$ smoothly on the space of symmetric matrices, i.e. ${\bf A}^{ab} = {\bf A}^{ba}$ and ${\bf B}^{ab} = {\bf B}^{ba}$.
\end{lemma}

\proof Let $\hat{\Omega}'\subset \hat{\Omega}$ be an open subset on which the restriction of the map $\II$ defined in Lemma~\ref{Lem:IIInv} on $\hat{\Omega}$ is injective. On this set we have (see Eq.~(\ref{Eq:MMmT}))
\begin{equation}
\omega^a =\frac{\partial E}{\partial I_a},\qquad
M_H\eta^a =\frac{\partial L}{\partial I_a},
\end{equation}
such that
\begin{equation}
{\bf A}^{ab} = \frac{\partial \omega^a}{\partial P_c}({\bf\hat{M}}^{-1})_c{}^b
 = \frac{\partial \omega^a}{\partial I_b} = \frac{\partial^2 E}{\partial I_a\partial I_b}.
\label{Eq:ARelation}
\end{equation}
Likewise,
\begin{equation}
{\bf B}^{ab} = \frac{\partial \eta^a}{\partial I_b} 
 = M_H^{-1}\frac{\partial^2 L}{\partial I_a\partial I_b},
\label{Eq:BRelation}
\end{equation}
which implies the statement of the lemma.
\qed

The determinants of the maps ${\bf A}$ and ${\bf B}$ will play a fundamental role in the following two subsections. More precisely, we shall consider the following conditions:

\begin{definition}
\label{Def:ABNonDegeneracy}
We shall say that a measurable subset $\hat{C}$ in $\hat{\Omega}$ satisfies the $A$-nondegeneracy condition if
\begin{equation}
\{ \det({\bf A}) = 0 \} \cap \hat{C}
\label{Eq:ANonDegeneracy}
\end{equation}
has zero Lebesgue-measure in $\Real^3$. Likewise, we say $\hat{C}$ satisfies the $B$-nondegeneracy condition if Eq.~(\ref{Eq:ANonDegeneracy}) with ${\bf A}$ replaced by ${\bf B}$ holds.
\end{definition}

\begin{remark}
Clearly, if $\hat{C}\subset \hat{\Omega}$ satisfies the $A$-nondegeneracy condition, so does any measurable subset of $\hat{C}$.
\end{remark}

\begin{remark}
\label{Rem:Analyticity}
For the Kerr problem which is the main focus of this article, the functions $\det({\bf A}), \det({\bf B}): \hat{\Omega}\to \Real$ are real analytic functions, as already noted before. In this case, the following proposition implies that either $\det({\bf A})$ is identically zero or otherwise the $A$-nondegeneracy condition is satisfied on the whole set $\hat{\Omega}$. As we will prove in the next section, $\det({\bf A})$ and $\det({\bf B})$ cannot be identically zero when $a_H\neq 0$, and hence it follows that both the $A$ and $B$-nondegeneracy conditions are satisfied on $\hat{\Omega}$ in the rotating Kerr case.
\end{remark}

\begin{proposition}
\label{Prop:ZeroLevelSetAnalyticFunction}
Let $U\subset \Real^d$ be an open connected subset of $\Real^d$ and let $F: U\to \Real$ be real analytic. Then, either $F$ is identically zero, or the set $\{ x\in U : F(x) = 0 \}$ has zero Lebesgue-measure in $\Real^d$.
\end{proposition}

\proof See, for instance, Ref.~\cite{sM20}.
\qed

For the next result we formulate the following well-known definition and result.

\begin{definition}
A three-tuple of frequencies ${\bf w}\in \Real^3$ is called resonant if there exists ${\bf k}\in \Integer^3\setminus \{ \ve{0} \}$ such that ${\bf k}\cdot {\bf w} = 0$. Otherwise, ${\bf w}$ is called non-resonant.
\end{definition}

\begin{lemma}
\label{Lem:Resonant}
Suppose $\hat{C}\subset \hat{\Omega}$ satisfies the $A$-nondegeneracy condition. Then, the frequencies $\boldsymbol{\omega}(\ve{P})$ are non-resonant for almost all $\ve{P}\in \hat{C}$.
\end{lemma}

\proof Since this result is well-known (cf.~\cite{Arnold-Book}) we only sketch the proof. First, it is not difficult to verify that the set $\mathcal{M}$ of resonant three-tuples ${\bf w}$ is a zero measure set in $\Real^3$. Next, we denote by $\boldsymbol{\omega}'$ the restriction of $\boldsymbol{\omega}$ on $\hat{C}' := \hat{C}\setminus \{ \det({\bf A}) = 0 \}$. Because $\hat{C}$ satisfies the $A$-nondegeneracy condition, the statement of the lemma follows if we can show that the inverse image of $\mathcal{M}$, $(\boldsymbol{\omega}')^{-1}(\mathcal{M})\subset \hat{C}'$, is also a zero-measure set in $\Real^3$.

To show this, we use the inverse function theorem and cover the set $\hat{C}'$ with a countable number of open sets $U_n$ on which $\boldsymbol{\omega}'$ is a local diffeomorphism. Since each set
$$
(\boldsymbol{\omega}')^{-1}(\mathcal{M}) \cap U_n 
$$
has zero measure and
$$
(\boldsymbol{\omega}')^{-1}(\mathcal{M}) \subset  \bigcup\limits_n (\boldsymbol{\omega}')^{-1}(\mathcal{M}) \cap U_n,
$$
the lemma follows.
\qed

\subsection{Strong Jeans theorem}
\label{SubSec:Jeans}

The strong Jeans theorem~\cite{dL62a,MoBoschWhite-Book} states that a stationary solution of the Liouville equation depends only on the constants of motion $E$, $L$ and $L_z$. A priori, this statement appears to be surprising since a function $F$ depending on the combination $\omega^3 q^2 - \omega^2 q^3$, say, is independent of time. However, the  requirement for $F$ to be $2\pi$-periodic in $q^2$ and $q^3$ implies that there cannot be a nontrivial dependency on any of such combinations, if the $A$-nondegeneracy condition holds. This is shown in the next theorem.

\begin{definition}
\label{Def:ABNonDegeneracyF}
Let $F\in L^p(\hat{\Gamma},\hat{\eta})$ with $p\geq 1$. We denote by
\begin{equation}
S(F) := \{ \ve{P}\in \hat{\Omega} : (\ve{q},\ve{P})\in \supp(F) \hbox{ for some $\ve{q}\in T^3$ } \}
\end{equation}
the support of $F$ in $\hat{\Omega}$. We say that $F$ satisfies the $A$-nondegeneracy condition if the set $S(F)$ satisfies the $A$-nondegeneracy condition. Likewise, $F$ satisfies the $B$-nondegeneracy condition if $S(F)$ satisfies the $B$-nondegeneracy condition.
\end{definition}

\begin{theorem}[Strong Jeans theorem]
\label{Thm:StrongJeans}
Suppose $F\in L^1(\hat{\Gamma},\hat{\eta})$ satisfies the $A$-nondegeneracy condition, and suppose in addition that it gives rise to a stationary DF. Then $F$ is independent of the angle variables $q^1$, $q^2$ and $q^3$. 
\end{theorem}

\proof  Denote by $\hat{F}_k$ the Fourier coefficients of $F$:
\begin{equation}
\hat{F}_k(\ve{P}) := \frac{1}{(2\pi)^{3/2}}\int\limits_{T^3} 
F(\ve{q},\ve{P}) e^{-i\ve{k}\cdot\ve{q}} d^3 q,\qquad
\ve{k}\in \Integer^3,\quad \ve{P}\in \hat{\Omega}.
\label{Eq:FkDef}
\end{equation}
According to the assumptions, $\hat{F}_k\in L^1(\hat{\Omega},\det({\bf\hat{M}})d^3 P)$ and $\supp(\hat{F}_k) \subset S(F)$ for all $\ve{k}\in \Integer^3$. Moreover, since $S(F)$ satisfies the $A$-degeneracy condition, Lemma~\ref{Lem:Resonant} implies that $\boldsymbol{\omega}(\ve{P})$ are non-resonant for almost all $\ve{P}\in S(F)$. Since
\begin{equation}
\widehat{U(t) F}_k(\ve{P}) = \frac{1}{(2\pi)^{3/2}}\int\limits_{T^3} 
F(\ve{q} - t\boldsymbol{\omega}(\ve{P}),\ve{P}) e^{-i\ve{k}\cdot\ve{q}} d^3 q
 = \frac{1}{(2\pi)^{3/2}}\int\limits_{T^3} 
F(\ve{q},\ve{P}) e^{-it \ve{k}\cdot\boldsymbol{\omega}(\ve{P})} e^{-i\ve{k}\cdot\ve{q}} d^3 q
 = e^{-it \ve{k}\cdot\boldsymbol{\omega}(\ve{P})}\hat{F}_k(\ve{P}),
\label{Eq:UtFhat}
\end{equation}
the stationarity assumption implies that
\begin{equation}
e^{-i t\ve{k}\cdot\boldsymbol{\omega}(\ve{P})}\hat{F}_k(\ve{P}) = \hat{F}_k(\ve{P})
\end{equation}
for all $t\in \Real$, $\ve{k}\in \Integer^3$ and almost all $\ve{P}\in \hat{\Omega}$. This equality is obviously satisfied for the zero mode $\ve{k} = (0,0,0)$. For $\ve{k}\in \Integer^3\setminus \{ (0,0,0) \}$, differentiate both sides with respect to $t$ and evaluate at $t=0$, giving
\begin{equation}
\ve{k}\cdot\boldsymbol{\omega}(\ve{P}) \hat{F}_k(\ve{P}) = 0
\end{equation}
for almost all $\ve{P}\in \hat{\Omega}$. Since $\ve{k}\cdot\boldsymbol{\omega}(\ve{P})\neq 0$ for almost all $\ve{P}\in S(F)$, this implies that $\hat{F}_k(\ve{P}) = 0$ for almost all $\ve{P}\in \hat{\Omega}$ and thus $\hat{F}_k = 0$ in $L^1(\hat{\Omega})$ for all $\ve{k}\in \Integer^3\setminus \{ \ve{0} \}$. Since an $L^1$-function is uniquely determined by its Fourier coefficients, this implies that
\begin{equation}
F(\ve{q},\ve{P}) = \frac{1}{(2\pi)^{3/2}}\hat{F}_0(\ve{P}) 
 = \frac{1}{(2\pi)^3}\int\limits_{T^3} F(\ve{q},\ve{P}) d^3 q,
\end{equation}
which is independent of $q$. Note that the right-hand side is the angle-average of $F$.
\qed

In complete analogy with the previous theorem, one has the following result which shows that invariance with respect to the Carter flow and the satisfaction of the $B$-nondegeneracy condition also imply that the DF is independent of the angle variables:

\begin{theorem}
\label{Thm:StrongJeansBis}
Suppose $F\in L^1(\hat{\Gamma},\hat{\eta})$ satisfies the $B$-nondegeneracy condition, and suppose in addition that $F$ is invariant with respect to the Carter flow. Then $F$ is independent of the angle variables $q^1$, $q^2$ and $q^3$. 
\end{theorem}

\subsection{Phase space mixing}
\label{SubSec:Mixing}

Phase space mixing can be interpreted as a dynamical version of the strong Jeans theorem, and states that the macroscopic observables associated with a (time-dependent) DF converge in time to those of the angle-averaged DF. Here, we define a "macroscopic observable" to be a time-dependent function of the form
\begin{equation}
N_g(t) := \int\limits_{\Sigma_0} f_t(x,p) g(x,p) \hat{\eta},\qquad t\in \Real,
\end{equation}
with $g: \Sigma_0\to \Real$ a suitable test function on $\Sigma_0$. In view of Eq.~(\ref{Eq:ftSolution}) this is equivalent to
\begin{equation}
N_g(t) = \int\limits_{\hat{\Gamma}} [U(t) F](\ve{q},\ve{P}) G(\ve{q},\ve{P})\hat{\eta},
\end{equation}
where $G := g\circ\hat{\Psi}_0$ denotes the action-angle representation of $g$. The mixing property consists in showing that the macroscopic observable $N_g(t)$ relaxes in time, that is, that the limit $\lim\limits_{t\to\infty} N_g(t)$ exists. The next theorem shows that under suitable regularity assumptions on $F$ and $G$ this is indeed the case provided that $F$ satisfies the $A$-nondegeneracy condition.

\begin{theorem}[Mixing]
\label{Thm:Mixing}
Let $1\leq p < \infty$ and $1 < q\leq\infty$ be such that $1/p + 1/q = 1$. Define $Y_\infty := C_b(\hat{\Gamma})$, the space of bounded continuous functions on $\hat{\Gamma}$, and $Y_q := L^q(\hat{\Gamma},\hat{\eta})$ for $q < \infty$. Suppose $F\in L^p(\hat{\Gamma},\hat{\eta})$ satisfies the $A$-nondegeneracy condition. Denoting by $\overline{F}$ its angle-average, then for all $G\in Y_q$ one has
\begin{equation}
\lim\limits_{t\to \infty}
\int\limits_{\hat{\Gamma}} [U(t) F](\ve{q},\ve{P}) G(\ve{q},\ve{P}) \hat{\eta}
 = \int\limits_{\hat{\Gamma}} \overline{F}(\ve{q},\ve{P}) G(\ve{q},\ve{P}) \hat{\eta}.
\label{Eq:Mixing}
\end{equation}
\end{theorem}

\proof The proof is based on a refinement of the arguments presented in appendix A of Ref.~\cite{pRoS20} (which only treated the case $p=1$ and assumed the frequency map $\boldsymbol{\omega}$ to be $C^2$ instead of $C^1$) which in turn, are based on work by C.~Mitchell~\cite{cM19}.

We start with the symmetric case $p=q=2$. Hence, let $F,G\in L^2(\hat{\Gamma},\hat{\eta})$ and consider their Fourier coefficients $\hat{F}_k,\hat{G}_k$, see Eq.~(\ref{Eq:FkDef}). Using the Cauchy-Schwarz inequality it follows that $\hat{F}_k,\hat{G}_k\in L^2(\hat{\Omega},\det(\hat{\bf M})d^3 P)$ and according to Parseval's identity,
\begin{equation}
\sum\limits_{\ve{k}\in \Integer^3} |\hat{F}_k(\ve{P})|^2 
 = \int\limits_{T^3} | F(\ve{q},\ve{P})|^2 d^3 q,\qquad
\sum\limits_{\ve{k}\in \Integer^3} |\hat{G}_k(\ve{P})|^2 
 = \int\limits_{T^3} | G(\ve{q},\ve{P})|^2 d^3 q,
\end{equation}
for almost all $\ve{P}\in \hat{\Omega}$, which implies that for each $\ve{k}\in \Integer^3$ the function $h_k := \hat{F}_k\hat{G}_k^*: \hat{\Omega}\to \Complex$ belongs to $L^1(\hat{\Omega},\det(\hat{\bf M})d^3 P)$ and satisfies
\begin{equation}
\sum\limits_{\ve{k}\in \Integer^3} \| h_k \|_{L^1(\hat{\Omega},\det(\hat{\bf M})d^3 P)}
 \leq \| F \|_{L^2(\hat{\Gamma},\hat{\eta})} \| G \|_{L^2(\hat{\Gamma},\hat{\eta})}.
\end{equation}
Using Parseval's identity again and the expression in Eq.~(\ref{Eq:UtFhat}) for the Fourier coefficients of $U(t) F$ one obtains
\begin{equation}
\lim\limits_{t\to \infty}
\int\limits_{\hat{\Gamma}} [U(t) F](\ve{q},\ve{P}) G(\ve{q},\ve{P})\hat{\eta}
 = \sum\limits_{k\in \Integer^3}\lim\limits_{t\to \infty}\int\limits_{\hat{\Omega}}
 h_k(\ve{P}) e^{-it\ve{k}\cdot\boldsymbol{\omega}(\ve{P})}\det(\hat{\bf M})d^3 P,
\end{equation}
where we have used the fact that the series converges absolutely to pass the limit below the series. Lemma~\ref{Lem:GRL} below, combined with the observation that the sets $\supp(h_k)\subset \supp(\hat{F}_k)\subset S(F)$ satisfy the $A$-nondegeneracy condition, implies that for each $\ve{k}\neq \ve{0}$ the integrand on the right-hand side converges to zero. Consequently,
\begin{equation}
\lim\limits_{t\to \infty}
\int\limits_{\hat{\Gamma}} [U(t) F](\ve{q},\ve{P}) G(\ve{q},\ve{P})\hat{\eta}
 = \int\limits_{\hat{\Omega}} \hat{F}_0(\ve{P})^*\hat{G}_0(\ve{P})\det(\hat{\bf M})d^3 P
 = (2\pi)^3
 \int\limits_{\hat{\Gamma}} \overline{F}(\ve{q},\ve{P})\overline{G}(\ve{q},\ve{P}) \hat{\eta},
\end{equation}
where we have used the identity $\hat{F}_0(\ve{P}) = (2\pi)^{3/2}\overline{F}(\ve{q},\ve{P})$ in the last step. This proves the theorem for $p=q=2$.

The proof for the remaining cases uses a density argument. According to H\"older's inequality,
$$
\langle F,G \rangle := \int\limits_{\hat{\Gamma}} F(\ve{q},\ve{P}) G(\ve{q},\ve{P}) \hat{\eta}
$$
is well-defined and satisfies $|\langle F,G \rangle | \leq \| F \|_{L^p} \| G \|_{L^q}$ for all $F\in L^p(\hat{\Gamma},\hat{\eta})$ and $G\in L^q(\hat{\Gamma},\hat{\eta})$. With this notation, the statement is equivalent to proving that
$$
\lim_{t\to \infty}\langle U(t) F - \overline{F}, G\rangle = 0
$$
for all $F\in L^p(\hat{\Gamma},\hat{\eta})$ and $G\in Y_q$. Since we already know that the theorem is true for $p=2$ it holds, in particular, for any $F,G\in C_0(\hat{\Gamma})$. We first extend the statement to $G\in Y_q$. If $F\in C_0(\hat{\Gamma})$ and $G\in C_b(\hat{\Gamma})$ we take $\chi\in C_0(\hat{\Gamma})$ which is invariant with respect to $U(t)$ and such that $\chi=1$ on the support of $F$. Then,
\begin{equation}
\langle U(t) F - \overline{F}, G\rangle = \langle U(t) F - \overline{F}, \chi G\rangle \to 0
\end{equation}
as $t\to \infty$, since $\chi G\in C_0(\hat{\Gamma})$. If $F\in C_0(\hat{\Gamma})$ and $G\in L^q(\hat{\Gamma})$ with $1 < q < \infty$ we take a sequence $G_n$ in $C_0(\hat{\Gamma})$ such that $G_n\to G$ in $L^q(\hat{\Gamma},\hat{\eta})$ and note:
\begin{eqnarray*}
| \langle U(t) F - \overline{F}, G\rangle | 
 &\leq& | \langle U(t) F - \overline{F}, G - G_n\rangle | 
 + | \langle U(t) F - \overline{F}, G_n \rangle | \\
 &\leq& 2\| F \|_{L^p} \| G - G_n \|_{L^q} + | \langle U(t) F - \overline{F}, G_n \rangle |,
\end{eqnarray*}
where in the second step we have used the unitarity of $U(t)$ and the  estimate $\| \overline{F} \|_{L^p} \leq \| F \|_{L^p}$ which follows from H\"older's inequality. By first choosing $n$ large, and then $t$ large, we can make the right-hand side arbitrarily small, which proves the theorem for $p > 1$, $F\in C_0(\hat{\Gamma})$ and $G\in L^q(\hat{\Gamma},\hat{\eta})$. Finally, let $F\in L^p(\hat{\Gamma},\hat{\eta})$ and let $F_n$ be a sequence in $C_0(\hat{\Gamma})$ such that $F_n\to F$ in $L^p(\hat{\Gamma},\hat{\eta})$. Then, for any $G\in Y_q$,
\begin{eqnarray*}
| \langle U(t) F - \overline{F}, G\rangle | 
&\leq& | \langle U(t)(F - F_n), G \rangle
 + | \langle U(t) F_n - \overline{F}_n, G\rangle |
 + |\langle \overline{F}_n - \overline{F}, G \rangle | \\
  &\leq& 2\| F - F_n \|_{L^p} \| G \|_{L^q} + | \langle U(t) F_n - \overline{F}_n, G\rangle |.
\end{eqnarray*}
Again, by first choosing $n$ sufficiently large and then $t$ large, the right-hand side is made arbitrarily small, and this concludes the proof of the theorem.
\qed

\begin{lemma}[Generalized Riemann-Lebesgue lemma]
\label{Lem:GRL}
Let $h\in L^1(\hat{\Omega},\det(\hat{\bf M})d^3 P)$ and assume $\supp(h)$ satisfies the $A$-nondegeneracy condition. Then, for all $\ve{k}\in \Integer^3\setminus \{ \ve{0} \}$,
\begin{equation}
\lim_{t\to\infty}\int\limits_{\hat{\Omega}} 
 h(\ve{P}) e^{-i t\ve{k}\cdot\boldsymbol{\omega}(\ve{P})} \det(\hat{\bf M})d^3 P = 0.
\end{equation}
\end{lemma}

\proof According to the assumptions, the set
$$
Z := \supp(h)\cap \{ \det{\bf A} = 0 \}
$$
is closed and has zero-measure. Hence, it is sufficient to prove the statement for $\hat{\Omega}$ replaced by $\hat{\Omega}_0 := \hat{\Omega}\setminus Z$. By density, we can approximate $h$ by functions $h_n\in C_0(\hat{\Omega}_0)$ with compact support $\supp(h_n)\subset\supp(h)\cap\hat{\Omega}_0$. Since $\boldsymbol{\omega}$ has no critical points on $\supp(h_n)$, we can cover the latter by a finite number of open sets $U_m\subset \hat{\Omega}_0$ on which the maps $W_m: U_m\to \Real^3$, $\ve{P}\mapsto \boldsymbol{\omega}(\ve{P})$ are injective. Using the variable substitution $\ve{w} := \boldsymbol{\omega}(\ve{P})$ and recalling the definition~(\ref{Eq:ABDef}) one finds
\begin{equation}
\int\limits_{U_m} h(\ve{P}) e^{-i t\ve{k}\cdot\boldsymbol{\omega}(\ve{P})} \det(\hat{\bf M})d^3 P 
= \int\limits_{W_m(U_m)} \left. \frac{h(\ve{P})}{|\det({\bf A}(\ve{P}))|} \right|_{\ve{P} = W_m^{-1}(\ve{w})}
e^{-i t\ve{k}\cdot\ve{w}} d^3 w.
\end{equation}
By the Riemman-Lebesgue lemma, the right hand side converges to zero as $t\to \infty$ for each $\ve{k}\in \Integer^3\setminus \{ \ve{0} \}$.
\qed

\begin{remark}
We see from the proof of the previous lemma that the decay rate depends on the smoothness of the functions $\ve{P}\mapsto \hat{F}_k(\ve{P})\hat{G}_k^*(\ve{P})/|\det({\bf A}(\ve{P}))|$. This in turn depends on the smoothness properties of the functions $F$ and $G$ along with those of the function $\ve{P}\mapsto 1/|\det({\bf A}(\ve{P}))|$. Therefore, the zeros of the determinant of ${\bf A}$ are expected to play an important role for the determination of the decay rate.
\end{remark}

\begin{remark}
Theorem~\ref{Thm:Mixing} allows one to provide an alternative proof of the strong Jeans theorem. Indeed, if $F\in L^p(\hat{\Gamma},\hat{\eta})$ satisfies the $A$-nondegeneracy condition and is stationary, such that $U(t) F = F$ for all $t\in \Real$, then Eq.~(\ref{Eq:Mixing}) implies that
\begin{equation}
\int\limits_{\hat{\Gamma}} \left[ F(\ve{q},\ve{P}) - \overline{F}(\ve{q},\ve{P}) \right]
G(\ve{q},\ve{P}) \hat{\eta} = 0 
\end{equation}
for all $G\in Y_q$, which implies that $F = \overline{F}$. In this sense the mixing theorem can be interpreted as a dynamical generalization of the strong Jeans theorem. However, note that Theorem~\ref{Thm:StrongJeans} holds under weaker assumptions. Indeed, it is sufficient that almost all frequencies are non-resonant (irrespectively whether or not the non-degeneracy condition holds). For example, it holds even if the frequencies $\boldsymbol{\omega}$ are constant and non-resonant, whereas this property is clearly not sufficient for phase mixing.
\end{remark}

In the next section we provide asymptotic expressions for the maps ${\bf A}$ and ${\bf B}$ defined in Eq.~(\ref{Eq:ABDef}) in the Keplerian limit, and we prove the validity of both the $A$ and $B$-nondegeneracy conditions for bound orbits which lie sufficiently far from the black hole, provided that $a_H\neq 0$. Together with Remark~\ref{Rem:Analyticity} this implies that the $A$- and $B$-nondegeneracy conditions holds for all bound orbits.

\section{Validity of the nondegeneracy conditions using the Keplerian limit}
\label{Sec:DeterminantConditions}

In this section we prove the validity of the $A$ and $B$-nondegeneracy conditions on $\hat{\Omega}$. In principle this could be done by first expressing the frequencies $\boldsymbol{\omega}$ and $\boldsymbol{\eta}$ defined in Eqs.~(\ref{Eq:omegaDef},\ref{Eq:etaDef}) in terms of Legendre's elliptic integrals using the expressions in Eqs.~(\ref{Eq:GG1Def}--\ref{Eq:GG3Def},\ref{Eq:HH1Def}--\ref{Eq:HH3Def}) and then differentiating the result with respect to the integrals of motion $(E,L_z,L)$. However, this would result in rather lengthy expressions for the matrix-valued maps ${\bf A}$ and ${\bf B}$ defined in Eq.~(\ref{Eq:ABDef}), and it is not immediately clear if those would be useful to check the conditions of Definition~\ref{Def:ABNonDegeneracyF}.

For this reason, in this section we pursue a slightly different goal and only compute the maps ${\bf A}$ and ${\bf B}$ in the Keplerian limit. In order to do so, we use the parametrization of the bound orbits in terms of the quantities $(\beta,p,e)\in \mathcal{E}_\alpha$ which are discussed in appendix~\ref{App:OrbitParametrization}. Recall that the Keplerian limit corresponds to $p\to\infty$ with $\beta$ and $e$ kept fixed. The main result of this section is the following:

\begin{theorem}
\label{Thm:Determinant}
Let $0 < \alpha < 1$, $0 < e_0 < e_1 < 1$ and $0 < \beta_0 < \beta_1 < 1$. Then, there exists $p_1 > 0$ sufficiently large such that the set
\begin{equation}
\mathcal{E}_\alpha^\infty
 := \{ (\beta,p,e) :  \beta_0 < |\beta| < \beta_1, p > p_1, e_0 < e < e_1 \}
\end{equation}
is contained in $\mathcal{E}_\alpha$ and such that for any $(\beta,p,e)\in \mathcal{E}_\alpha^\infty$ the corresponding quantity ${\bf P} = (E,L_z,L)$ satisfies
\begin{equation}
\det{\bf A}({\bf P}) > 0,\qquad |\det{\bf B}({\bf P})| > 0.
\end{equation}
\end{theorem}

Together with the observations made in remark~\ref{Rem:Analyticity}, this theorem implies the following important result:

\begin{corollary}
\label{Cor:Mixing}
Suppose $0 < \alpha < 1$. Then, both the $A$- and $B$-nondegeneracy conditions are satisfied on $\hat{\Omega}$.
\end{corollary}

We prove Theorem~\ref{Thm:Determinant} in several steps. In a first step we collect a few useful formulas that allow one to express the action variables in terms of the quantities $(\beta,p,e)$. Next, we show that the roots of the polynomials $R(r)$ defined in Eq.~(\ref{Eq:rPlaneRestrictionBis}) and those of the polynomial $q(\zeta)$ defined in Eq.~(\ref{Eq:qDef}) are analytic in the parameter $\mu := 1/\sqrt{p}$ in a vicinity of $\mu = 0$. This allows one to express all the relevant quantities in power series of $\mu$ which converge uniformly for small enough $|\mu|$. The next step consists in expanding the action variables $I_1$, $I_2$ and $I_3$ in terms of $\mu$ and to show that for small enough $\mu > 0$ the map $(\mu,\beta,e)\mapsto (I_1,I_2,I_3)$ is invertible. In the next step one computes the expansions of $E$ and $L$ and expresses the lowest-order terms as a function of $(I_1,I_2,I_3)$. This allows one to compute the frequencies $\omega^a = \partial E/\partial I_a$ and the matrix $A^{ab} = \partial^2 E/(\partial I_a\partial I_b)$ (see Eqs.~(\ref{Eq:ARelation},\ref{Eq:BRelation})) in the Keplerian limit, up to the desired order of accuracy in $\mu$, and similarly for $\eta^a = \partial L/\partial I_a$ and $B^{ab} = \partial^2 L/(\partial I_a\partial I_b)$. Finally, by means of the resulting expansions for $A^{ab}$ and $B^{ab}$ one shows that their determinants are nonzero for small enough $\mu > 0$.

\subsection{Action variables in terms of the quantities $(\beta,p,e)$}

In appendix~\ref{App:OrbitParametrization} we show that the orbits can be parametrized by the quantities $(\beta,p,e)\in \mathcal{E}_\alpha$ instead of the constants of motion $(\beta,\lambda,\varepsilon)\in \mathcal{D}_\alpha$. These quantities allow one to determine  the four roots $x_1$, $x_2$, $x_3$ and $x_4$ (or, equivalently, $x_1$, $x_2$, $w_+ = x_3 + x_4$ and $w_\times = x_3 x_4$) of $R(r)$ and $(\varepsilon,\lambda)$ in an explicit manner. From this, one can also compute the roots $\zeta_1$ and $\zeta_2$ of the polynomial $q(\zeta)$ determining the polar motion and their ratio $k_1 = \zeta_1/\zeta_2$. From Eqs.~(\ref{Eq:z12Rel1},\ref{Eq:z12Rel2}) one finds, taking into account that $\zeta_1^2 \leq \zeta_2^2$ and using Eqs.~(\ref{Eq:ConsRootRel1},\ref{Eq:ConsRootRel3},\ref{Eq:ConsRootRel4}) the following two expressions:
\begin{eqnarray}
\zeta_1^2 &=& 
\frac{2w_\times}{\alpha^2}
\frac{1}{Z + \sqrt{Z^2 - \frac{4(1-e^2)}{p^2} w_\times}},
\qquad Z := 1 + \frac{2w_+}{p} + \frac{1-e^2}{p^2} w_\times,
\label{Eq:zeta1}\\
k_1^2 &=& \frac{4(1-e^2)w_\times}{p^2}
\frac{1}{\left[ Z + \sqrt{Z^2 - \frac{4(1-e^2)}{p^2} w_\times} \right]^2}.
\label{Eq:k1}
\end{eqnarray}
For the analysis in this section, it is convenient to express the action variables $I_1$, $I_2$ and $I_3$ defined in Eqs.~(\ref{Eq:DefI1}--\ref{Eq:DefI3}) as follows. The integral defining $I_2$ is rewritten in terms of the angle $\phi$ defined in Eqs.~(\ref{Eq:thetaphi},\ref{Eq:pthetaphi}), while the integral defining $I_3$ is written in terms of the new angle $\theta$ defined by
\begin{equation}
x(\theta) := \frac{x_1 + x_2}{2} + \frac{x_2 - x_1}{2}\sin\theta = \frac{p}{1-e^2}(1 + e\sin\theta).
\end{equation}
Recalling that $\zeta_1 = \cos\vartheta_1$ and using Eq.~(\ref{Eq:PolyRRoots}) this yields
\begin{equation}
I_1 = M_H m(\beta\lambda + \alpha\varepsilon),
\qquad
I_2 = M_H m\sqrt{1-\beta^2}\lambda\zeta_1 {\cal K}_2(\alpha,\beta,e,p),
\qquad
I_3 = M_H m\sqrt{1-\varepsilon^2} \frac{p e^2}{1-e^2}  {\cal K}_3(\alpha,\beta,e,p),
\label{Eq:IKeplerian}
\end{equation}
with the integrals
\begin{eqnarray}
{\cal K}_2(\alpha,\beta,e,p) &:=& \frac{1}{\pi} \int\limits_{-\pi/2}^{\pi/2} 
\frac{\sqrt{1 - k_1^2\sin^2\phi}}{1 - \zeta_1^2\sin^2\phi} \cos^2\phi d\phi,
\label{Eq:K2Def}\\
{\cal K}_3(\alpha,\beta,e,p) &:=& \frac{1}{\pi} \int\limits_{-\pi/2}^{\pi/2} 
\frac{ \sqrt{ (1+e\sin\theta)^2 - \frac{1-e^2}{p}(1+ e\sin\theta)w_+  + \frac{(1-e^2)^2}{p^2} w_\times} }{(1+e\sin\theta)^2 - 2\frac{1-e^2}{p} (1+e\sin\theta) + \frac{(1-e^2)^2}{p^2}\alpha^2}
\cos^2\theta d\theta.
\label{Eq:K3Def}
\end{eqnarray}

Before we proceed, it is instructive to recall the Kepler case, which can formally be obtained by taking the leading-order contribution for $p\to \infty$ in the above expressions (this limit will be performed in a rigorous manner in the following subsections). In this case, one obtains $w_+ = 2$, $w_\times = \kappa^2$, $\varepsilon = 1 - (1-e^2)/(2p)$, $\lambda = \sqrt{p}$, $\zeta_1 = \sqrt{1 - \beta^2}$, $k_1 = 0$, such that\footnote{The following integrals will be useful in this section:
$$
\frac{1}{\pi}\int\limits_{-\pi/2}^{\pi/2} \frac{\cos^2\theta d\theta}{1 + e\sin\theta} 
 = \frac{1}{\pi}\int\limits_{-\pi/2}^{\pi/2} \frac{\cos^2\theta d\theta}{1 - e^2\sin^2\theta} 
 = \frac{1}{1 + \sqrt{1-e^2}},\qquad
\frac{1}{\pi}\int\limits_{-\pi/2}^{\pi/2} \frac{\cos^2\theta d\theta}{(1 + e\sin\theta)^2} 
 = \frac{1}{\sqrt{1-e^2}}\frac{1}{1 + \sqrt{1-e^2}}.
$$
\label{Footnote:Integrals}
}
\begin{equation}
I_1 = M_H m\beta\sqrt{p},\qquad
I_2 = M_H m(1 - |\beta|)\sqrt{p},\qquad
I_3 = M_H m\sqrt{p}\left[ \frac{1}{\sqrt{1-e^2}} - 1 \right].
\label{Eq:KeplerI}
\end{equation}
It is useful to replace $I_1$, $I_2$ and $I_3$ with the dimensionless quantities
\begin{equation}
j_1 := \frac{I_1}{M_H m},\qquad
j_2 := \frac{|I_1| + I_2}{M_H m},\qquad
j_3 := \frac{|I_1| + I_2 + I_3}{M_H m},
\label{Eq:jDef}
\end{equation}
such that $(j_1,j_2,j_3) = \sqrt{p}(\beta,1,(1-e^2)^{-1/2})$ and $\varepsilon = 1 - 1/(2j_3^2)$. The fundamental frequencies are
\begin{equation}
\boldsymbol{\omega} = \frac{\partial E}{\partial {\bf I}} 
 = {\bf S}\frac{\partial (m\varepsilon)}{\partial {\bf j}}
 = \omega_{Kepler}\left( \begin{array}{l}
\sign(\beta) \\
1 \\
1
\end{array} \right),
\label{Eq:FreqFormula}
\end{equation}
where we have defined
\begin{equation}
{\bf S} := \frac{1}{M_H m}\left( \begin{array}{ccc} 
1 & \sign(\beta) & \sign(\beta) \\
 0 & 1 & 1 \\
0 & 0 & 1
\end{array} \right),\qquad
\omega_{Kepler} := \frac{1}{M_H}\left[ \frac{1-e^2}{p} \right]^{3/2}.
\label{Eq:SomegaKepler}
\end{equation}

\subsection{Analytic dependency of the roots on the square root of the inverse semi-latus rectum}

The expression for $u_+$ in Eq.~(\ref{Eq:u+Schwarzschild}) in the Schwarzschild limit motivates the following ansatz for large values of $p$ or small values of $\alpha > 0$:
\begin{equation}
u_+ = \frac{p-4}{2(1+e)}\left[ 1 + \frac{\alpha}{\sqrt{p}} h_+ \right].
\label{Eq:u+Ansatz}
\end{equation}
Introduced into Eq.~(\ref{Eq:u+}) this yields the following equation for $h_+$:
\begin{equation}
h_+ - 2\beta A(\alpha,\beta,e,\mu)\sqrt{1 + \alpha\mu B(\alpha,\beta,e,\mu) h_+} 
 - \alpha\mu C(\alpha,\beta,e,\mu) = 0,
\label{Eq:h+}
\end{equation}
where we recall that $\mu = 1/\sqrt{p}$, $\kappa = \alpha\sqrt{1-\beta^2}$, and where we have set
\begin{eqnarray}
A(\alpha,\beta,e,\mu) &:=& \frac{1}{1-\alpha^2\mu^2}
\sqrt{\frac{1-(1-e^2)\mu^2}{(1-4\mu^2)\left[ 1 - \frac{1}{2}\kappa^2(3+e)\mu^2\right]}}
B(\alpha,\beta,e,\mu)^{-1/2},
\\
B(\alpha,\beta,e,\mu) &:=& \left[ 
1 + \frac{\mu^2(1-e^2)(1-\kappa^2\mu^2)}{(1-4\mu^2)[1-(1-e^2)\mu^2][1 - \frac{1}{2}\kappa^2(3+e)\mu^2]}
\right]^{-1},\\
C(\alpha,\beta,e,\mu) &:=& \frac{1}{1-\alpha^2\mu^2}
\frac{1}{1 - \frac{1}{2}\kappa^2(3+e)\mu^2}
\nonumber\\
&\times&\left\{
1 + \frac{1-\beta^2}{2}(3+e)(1-\alpha^2\mu^2)
 + \mu^2\frac{1-e^2 + (1-\beta^2)\left[ 3 + e^2 - \alpha^2(1-e^2)\mu^2 \right] }{1 - 4\mu^2}
\right\}.\qquad
\end{eqnarray}
Note that $A$, $B$ and $C$ are analytic functions of their arguments which are well-defined and positive as long as $-1 < \alpha < 1$, $-1 < \beta < 1$, $0 < e < 1$ and $|\mu| < 1/2$. Furthermore, these functions are even in $\mu$ and they satisfy $A(\alpha,\beta,e,0) = B(\alpha,\beta,e,0) = 1$ and $C(\alpha,\beta,e,0) = 1 + (1-\beta^2)(3+e)/2$, such that $h_+ = 2\beta$ when $\mu = 0$. An explicit expression for $h_+$ is obtained by squaring Eq.~(\ref{Eq:h+}) and solving the resulting quadratic equation. Taking into account that $h_+ = 2\beta$ for $\mu=0$, this yields the following explicit expression for $h_+$:
\begin{equation}
h_+ = \alpha\mu\left( C + 2\beta^2 A^2 B \right) 
 + 2\beta A\sqrt{1 + \alpha^2\mu^2\left( BC + \beta^2 A^2 B^2 \right)}.
\label{Eq:h+Explicit}
\end{equation}
As a consequence of this, we can formulate:

\begin{lemma}
\label{Lem:AnalyticExp1}
Let $\alpha\in [0,1)$. Then, $h_+$, $w_+$, $w_\times$, $\zeta_1$, $k_1$, $\varepsilon$ and $\mu\lambda$ are analytic functions of $\beta$, $e$ and $\mu$ as long as $(\beta,e)\in (-1,1)\times (0,1)$ and $\mu$ is restricted to a small enough open neighborhood of $\mu=0$.
\end{lemma}

\proof The statement for $h_+$ follows directly from Eq.~(\ref{Eq:h+Explicit}) and the aforementioned properties of the functions $A$, $B$ and $C$. Next, using Eqs.~(\ref{Eq:u+Ansatz},\ref{Eq:u+Def}) yields
\begin{equation}
w_+ = \frac{2}{1-4\mu^2}\frac{1 - \kappa^2\mu^2}{1 - \frac{1}{2}\kappa^2(3+e)\mu^2}
\frac{1}{1 + \alpha\mu h_+},
\label{Eq:w+Fromh+}
\end{equation}
which implies the statement for $w_+$. Finally, the statements for $w_\times$, $\zeta_1$, $\zeta_2$, and $(\varepsilon,\mu\lambda)$ follow from this using Eqs.~(\ref{Eq:wmx}), (\ref{Eq:zeta1}), (\ref{Eq:k1}), and (\ref{Eq:epsilonlambda2}), respectively.
\qed

\begin{remark}
To second order in $\mu$ one obtains
\begin{equation}
h_+ = 2\beta + \frac{1}{2}\alpha\left[ 5 + e + \beta^2(1-e) \right]\mu 
 + \beta\left[ 4 + 4\alpha^2 + \kappa^2(2+e) \right]\mu^2 + {\cal O}(\mu^3),
\end{equation}
which together with Eqs.~(\ref{Eq:w+Fromh+},\ref{Eq:wmx},\ref{Eq:zeta1},\ref{Eq:k1}) yields $k_1^2 = {\cal O}(\mu^4)$ and
\begin{eqnarray}
w_+ &=& 
 2\left[ 1 - 2\alpha\beta\mu + (4 + \alpha^2 - 3\kappa^2 )\mu^2 
 - 3\alpha\beta(4-\kappa^2)\mu^3 + {\cal O}(\mu^4) \right],
\label{Eq:w+Kepler}\\
w_\times &=& \kappa^2\left[ 1 - 2\alpha\beta\mu + (4 + \alpha^2 - 2\kappa^2)\mu^2 
 -\alpha\beta(12-\kappa^2)\mu^3 + {\cal O}(\mu^4) \right],
\label{Eq:wxKepler}\\
\zeta_1 &=& \sqrt{1-\beta^2}\left[ 1 - \alpha\beta\mu - \frac{1}{2}\kappa^2\mu^2 
 + 2\alpha\beta\mu^3 +  {\cal O}(\mu^4) \right].
\label{Eq:zeta1Kepler}
\end{eqnarray}
Combined with Eq.~(\ref{Eq:epsilonlambda2}) this also gives
\begin{eqnarray}
\varepsilon &=& 1 - \frac{1}{2}(1-e^2)\mu^2 + \frac{3}{8}(1-e^2)^2\mu^4 
 - \alpha\beta(1-e^2)^2\mu^5 + {\cal O}(\mu^6),
\label{Eq:epsilonExpansion}\\
\lambda &=& \frac{1}{\mu}\left[ 1 - \alpha\beta\mu + \frac{1}{2}(3 + e^2 - \kappa^2)\mu^2 
 - \frac{1}{2}\alpha\beta(5 + 3e^2)\mu^3 + {\cal O}(\mu^4) \right].
\label{Eq:lambdaExpansion}
\end{eqnarray}
\end{remark}

\subsection{Expansion of the action variables}

The next step consists in expanding the action variables $I_1$, $I_2$ and $I_3$ defined in Eqs.~(\ref{Eq:DefI1}--\ref{Eq:DefI3}) in powers of $\mu$. For the following we denote by $U_0$ the open set $U_0 :=  [ (-\beta_1,-\beta_0) \cup (\beta_0,\beta_1) ] \times (e_0,e_1)$ with $0 < \beta_0 < \beta_1 < 1$  and $0 < e_0 < e_1 < 1$ the same constants as in the hypothesis of Theorem~\ref{Thm:Determinant}.  As a consequence of Lemma~\ref{Lem:AnalyticExp1} one has:

\begin{lemma}
\label{Lem:AnalyticExp2}
Let $\alpha\in [0,1)$. Then, $\mu I_1$, $\mu I_2$, $\mu I_3$ are analytic functions of $\beta$, $e$ and $\mu$ as long as $(\beta,e)\in U_0$ and $\mu$ is restricted to a small enough open neighborhood of $\mu=0$.
\end{lemma}

\proof The statement for $I_1$ is a direct consequence of Lemma~\ref{Lem:AnalyticExp1} and the first identity in Eq.~(\ref{Eq:IKeplerian}). Next, to prove the statement for $I_3$ we note that, again as a consequence of Lemma~\ref{Lem:AnalyticExp1} and the expansions~(\ref{Eq:w+Kepler},\ref{Eq:wxKepler}), the function ${\cal K}_3(\alpha,\beta,e,p=\mu^{-2})$ is analytic in $\beta$, $e$ and $\mu$ as long as $(\beta,e)\in U_0$ and $\mu$ is restricted to a neighborhood of $\mu=0$. This observation, together with the fact that $\sqrt{1-\varepsilon^2} = {\cal O}(\mu)$ implies the statement for $I_3$. Finally, in order to analyze $I_2$, we use again Lemma~\ref{Lem:AnalyticExp1} and recall that $k_1^2 =  {\cal O}(\mu^4)$ and $\zeta_1^2 = (1-\beta^2)(1 + {\cal O}(\mu))$, which implies that ${\cal K}_2(\alpha,\beta,e,p=\mu^{-2})$ is analytic in $\beta$, $e$ and $\mu$ as long as $(\beta,e)\in U_0$ and $\mu$ is restricted to a vicinity of $\mu=0$.\footnote{Note that it is at this point that we need to exclude $\beta=0$ from our analysis, since in this case $\zeta_1 = 1$ in the limit $\mu = 0$ (see Eq.~(\ref{Eq:zeta1Kepler})) such that the denominator in the integrand of Eq.~(\ref{Eq:K2Def}) becomes zero when $\theta=\pm \pi/2$.} Now the statement for $I_2$ follows from these observations and the known behavior of $\lambda$ and $\zeta_1$ from Lemma~\ref{Lem:AnalyticExp1}.
\qed

\begin{remark}
Since $k_1^2 =  {\cal O}(\mu^4)$ it follows from Eq.~(\ref{Eq:K2Def}) that
\begin{equation}
{\cal K}_2(\alpha,\beta,e,\mu^{-2}) = \frac{1}{1 + \sqrt{1 - \zeta_1^2}} + {\cal O}(\mu^4),
\end{equation}
which can easily be expanded up to third order in $\mu$ using Eq.~(\ref{Eq:zeta1Kepler}). Likewise, it follows from Eq.~(\ref{Eq:K3Def}) that
\begin{equation}
{\cal K}_3(\alpha,\beta,e,\mu^{-2}) = \frac{1}{1 + \sqrt{1-e^2}}
\left[ 1 + \sqrt{1-e^2}\mu^2(1 + 2\alpha\beta\mu) + {\cal O}(\mu^4) \right],
\end{equation}
where we have used footnote~\ref{Footnote:Integrals} to perform this calculation.
\end{remark}

From the above, one finds for the variables $(j_1,j_2,j_3)$ defined in Eq.~(\ref{Eq:jDef}) the following expansions:
\begin{eqnarray}
j_1 &=& \frac{\beta}{\mu}\left\{ 1 + \frac{\alpha}{\beta}(1-\beta^2)\mu
 + \frac{1}{2}(3+e^2-\kappa^2)\mu^2 
 - \frac{\alpha}{2\beta}\left[ 1 - e^2 + (5 + 3e^2)\beta^2 \right]\mu^3 + {\cal O}(\mu^4) \right\},
\label{Eq:j1}\\
j_2 &=& \frac{1}{\mu}\left[ 1 + \frac{1}{2}(3+e^2)\mu^2 - \alpha\beta(3+e^2)\mu^3 
  + {\cal O}(\mu^4) \right],
\label{Eq:j2}\\
 j_3 &=& \frac{1}{\sqrt{1-e^2}}\frac{1}{\mu}\left[
 1 + \frac{3}{2}\sqrt{1-e^2}\left( 2 - \sqrt{1-e^2} \right)\mu^2 
 - \alpha\beta\sqrt{1-e^2}\left( 2 + \sqrt{1-e^2} \right)\mu^3 + {\cal O}(\mu^4) \right].
\label{Eq:j3}
\end{eqnarray}

\begin{lemma}
\label{Lem:JMapInvertibility}
Let $\alpha\in [0,1)$. Then, for $\mu_1 > 0$ sufficiently small, the map ${\cal J}_\alpha: (0,\mu_1)\times U_0\to \Real^3$, $(\mu,\beta,e)\mapsto (j_1,j_2,j_3)$ is injective.
\end{lemma}

\proof The idea is to write the map ${\cal J}_\alpha$ in the form ${\cal J}_\alpha = {\cal J}_{Kepler}\circ {\cal L}_\alpha$ with ${\cal J}_{Kepler}$ the Kepler map defined by Eqs.~(\ref{Eq:KeplerI},\ref{Eq:jDef}) and to prove that both ${\cal J}_{Kepler}$ and ${\cal L}_\alpha$ are injective when $\mu_1 > 0$ is small enough.

The Kepler map ${\cal J}_{Kepler}: D_1\to D_2$ is defined by 
\begin{equation}
{\cal J}_{Kepler}(\mu,\beta,e) := \frac{1}{\mu}\left( \begin{array}{l}
\beta \\
1 \\
\frac{1}{\sqrt{1-e^2}} 
\end{array} \right),\qquad
(\mu,\beta,e)\in D_1
\end{equation}
with domain $D_1 := (0,\infty)\times (-1,1)\times (0,1)$ and image
\begin{equation}
D_2 := \{ (j_1,j_2,j_3)\in \Real^3 : j_2 > 0, |j_1| < j_2 < j_3 \}.
\end{equation}
Clearly it is invertible; its inverse is given by
\begin{equation}
{\cal J}_{Kepler}^{-1}(j_1,j_2,j_3) 
= \left( \begin{array}{r}
\frac{1}{j_2} \\
\frac{j_1}{j_2} \\
\sqrt{1 - \frac{j_2^2}{j_3^2}}
\end{array} \right),\qquad
(j_1,j_2,j_3)\in D_2.
\end{equation}
Since $I_2$ and $I_3$ are positive it follows that the image of ${\cal J}_\alpha$ lies in $D_2$ (see Eq.~(\ref{Eq:jDef})); hence the map ${\cal L}_\alpha := {\cal J}_{Kepler}^{-1}\circ {\cal J}_\alpha: (0,\mu_1)\times U_0\to \Real^3$ is a well defined differentiable map for sufficiently small $\mu_1 > 0$. Using Eqs.~(\ref{Eq:j1},\ref{Eq:j2},\ref{Eq:j3}) one finds
\begin{equation}
{\cal L}_\alpha(\mu,\beta,e) = \left( \begin{array}{l}
\mu \\
\beta + \alpha(1-\beta^2)\mu\\
e
\end{array} \right) + {\cal O}(\mu^2).
\end{equation}
Its differential satisfies
\begin{equation}
D{\cal L}_\alpha(\mu,\beta,e) = \left( \begin{array}{ccc}
1 & 0 & 0 \\
\alpha(1-\beta^2) & 1 & 0 \\
0 & 0 & 1
\end{array}\right) + {\cal O}(\mu),
\label{Eq:DL}
\end{equation}
and hence
\begin{equation}
\sup\limits_{(\mu,\beta,e)\in (0,\mu_1)\times U_0 } 
\left\| D{\cal L}_\alpha(\mu,\beta,e) - \identy_3 \right\| < 1
\end{equation}
for sufficiently small $\mu_1 > 0$. This condition implies the injectivity of the map ${\cal L}_\alpha: (0,\mu_1)\times U_0\to \Real^3$.
\qed

\begin{remark}
Using the decomposition ${\cal J}_\alpha = {\cal J}_{Kepler}\circ {\cal L}_\alpha$,
\begin{equation}
(D{\cal J}_{Kepler})^{-T} = \mu\left( \begin{array}{rrr}
0 & 1 & 0 \\
-\mu & -\beta & -e^{-1}(1-e^2) \\
0 &  0 & e^{-1}(1-e^2)^{3/2}
\end{array} \right)
\end{equation}
and Eq.~(\ref{Eq:DL}), one finds that
\begin{equation}
\frac{\partial}{\partial j_a} = -{\cal O}(\mu^2)\frac{\partial}{\partial \mu} 
 + {\cal O}(\mu)\frac{\partial}{\partial\beta} + {\cal O}(\mu)\frac{\partial}{\partial e}, 
\end{equation}
and hence differentiating a power series in $\mu$ with respect to $j_a$ augments its order by at least one, that is, $\partial/\partial j_a {\cal O}(\mu^n) = {\cal O}(\mu^{n+1})$ for all $n\in \Natural$ and $a = 1,2,3$.
\end{remark}

\subsection{Expansion of energy in terms of the action variables}

Using the results from the previous subsection, we are ready to prove the following proposition.

\begin{proposition}[Expansion of $E$ in terms of action variables]
\label{Prop:EExpansion}
Let $\alpha\in [0,1)$ and suppose $\mu_1 > 0$ is sufficiently small such that the map ${\cal J}_\alpha: (0,\mu_1)\times U_0\to \Real^3$ from Lemma~\ref{Lem:JMapInvertibility} is injective. Then,
\begin{equation}
\varepsilon = 1 - \frac{1}{2j_3^2} - \frac{3}{j_2 j_3^3} + \frac{15}{8j_3^4} 
 + \frac{2\alpha j_1}{j_2^3 j_3^3} + {\cal O}(\mu^6),
\label{Eq:EExpansion}
\end{equation} 
for all $(j_1,j_2,j_3)$ lying in the image of ${\cal J}_\alpha$.
\end{proposition}

\begin{remark}
The term of order ${\cal O}(j^{-2}) = {\cal O}(\mu^2)$ is the Kepler term (see below Eq.~(\ref{Eq:jDef})), whereas the next-order correction terms of order ${\cal O}(j^{-4}) = {\cal O}(\mu^4)$ are the first relativistic corrections. As we will see, the dominant term that is responsible for the mixing property is the fifth-order correction term $2\alpha j_1/(j_2^3 j_3^3)$.
\end{remark}

\proofof{Proposition~\ref{Prop:EExpansion}}
From Eq.~(\ref{Eq:j3}) we find
\begin{equation}
\frac{1}{j_3} = \mu\sqrt{1-e^2}\left[
 1 - \frac{3}{2}\sqrt{1-e^2}\left( 2 - \sqrt{1-e^2} \right)\mu^2 
 + \alpha\beta\sqrt{1-e^2}\left( 2 + \sqrt{1-e^2} \right)\mu^3 + {\cal O}(\mu^4) \right].
\label{Eq:j3inv}
\end{equation}
This can be used to eliminate the term which is quadratic in $\mu$ in Eq.~(\ref{Eq:epsilonExpansion}). Specifically, we find
\begin{equation}
\varepsilon - 1 + \frac{1}{2j_3^2}
 = \left[ \frac{15}{8}(1-e^2)^2 - 3(1-e^2)^{3/2} \right]\mu^4 + 2\alpha\beta(1-e^2)^{3/2}\mu^5 
  + {\cal O}(\mu^6).
\end{equation}
Next, we use Eq.~(\ref{Eq:j3inv}) again, $1/j_2 = \mu + {\cal O}(\mu^3)$ and $j_1/j_2 = \beta + {\cal O}(\mu)$ in order to eliminate the quartic and fifth-order terms. This yields
\begin{equation}
\varepsilon - 1 + \frac{1}{2j_3^2} - \frac{15}{8j_3^4} + \frac{3}{j_2 j_3^3}
 - \frac{2\alpha j_1}{j_2^3 j_3^3} = {\cal O}(\mu^6),
\end{equation}
which concludes the proof of the proposition.
\qed

Using Eq.~(\ref{Eq:FreqFormula}) and taking into account the previous remark we can compute the corresponding expansion for the fundamental frequencies $\boldsymbol{\omega}$, which yields
\begin{equation}
\boldsymbol{\omega} = {\bf S} \frac{\partial E}{\partial\ve{j}}
 = m {\bf S}\left( \begin{array}{l}
 \frac{2\alpha}{j_2^3 j_3^3} \\
 \frac{3}{j_2^2 j_3^3}\left[ 1 - \frac{2\alpha j_1}{j_2^2} \right] \\
 \frac{1}{j_3^3} - \frac{15}{2j_3^5} + \frac{3}{j_2 j_3^4}\left[ 3 - \frac{2\alpha j_1}{j_2^2} \right]
 \end{array} \right) + {\cal O}(\mu^7),
\end{equation}
where the matrix ${\bf S}$ is defined in Eq.~(\ref{Eq:SomegaKepler}) and where $\ve{j} = (j_1,j_2,j_3)$. Re-expressing this result in terms of $(\beta,p,e)$ using Eqs.~(\ref{Eq:j1},\ref{Eq:j2},\ref{Eq:j3}) yields
\begin{equation}
\boldsymbol{\omega} = 
\left( \begin{array}{l} 
\sign\beta[ \omega_r + \Delta\omega_{perihelion}] + \Delta\omega_{LT} \\
\omega_r + \Delta\omega_{perihelion} \\
\omega_r
\end{array} \right),
\label{Eq:FreqKeplerianLimit}
\end{equation}
where
\begin{eqnarray}
\omega_r &=& \omega_{Kepler}\left[ 1 
- 3\frac{1-e^2}{p}\left( 1 - \frac{\alpha\beta}{\sqrt{p}} \right) 
 + {\cal O}\left(\frac{1}{p^2} \right) \right],
\\
\Delta\omega_{perihelion} &=& \frac{3\omega_{Kepler}}{p}
\left[ 1 - \frac{2\alpha\beta}{\sqrt{p}} + {\cal O}\left(\frac{1}{p} \right) \right],
\\
\Delta\omega_{LT} &=& \frac{2\alpha \omega_{Kepler}}{p^{3/2}}
\left[ 1 + {\cal O}\left(\frac{1}{\sqrt{p}} \right) \right],
\end{eqnarray}
and where we recall the frequency $\omega_{Kepler}$ of the Kepler trajectories defined in Eq.~(\ref{Eq:SomegaKepler}). Hence, as expected, in the limit $p\to \infty$ the three frequencies $\omega_1$, $\omega_2$, $\omega_3$ have equal magnitude. The next-order correction term $1/p$ yields a difference between the frequencies corresponding to the polar and radial motions, given by $\Delta\omega_{perihelion}$, and it describes the perihelion precession. The next-order correction term of oder $1/p^{3/2}$ breaks the degeneracy between the frequencies associated with the azimuthal and polar motion. It describes the Lense-Thirring effect which yields a precession of the line of nodes (the line connecting the points at which the trajectory crosses the equatorial plane) which has frequency $\Delta\omega_{LT}$ (see e.g. section~10.4 in Ref.~\cite{PoissonWill-Book}).

Computing the Hessian $D^2\varepsilon$ of $\varepsilon$ with respect to $\ve{j}$ one finds
\begin{equation}
{\bf A} = {\bf S}(m D^2\varepsilon) {\bf S}^T = m {\bf S}\left[
\left(\begin{array}{c|c|c}
 0 & -\frac{6\alpha}{j_2^4 j_3^3} & -\frac{6\alpha}{j_2^3 j_3^4} \\
\hline
-\frac{6\alpha}{j_2^4 j_3^3} & -\frac{6}{j_2^3 j_3^3} + \frac{24\alpha j_1}{j_2^5 j_3^3}
& -\frac{9}{j_2^2 j_3^4} + \frac{18\alpha j_1}{j_2^4 j_3^4} \\
\hline
 -\frac{6\alpha}{j_2^3 j_3^4} & -\frac{9}{j_2^2 j_3^4} + \frac{18\alpha j_1}{j_2^4 j_3^4} 
 & -\frac{3}{j_3^4} - \frac{36}{j_2 j_3^5} + \frac{75}{2j_3^6} + \frac{24\alpha j_1}{j_2^3 j_3^5}
\end{array} \right) + {\cal O}(\mu^8) \right ]{\bf S}^T.
\end{equation}
Note that the dominant term in the matrix comes from the contribution $-3/j_3^4$ in its $33$-component, which is of the order $\mu^4$. Since $\det({\bf S}) = 1/(M_H m)^3$ it follows that
\begin{equation}
\det({\bf A}) = \frac{108\alpha^2}{M_H^6 m^3}
\frac{1}{j_2^8 j_3^{10}}\left[ 1 +  {\cal O}(\mu) \right].
\end{equation}
As a direct consequence, we have the following lemma which proves the statement of Theorem~\ref{Thm:Determinant} for ${\bf A}$:

\begin{lemma}
Let $\alpha\in (0,1)$. Then, for $\mu_1 > 0$ sufficiently small, it follows that $\det({\bf A}) > 0$ for all $(\mu,\beta,e)\in (0,\mu_1)\times U_0$.
\end{lemma}

\subsection{Expansion of Carter constant in terms of the action variables}

In this final subsection we repeat the steps performed in the previous subsection for the constant of motion $L$ instead of $E$ and verify the validity of the $B$-nondegeneracy condition.

\begin{proposition}[Expansion of $L$ in terms of action variables]
\label{Prop:LExpansion}
Let $\alpha\in [0,1)$ and suppose $\mu_1 > 0$ is sufficiently small such that the map ${\cal J}_\alpha: (0,\mu_1)\times U_0\to \Real^3$ from Lemma~\ref{Lem:JMapInvertibility} is injective. Then,
\begin{equation}
\lambda = j_2  - \frac{\alpha j_1}{j_2} + \frac{\alpha^2}{2j_2}\left( 1 - \frac{j_1^2}{j_2^2} \right)
 + \frac{\alpha j_1}{2j_2}\left[ \frac{1}{j_3^2} 
 + \frac{\alpha^2}{j_2^2}\left( 1 - \frac{j_1^2}{j_2^2} \right) \right]
+ {\cal O}(\mu^3),
\label{Eq:LExpansion}
\end{equation} 
for all $(j_1,j_2,j_3)$ lying in the image of ${\cal J}_\alpha$.
\end{proposition}

\proof We use the same strategy as in the proof of Proposition~\ref{Prop:EExpansion}, and successively eliminate the $\mu$-terms in the expansion~(\ref{Eq:lambdaExpansion}) of $\lambda$ up to the required order.

First, using Eq.~(\ref{Eq:j2}) one finds
\begin{equation}
\lambda - j_2 = -\alpha\beta - \frac{1}{2}\kappa^2\mu 
 + \frac{1}{2}\alpha\beta(1-e^2)\mu^2 + {\cal O}(\mu^3).
\end{equation}
In a next step we use Eqs.~(\ref{Eq:j1},\ref{Eq:j2}) which give
\begin{equation}
\frac{j_1}{j_2} = \beta + \alpha(1-\beta^2)\mu - \frac{1}{2}\kappa^2\beta\mu^2
 + {\cal O}(\mu^3),
\label{Eq:Fracj1j2}
\end{equation}
and one obtains
\begin{equation}
\lambda - j_2 +\frac{\alpha j_1}{j_2} = \frac{1}{2}\kappa^2\mu 
 + \frac{1}{2}\alpha\beta\left[ 1-e^2 - \kappa^2 \right]\mu^2 + {\cal O}(\mu^3).
\end{equation}
Using $j_2^{-1} = \mu[ 1 + {\cal O}(\mu^2) ]$ and again Eq.~(\ref{Eq:Fracj1j2}), and recalling that $\kappa^2 = \alpha^2(1-\beta^2)$ yields
\begin{equation}
\lambda - j_2 + \frac{\alpha j_1}{j_2} - \frac{\alpha^2}{2j_2}\left( 1 - \frac{j_1^2}{j_2^2} \right)
= \frac{1}{2}\alpha\beta\left[ 1-e^2 + \alpha^2(1-\beta^2) \right]\mu^2 + {\cal O}(\mu^3).
\end{equation}
Now the claim follows using Eq.~(\ref{Eq:Fracj1j2}) once again and noting that $(1-e^2)\mu^2 = 1/j_3^2$.
\qed

From this, we compute
\begin{eqnarray}
\boldsymbol{\eta} &=& {\bf S} \frac{\partial L}{\partial\ve{j}}
 = M_H m {\bf S}\left( \begin{array}{l}
 -\frac{\alpha}{j_2} - \frac{\alpha^2 j_1}{j_2^3} + \frac{\alpha}{2j_2}\left[ \frac{1}{j_3^2}
 + \frac{\alpha^2}{j_2^2}\left(1 - \frac{3j_1^2}{j_2^2} \right) \right]
 \\
 1 + \frac{\alpha j_1}{j_2^2} - \frac{\alpha^2}{2j_2^2}\left( 1 - \frac{3j_1^2}{j_2^2} \right)
 - \frac{\alpha j_1}{2j_2^2}\left[ \frac{1}{j_3^2} + \frac{\alpha^2}{j_2^2} 
 \left( 3 - \frac{5j_1^2}{j_2^2} \right) \right]
\\
-\frac{\alpha j_1}{j_2 j_3^3}
\end{array} \right) + {\cal O}(\mu^4)
\nonumber\\
 &=& M_H m {\bf S}\left( \begin{array}{l}
  -\alpha\mu - \alpha^2\beta\mu^2 + \frac{1}{2}\alpha[4 - \alpha^2(1+\beta^2)]\mu^3
\\
1 + \alpha\beta\mu + \frac{\alpha^2}{2}(1+\beta^2)\mu^2 - \alpha\beta(2 - \alpha^2)\mu^3
\\
-\alpha\beta(1-e^2)^{3/2}\mu^3
\end{array} \right) + {\cal O}(\mu^4),
\end{eqnarray}
and
\begin{equation}
{\bf B} = {\bf S}(D^2 L) {\bf S}^T = M_H m {\bf S}\left[
\left(\begin{array}{c|c|c}
 -\frac{\alpha^2}{j_2^3} - \frac{3\alpha^3 j_1}{j_2^5} 
  & \frac{\alpha}{j_2^2} + \frac{3\alpha^2 j_1}{j_2^4} - \frac{\alpha}{2j_2^2}\tau_{3,15}
  &  -\frac{\alpha}{j_2 j_3^3} \\
\hline
\frac{\alpha}{j_2^2} + \frac{3\alpha^2 j_1}{j_2^4} - \frac{\alpha}{2j_2^2}\tau_{3,15}
& -\frac{2\alpha j_1}{j_2^3} + \frac{\alpha^2}{j_2^3}\left( 1 - \frac{6j_1^2}{j_2^2} \right)
 + \frac{\alpha j_1}{j_2^3}\tau_{6,15}
& \frac{\alpha j_1}{j_2^2 j_3^3} \\
\hline
 -\frac{\alpha}{j_2 j_3^3} & \frac{\alpha j_1}{j_2^2 j_3^3} & \frac{3\alpha j_1}{j_2 j_3^4}
\end{array} \right) + {\cal O}(\mu^5) \right ]{\bf S}^T,
\end{equation}
where we have abbreviated $\tau_{n,m} := 1/j_3^2 + \alpha^2(n - mj_1^2/j_2^2)/j_2^2$. Taking into account that the dominant terms in the matrix are the contributions $\alpha/j_2^2$ appearing in the $12$- and $21$-components and $-2\alpha j_1/j_2^3$ appearing in the $22$-component, and recalling that $\det({\bf S}) = 1/(M_H m)^3$, it follows that
\begin{equation}
\det({\bf B}) = -\frac{3\alpha^3}{M_H^3 m^3}\frac{j_1}{j_2}
\frac{1}{j_2^4 j_3^4}\left[ 1 +  {\cal O}(\mu) \right].
\end{equation}
As a direct consequence, we have the following lemma which proves the statement of Theorem~\ref{Thm:Determinant} for ${\bf B}$:

\begin{lemma}
Let $\alpha\in (0,1)$. Then, for $\mu_1 > 0$ sufficiently small, it follows that $|\det({\bf B})| > 0$ for all $(\mu,\beta,e)\in (0,\mu_1)\times U_0$.
\end{lemma}

\section{Conclusions}
\label{Sec:Conclusions}

We have analyzed the phase space mixing of a relativistic, collisionless kinetic gas whose individual gas particles follow spatially bound future-directed timelike geodesics in the exterior of a Kerr black hole of mass $M_H$ and with rotation parameter $a_H$ such that $a_H^2 < M_H^2$. Our main Theorem~\ref{Thm:Mixing} shows that mixing takes place for any DF whose initial data lies in $L^p$ ($1\leq p < \infty$) and test function lying in a suitable function space (which is the dual of $L^p$ when $p >1$), as long as the $A$-nondegeneracy condition is satisfied (see Definitions~\ref{Def:ABNonDegeneracy} and~\ref{Def:ABNonDegeneracyF}). Theorem~\ref{Thm:Determinant} and Corollary~\ref{Cor:Mixing} show that this condition holds if $a_H\neq 0$.

Our proof of Theorem~\ref{Thm:Mixing} exploits the integrability property of the free-particle Hamiltonian on the cotangent bundle associated with the spacetime manifold, and it relies on the construction of generalized action-angle variables $(J_\alpha,Q^\alpha)$ parametrizing the region of phase space $\Gamma_{bound}$ corresponding to bound orbits in the Kerr black hole exterior. More precisely, the $J$-variables label, locally, the invariant sets of topology $\Real\times T^3$ in $\Gamma_{bound}$ while the $Q$-variables provide global coordinates on each of these sets. We have argued that it is convenient to define $J_0$ as a function of the free-particle Hamiltonian, whereas the action variables $\ve{J} := (J_1,J_2,J_3)$ are topological invariants associated with each $S^1$-factor of $T^3$ which are dual to the angle variables $\ve{Q} := (Q^1,Q^2,Q^3)$. Whether or not the $J$-variables provide a globally well-defined labeling of the invariant sets foliating $\Gamma_{bound}$ constitutes an open problem which requires a proof that the transformation $\hat{\II}: \ve{P}\mapsto \ve{J}$ which maps the constants of motion $\ve{P} = (E,L_z,L)$ to the action variables $\ve{J}$ is globally invertible. Although it can be shown that $\hat{\II}$ is locally invertible (cf. Lemma~\ref{Lem:IIInv}) and globally invertible in the Keplerian limit (cf. Lemma~\ref{Lem:JMapInvertibility}) or in the Schwarzschild limit $a_H = 0$, understanding its global invertibility in the general case requires further work, showing, for instance, that this map is proper. To circumvent this problem, we have labeled the invariant sets by the constants of motion $(P_\alpha) = (m,\ve{P})$ instead of $(J_\alpha)$ and worked with the globally-defined (though non-canonical) coordinates $(P_\alpha,Q^\alpha)$ on $\Gamma_{bound}$. Based on these coordinates, the free-particle flow on $\Gamma_{bound}$ still has a simple form which allows one to easily show that the propagator associated with the Vlasov equation is a strongly continuous group. The mixing Theorem~\ref{Thm:Mixing} can then be established using standard Fourier methods generalizing previous work~\cite{cM19,pRoS20}. An immediate corollary of our mixing theorem is the strong Jeans theorem for our setting (see Theorem~\ref{Thm:StrongJeans}), which states that a stationary DF is a function of the integrals of motion only.

Although Theorem~\ref{Thm:Mixing} provides sufficient condition for phase space mixing to take place under rather weak regularity assumptions on the DF and the test function, it provides no information on the decay rates. We have made only brief remarks on what would be involved in showing decay, following the generalized Riemann-Lebesgue lemma~\ref{Lem:GRL}. However, the proof indicates that, apart from the requirement of stronger regularity, the eigenvalues and eigenvector of the matrix ${\bf A}$ may play an important role when analyzing the decay properties. In this regard, we mention that decay results for toy models have recently been obtained in Refs.~\cite{sCjL21,mMpRhB22}, based on the vector field method.

To prove Theorem~\ref{Thm:Determinant} we have shown that the integral of motion corresponding to the energy $E$ of the particle can be expressed in terms of the action variables $\ve{J}$ in the Keplerian limit, and that this dependency can be represented in terms of a power law to any desired accuracy of the parameter $\mu$ which is related to the inverse square root of the semi-latus rectum of the orbit. The gradient of $E$ with respect to $\ve{J}$ yields the fundamental frequencies and its Hessian is the matrix ${\bf A}$ whose determinant is relevant for the $A$-nondegeneracy condition and the mixing. The resulting power-law expansion for the frequency (see Eq.~(\ref{Eq:FreqKeplerianLimit})) has an interesting interpretation: to leading order, the three frequencies are equal in magnitude, which reflects the fact that the Kepler orbits are closed. The next-to-leading term removes the degeneracy between the frequencies associated with the radial and polar motions and describes the perihelion shift (the orbit still taking place within a plane if truncated at this order). The effects of the rotation of the black hole appear to the next-to-next-to-leading order and break the degeneracy between the frequencies associated with the azimuthal and polar motions and describes the Lense-Thirring effect. It is at this order that the determinant of ${\bf A}$ is non-zero and that mixing becomes perceivable. An analogous expansion can be performed for the variable $L$, which represents the square root of the Carter constant. Its Hessian with respect to $\ve{J}$ yields the matrix ${\bf B}$ whose determinant is relevant for the $B$-nondegeneracy condition which is the fundamental hypothesis in Theorem~\ref{Thm:StrongJeansBis}, stating that the imposition of the Carter symmetry implies that the DF is a function of the integrals of motion only. Our power-law expression for $E$ should have an interest beyond the problems analyzed in this work, especially for astrophysical problems in which a star is orbiting the black hole at a radius much larger than its Schwarzschild radius.

The validity of the $A$- and $B$-nondegeneracy conditions in the Kepler limit, combined with the analytic dependency of the fundamental frequencies on the constants of motion and Proposition~\ref{Prop:ZeroLevelSetAnalyticFunction}, allows one to conclude that these conditions are actually valid on the whole space $\Gamma_{bound}$, as long as $a_H\neq 0$. This implies that a kinetic gas configuration propagating in the background of a rotating Kerr black hole exterior settles down to a stationary, axisymmetric configuration which is invariant with respect to the Carter flow. Such configurations have recently been constructed and their properties analyzed, see Refs.~\cite{cGoS22c,cGoS22a}. Furthermore, it has been shown that similar configurations yield stationary and axisymmetric solutions of the full Einstein-Vlasov system bifurcating from the Kerr solution~\cite{fJ22}. It should be interesting to investigate the role of phase space mixing for the stability of these self-gravitating solutions with respect to small perturbations.


\acknowledgments

We are indebted to many colleagues and friends for stimulating discussions. We particularly thank Eloy Ay\'on-Beato, Emilio Tejeda, Hanne Van den Bosch, and Thomas Zannias for enlightening discussions. We also thank Emilio Tejeda for bringing to our attention the possible connection between the frequency shift and the Lense-Thirring effect, Thomas Zannias for comments on a previous version of this manuscript, and Hanne Van den Bosch for pointing out to us Ref.~\cite{sM20}. This research was supported in part by CONACyT Network Project No. 376127 ``Sombras, lentes y ondas gravitatorias generadas por objetos compactos astrof\'isicos", and by a CIC Grant to Universidad Michoacana. We also thank the Erwin Schr\"odinger International Institute for Mathematics and Physics, where part of this work was completed, for hospitality. PR received support from the Center for Mathematical Modeling (Universidad de Chile CNRS IRL 2807) through ANID/Basal projects \#FB210005 , \#ACE210010, and by FONDECYT-ANID postdoctoral grant \#3220767. P.R acknowledges partial support from ``Junior research fellowship" and the thematic programme conference ``Mathematical Perspectives of Gravitation beyond the Vacuum Regime" from Erwin Schr\"odinger International Institute for Mathematics and Physics University of Vienna.

\appendix
\section{Polar motion}
\label{App:Polar}

As follows from Eq.~(\ref{Eq:thetaPlaneRestriction}) the polar motion is confined to the set in the $(\vartheta,p_\vartheta)$-plane determined by the equation
\begin{equation}
p_\vartheta^2 + K(\vartheta) = L^2,
\end{equation}
where here we rewrite the function $K: (0,\pi)\to \Real$ in the form
\begin{equation}
K(\vartheta) = \hat{L}_z^2 + L_z^2\cot^2\vartheta + a_H^2(m^2-E^2)\cos^2\vartheta,
\end{equation}
where we recall that $\hat{L}_z = L_z - a_H E$, see Eq.~(\ref{Eq:Lhatz}). For $L_z\neq 0$ the function $K$ diverges as $\vartheta\to 0,\pi$. When $a_H^2(E^2-m^2) \leq L_z^2$ this function has a global minimum at $\vartheta = \pi/2$, where $K(\pi/2) = \hat{L}_z^2$, so in this case the polar motion is described by a closed curve in the $(\vartheta,p_\vartheta)$-plane for each $L > |\hat{L}_z|$.

When $a_H^2(E^2-m^2) > L_z^2$, the function $K$ has a local maximum at $\vartheta = \pi/2$ and two global minima at $\vartheta = \vartheta_*$ and $\vartheta = \pi - \vartheta_*$, where $\vartheta_*\in (0,\pi/2)$ is determined by the equation
\begin{equation}
\sin^2\vartheta_* = \frac{|L_z|}{|a_H|\sqrt{E^2 - m^2}}.
\end{equation}
The value of $K$ at these minima is
\begin{equation}
K(\vartheta_*) = K(\pi - \vartheta_*) = \hat{L}_z^2 - \left( |a_H|\sqrt{E^2 - m^2} - |L_z| \right)^2.
\end{equation}
For the scenarios considered in the present article, $E^2 < m^2$, since we only consider bound orbits; hence only the first case where $K$ has a global minimum at $\vartheta = \pi/2$ is relevant.

\section{Radial motion}
\label{App:Radial}

In this appendix, we discuss the qualitative properties of the effective potential $W_+$ defined in Eq.~(\ref{Eq:Wpm}) describing the radial motion of future-directed timelike geodesics in the Kerr exterior spacetime. For related discussions, see for example~\cite{Chandrasekhar-Book,ONeill-Book,eTpTjM13,fJ22}.

In terms of the dimensionless variables
\begin{equation}
x := \frac{r}{M_H},\qquad
\varepsilon := \frac{E}{m},\qquad
\lambda := \frac{L}{M_H m},\qquad
\hat{\lambda}_z := \frac{\hat{L}_z}{M_H m},
\label{Eq:DimLessVariables}
\end{equation}
and $\alpha := a_H/M_H$, $\beta := \hat{L}_z/L$ we have $W_+(r) = m w_{\alpha,\beta,\lambda}(x)$ with the smooth function $w_{\alpha,\beta,\lambda}: (x_+,\infty)\to \Real$ defined by
\begin{equation}
w_{\alpha,\beta,\lambda}(x) = \frac{\alpha\beta\lambda}{x^2} 
 + \sqrt{\left( 1 - \frac{2}{x} + \frac{\alpha^2}{x^2} \right)\left( 1 + \frac{\lambda^2}{x^2}\right)},
\qquad
x > x_+ := 1 + \sqrt{1-\alpha^2}.
\label{Eq:wDef}
\end{equation}
Notice that $\lim_{x\to x_+} w_{\alpha,\beta,\lambda}(x) = \alpha\beta\lambda/x_+^2$ which can have either sign, while for large $x$,
\begin{equation}
w_{\alpha,\beta,\lambda}(x) = 1 - \frac{1}{x} + {\cal O}\left( \frac{1}{x^2} \right).
\end{equation}

\subsection{Spherical orbits}

\begin{figure}[ht]
\centerline{\resizebox{10.0cm}{!}{\includegraphics{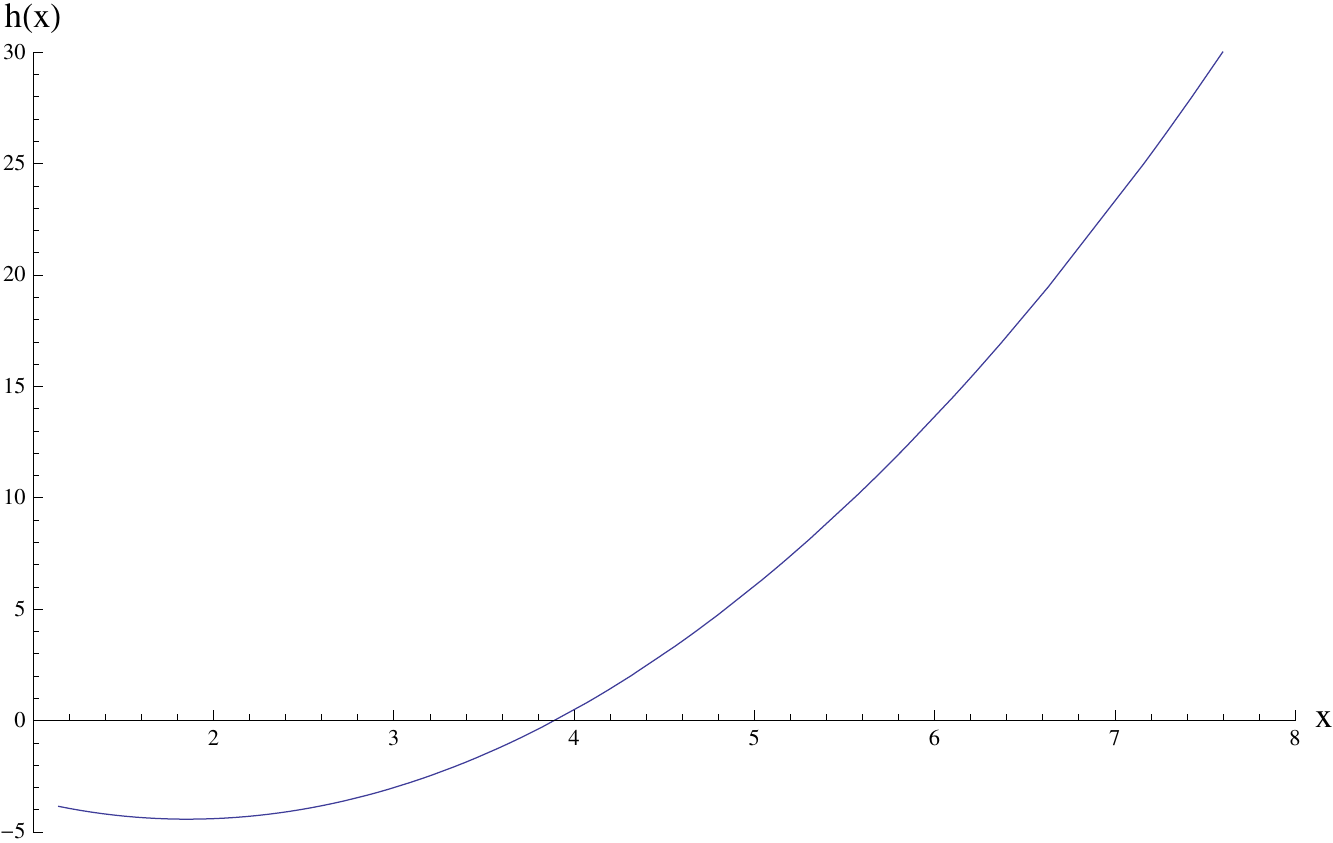}}}
\caption{The function $h: (x_+,\infty)\to \Real$ for the parameter values $\alpha = 0.99$ and $\beta = -0.95$. Note that $h$ is not monotonous for these values; nevertheless it has a unique zero.}
\label{Fig:h}
\end{figure}

In order to determine the qualitative properties of $w_{\alpha,\beta,\lambda}$ we first discuss its critical points, corresponding to spherical orbits (that is, orbits taking place within a sphere of constant radial coordinate $r$). For this, suppose $x > x_+$ is a critical point of $w_{\alpha,\beta,\lambda}$. Then,
\begin{eqnarray}
0 &=& 
 x^3\sqrt{(x^2 - 2x + \alpha^2)(x^2 + \lambda^2)}\frac{d}{dx} w_{\alpha,\beta,\lambda}(x) 
\nonumber\\
 &=& -2\alpha\beta\lambda\sqrt{(x^2 - 2x + \alpha^2)(x^2 + \lambda^2)} 
  + x^2(x - \alpha^2) - (x^2 - 3x + 2\alpha^2)\lambda^2.
\label{Eq:wCritical}
\end{eqnarray}
Taking the square on both sides and eliminating the square root one obtains a bi-quadratic equation for $\lambda$ of the form
\begin{equation}
A\lambda^4 - 2B\lambda^2 + C = 0,
\label{Eq:QuadraticLambda}
\end{equation}
with coefficients
\begin{eqnarray*}
A &=& x^2(x-3)^2 - 4\alpha^2 x + 4\alpha^2(1-\beta^2)\overline{\Delta}(x),\\
B &=& x^4(x-3) + \alpha^2 x^3(x+1) - 2\alpha^2(1-\beta^2)x^2\overline{\Delta}(x),\\
C &=& x^4(x-\alpha^2)^2,
\end{eqnarray*}
where for convenience we have set $\overline{\Delta}(x) := x^2 - 2x + \alpha^2$. The discriminant yields
\begin{equation}
D := B^2 - AC = 4\alpha^2\beta^2 x^4\overline{\Delta}(x)^2\left[ x - \alpha^2(1-\beta^2) \right],
\end{equation}
and is positive since $x > x_+ \geq 1$ and $0\leq \alpha^2(1-\beta^2)\leq 1$. Therefore, Eq.~(\ref{Eq:QuadraticLambda}) has two real solutions for $\lambda^2$, given by $\lambda^2 = (B\pm \sqrt{D})/A$. Observing the fact that
\begin{equation}
B + x^2 A = x^2\overline{\Delta}(x)\left[ x(x-3) + 2\alpha^2(1-\beta^2) \right]
\end{equation}
and the factorization
\begin{equation}
A = \left[ x(x-3) + 2\alpha^2(1-\beta^2) + 2\alpha\beta\sqrt{x - \alpha^2(1-\beta^2)} \right]
\left[ x(x-3) + 2\alpha^2(1-\beta^2) - 2\alpha\beta\sqrt{x - \alpha^2(1-\beta^2)} \right],
\end{equation}
one obtains
\begin{eqnarray}
\lambda^2 &=& x^2\left[ -1 + \frac{\overline{\Delta}(x)}
{x(x-3) + 2\alpha^2(1-\beta^2) \mp 2\alpha\beta\sqrt{x - \alpha^2(1-\beta^2)} } \right]
\nonumber\\
 &=& \frac{ x^2\left[ \sqrt{x - \alpha^2(1 - \beta^2)} \pm \alpha\beta \right]^2 }
 {x(x-3) + 2\alpha^2(1-\beta^2) \mp 2\alpha\beta\sqrt{x - \alpha^2(1-\beta^2)} }.
\label{Eq:lambda2}
\end{eqnarray}
Since $\sqrt{x - \alpha^2(1-\beta^2)} > \sqrt{1 - \alpha^2 + \alpha^2\beta^2} \geq |\alpha\beta|$ the numerator in Eq.~(\ref{Eq:lambda2}) is always positive, and hence the denominator needs to be positive as well for $\lambda^2 > 0$ to be well-defined. By introducing Eq.~(\ref{Eq:lambda2}) back into Eq.~(\ref{Eq:wCritical}) one can check that the correct solution is the one belonging to the lower sign. Therefore, we conclude that
\begin{equation}
\lambda = x\frac{\sqrt{x - \alpha^2(1 - \beta^2)} - \alpha\beta}{\sqrt{h(x)}},\qquad
h(x) > 0,
\label{Eq:lambda}
\end{equation}
with the smooth function $h: (x_+,\infty)\to \Real$ defined by
\begin{equation}
h(x) := x(x-3) + 2\alpha^2(1-\beta^2) + 2\alpha\beta\sqrt{x - \alpha^2(1-\beta^2)},\qquad
x > x_+.
\label{Eq:hDef}
\end{equation}
The next lemma shows the relevant behavior for the function $h$ that will be used in the following:

\begin{lemma}
\label{Lem:ph}
The function $h: (x_+,\infty)\to \Real$ defined by Eq.~(\ref{Eq:hDef}) has a unique zero at some point $x_{ph} = x_{ph}(\alpha,\beta) > x_+$, and it is strictly negative on the interval $(x_+,x_{ph})$ and strictly positive on the interval $(x_{ph},\infty)$ (see Fig.~\ref{Fig:h}).
\end{lemma}

\proof First, we observe that
\begin{eqnarray*}
\lim\limits_{y\to x_+} h(y) &=& -x_+ + \alpha^2 
 + 2\alpha\beta\left( \sqrt{x_+ - \alpha^2(1-\beta^2)} - \alpha\beta \right)\\
 &\leq& -(x_+ - \alpha^2) + 2\alpha^2\beta^2
  \left( \sqrt{ 1 + \frac{x_+-\alpha^2}{\alpha^2\beta^2}} - 1 \right) < 0,
\end{eqnarray*}
by virtue of the inequality $\sqrt{1 + c^2} - 1 < c^2/2$ which holds for all $c > 0$. On the other hand,
$$
\lim\limits_{y\to \infty} h(y) = +\infty;
$$
hence $h$ has at least one zero on the interval $(x_+,\infty)$. It remains to show that this zero is unique. For this we compute
$$
\frac{dh}{dy}(y) = 2y - 3 + \frac{\alpha\beta}{\sqrt{y - \alpha^2(1-\beta^2)}}.
$$
Combining this with Eq.~(\ref{Eq:hDef}) we obtain the identity
$$
y^3\frac{d}{dy}\left[ \frac{h(y)}{y^2} \right] = y\frac{dh}{dy}(y) - 2h(y)
 = \left[ 3y - 4\alpha^2(1-\beta^2)\right]
\left[ 1 - \frac{\alpha\beta}{\sqrt{y - \alpha^2(1-\beta^2)}} \right],\qquad
y > x_+.
$$
The second parenthesis on the right-hand side is positive since $\sqrt{y - \alpha^2(1-\beta^2)} > |\alpha\beta|$. The first parenthesis is negative for $y < y_1 := 4\alpha^2(1-\beta^2)/3$ and positive for $y > y_1$. Therefore, the function $h(y)/y^2$ decreases for $x_+ < y < y_1$ (as far as this interval is non-empty) and increases for $y > y_1$. Since $h(y)/y^2$ is negative for $y$ close to $x_+$ this implies that the function $h$ can have only one zero.
\qed

\begin{remark}
$r = M_H x_{ph}(\alpha,\beta)$ corresponds to the radius of spherical photon orbits. 
\end{remark}

Concluding from what we have obtained so far, the function $w_{\alpha,\beta,\lambda}$ has a critical point at $x$ only if $x > x_{ph}(\alpha,\beta)$. In this case, the dimensionless angular momentum and energy of the orbit are given by
\begin{eqnarray}
\lambda_{sph}(x) &=& x\frac{\sqrt{x - \alpha^2(1-\beta^2)} - \alpha\beta}{\sqrt{h(x)}},
\label{Eq:lambdasph}\\
\varepsilon_{sph}(x) &=& \frac{x^2 - 2x + \alpha^2(1-\beta^2)
 + \alpha\beta\sqrt{x - \alpha^2(1-\beta^2)}}{x\sqrt{h(x)}}
 = \frac{\sqrt{h(x)}}{x} + \frac{\sqrt{x - \alpha^2(1-\beta^2)}}{x^2}\lambda_{sph}(x),
\label{Eq:epsilonsph}
\end{eqnarray}
respectively. For large $x$ one finds
\begin{eqnarray}
\lambda_{sph}(x) &=& \sqrt{x}
\left[ 1 - \frac{\alpha\beta}{\sqrt{x}} + {\cal O}\left( \frac{1}{x} \right) \right],
\label{Eq:lambdasphAsym}\\
\varepsilon_{sph}(x) &=& 1 - \frac{1}{2x} +  {\cal O}\left( \frac{1}{x^2} \right),
\label{Eq:epsilonsphAsym}
\end{eqnarray}
and the leading order terms coincide with the Newtonian expressions (taking into account the rest energy of the particle), as expected. Note also that $\lambda_{sph}(x)$ and $\varepsilon_{sph}(x)$ are positive for all $x > x_{ph}(\alpha,\beta)$.

\subsection{Global behaviour of the functions $\lambda_{sph}$ and $\varepsilon_{sph}$}

Next, we analyze the behavior of the functions $\lambda_{sph}$ and $\varepsilon_{sph}$ as $x$ increases from $x_{ph}$ to $\infty$. First, we observe that $\lambda_{sph}(x)\to\infty$ as $x\to x_{ph}$ or $x\to \infty$; hence $\lambda_{sph}$ must have a global minimum at some $x = x_{ms}$. In order to determine this minimum we compute
\begin{equation}
\lambda_{sph}'(x) = \frac{G(x)}{2h(x)^{3/2}\sqrt{x - \alpha^2(1-\beta^2)}},\qquad
G(x) = x\overline{\Delta}(x) - [h(x) - x(x-1)]^2.
\label{Eq:GDef}
\end{equation}
The next lemma gives the required behavior of the function $G$.

\begin{lemma}
\label{Lem:ms}
The function $G: [x_{ph},\infty)\to \Real$ defined in Eq.~(\ref{Eq:GDef}) has a unique zero $x_{ms} = x_{ms}(\alpha,\beta)$ in the interval $(x_{ph},\infty)$. Further, $G$ is negative for $x < x_{ms}$ and positive for $x > x_{ms}$.
\end{lemma}

\proof First, note that $G$ is smooth, diverges to $\infty$ as $x\to \infty$, and satisfies
$$
G(x_{ph}) = \left[ x\overline{\Delta}(x) - x^2(x-1)^2 \right]_{x = x_{ph}}
 = -x_{ph}\left[ (x_{ph}-1)^3 + 1 - \alpha^2 \right] < 0.
$$
Hence, it has at least one zero in the interval $(x_{ph},\infty)$. Next, a short calculation reveals that
\begin{equation}
G'(x) = 3\left[ 2h(x) - \overline{\Delta}(x) \right].
\end{equation}
Suppose now $x$ is a zero of $G$. Since
\begin{equation}
h(x) - x(x-1) = -2\sqrt{x - \alpha^2(1-\beta^2)}
\left[ \sqrt{x - \alpha^2(1-\beta^2)} - \alpha\beta \right] < 0,
\label{Eq:hIdentity}
\end{equation}
it follows from the definition of $G$ in Eq.~(\ref{Eq:GDef}) that
\begin{equation}
h(x) = x(x-1) - \sqrt{x\overline{\Delta}(x)},
\label{Eq:hms}
\end{equation}
and thus
\begin{equation}
\frac{1}{3}\left. G'(x) \right|_{G(x) = 0} = x^2 - \alpha^2 - 2\sqrt{x\overline{\Delta}(x)} =: H(x).
\end{equation}
Finally, we claim that $H(x) > 0$ for all $x > x_+$, which implies that the function $G$ can only cross zero from below and hence can have only one zero.

To prove that $H(x)$ is positive for all $x > x_+$ we first notice that $H(x_+) = x_+^2 - \alpha^2 > 0$ and that $H(x)\to \infty$ as $x\to \infty$. Next, let $x_0$ be a global minimum of $H$ on the interval $[x_+,\infty)$. Since
\begin{equation}
H'(x) = 2x - \frac{1}{x\overline{\Delta}(x)}(3\overline{\Delta}(x) + 2x - 2\alpha^2),
\end{equation}
such that $\lim_{x\to x_+} H'(x) = -\infty$ and $\lim_{x\to \infty} H'(x) = \infty$, this minimum cannot be located at $x = x_+$. Hence, $x_0 > x_+$ satisfies $H'(x_0) = 0$ and 
\begin{equation}
H(x_0) = x_0^2 - \alpha^2 - \frac{1}{x_0}(3x_0^2 - 4x_0 + \alpha^2)
 = \frac{1}{x_0}\left[ (x_0-1)^3 + (x_0 + 1)(1-\alpha^2) \right] > 0,
\end{equation}
which proves that $H(x) > 0$ for all $x\geq x_+$. This concludes the proof of the Lemma.
\qed

\begin{remark}
For equatorial orbits $\beta^2 = 1$, the equation determining $x_{ms}$ simplifies to $2(\sqrt{x_{ms}} - \alpha\beta) = \sqrt{\overline{\Delta}(x_{ms})}$ which implies $h(x_{ms}) = 3(\sqrt{x_{ms}} - \alpha\beta)^2$.
\end{remark}

\begin{figure}[ht]
\centerline{\resizebox{8.5cm}{!}{\includegraphics{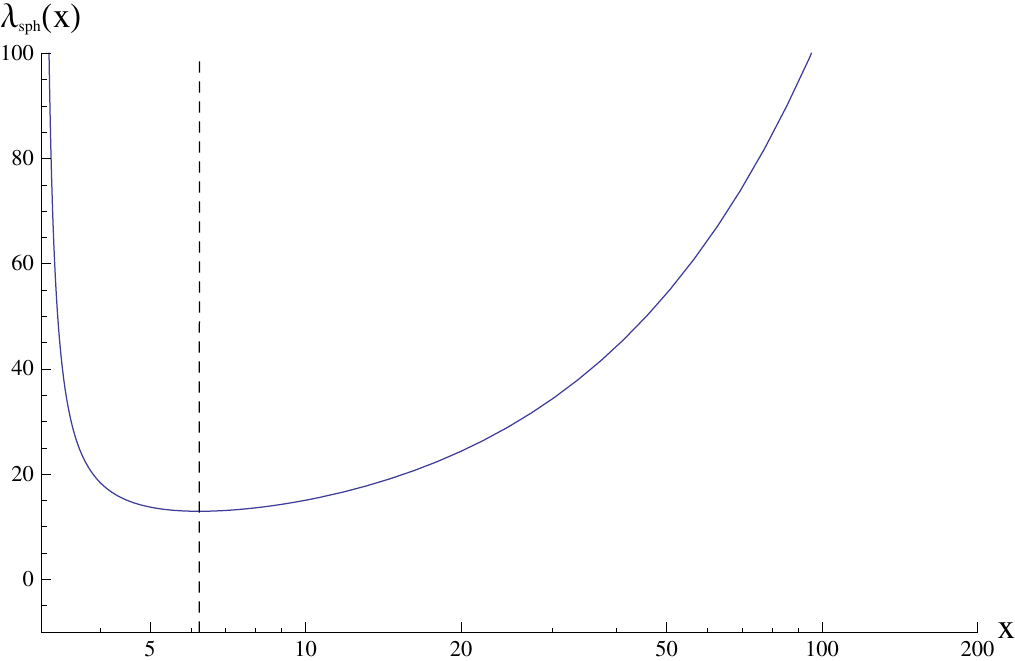}}
\resizebox{8.5cm}{!}{\includegraphics{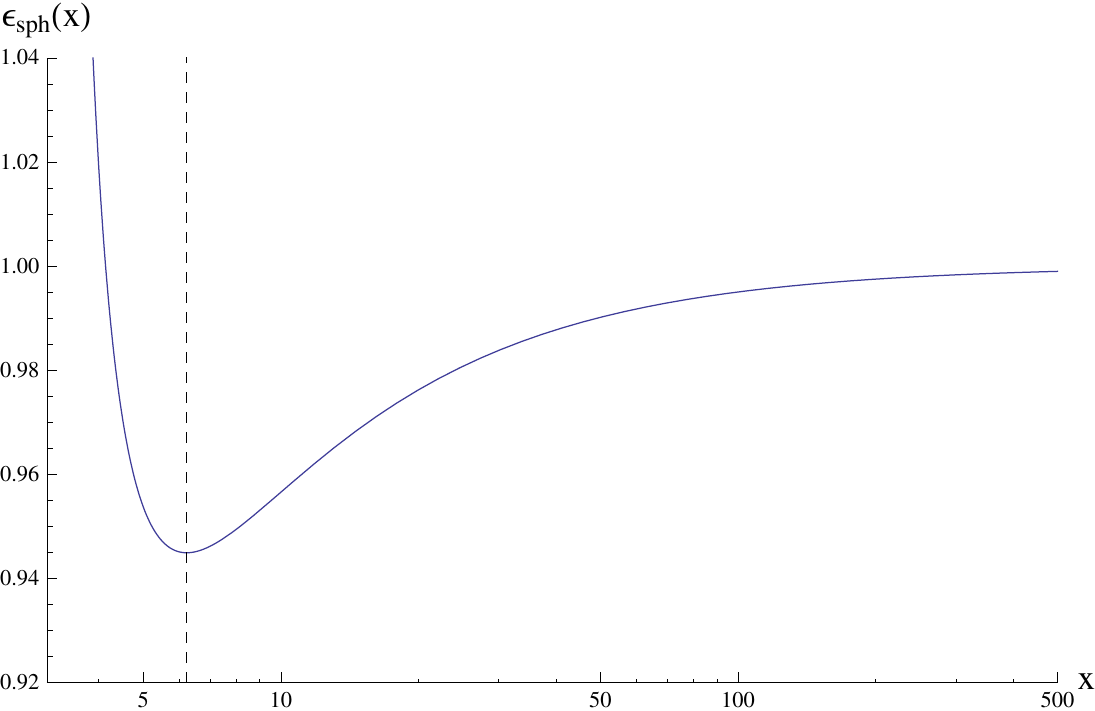}}
}
\caption{Plot of the functions $\lambda_{sph},\varepsilon_{sph}: (x_{ph},\infty)\to \Real$ for the parameter values $\alpha = 0.12$ and $\beta = -0.61$. The vertical dashed lines indicate the location of the marginally stable orbit $x_{ms} \approx 6.223$ for these values, which correspond to the minima of both functions.}
\label{Fig:sph}
\end{figure}

It follows from the previous Lemma that as $x$ increases from $x_{ph}$ to $x_{ms}$ and then from $x_{ms}$ to $\infty$ the function $\lambda_{sph}: (x_{ph},\infty) \to \Real$ decreases monotonically from $\infty$ to its global minimum $\lambda_{ms} := \lambda_{sph}(x_{ms})$ and then increases again monotonically to $\infty$. Next, a somehow long but straightforward calculation reveals that
\begin{equation}
\varepsilon_{sph}'(x) = \frac{\sqrt{x - \alpha^2(1-\beta^2)}}{x^2}\lambda_{sph}'(x),
\label{Eq:epsilonlambdaDeriv}
\end{equation}
which shows that as $x$ increases from $x_{ph}$ to $x_{ms}$ and then from $x_{ms}$ to $\infty$ the function $\varepsilon_{sph}: (x_{ph},\infty) \to \Real$ decreases monotonically from $\infty$ to $\varepsilon_{ms} := \varepsilon_{sph}(x_{ms})$ and then increases monotonically to $1$. The typical behavior of the functions $\lambda_{sph}$ and $\varepsilon_{sph}$ are shown in Fig.~\ref{Fig:sph}.

\subsection{Global behaviour of the effective potential $w_{\alpha,\beta,\lambda}$}

As a consequence of the inferred behavior for the function $\lambda_{sph}: (x_{ph},\infty)\to \Real$, given any value for $\lambda > \lambda_{ms}$, there exist precisely two spherical orbits at $x_{max}\in (x_{ph},x_{ms})$ and $x_{min}\in (x_{ms},\infty)$, respectively, such that
\begin{equation}
\lambda_{sph}(x_{min}) = \lambda_{sph}(x_{max}) = \lambda.
\end{equation}
A long calculation shows that
\begin{equation}
\left. w_{\alpha,\beta,\lambda}''(x) \right|_{\lambda = \lambda_{sph}(x)}
 = \frac{G(x)}{x^3\sqrt{h(x)} \overline{\Delta}(x)},
\end{equation}
with $G$ the function defined in Eq.~(\ref{Eq:GDef}). Therefore, $x_{max}$ corresponds to a local maximum of $w_{\alpha,\beta,\lambda}$ and $x_{min}$ to a local minimum. For $\lambda < \lambda_{ms}$ the function $w_{\alpha,\beta,\lambda}$ has no critical points and is monotonously increasing. We summarize our finding in the following

\begin{lemma}[Global behaviour of the effective potential]
\label{Lem:EffectivePotential}
Let $\alpha\in [0,1)$ and $\beta\in [-1,1]$ and denote by $\lambda_{ms} = \lambda_{ms}(\alpha,\beta)$ the global minimum of the function $\lambda_{sph}$ defined in Eq.~(\ref{Eq:lambdasph}). Then, the effective potential $w_{\alpha,\beta,\lambda}$ defined in Eq.~(\ref{Eq:wDef}) has the following qualitative behaviour:
\begin{enumerate}
\item[(i)] If $0 < \lambda <  \lambda_{ms}$, it is monotonously increasing.
\item[(ii)] If $\lambda > \lambda_{ms}$, it has a local maximum at $x_{max}\in (x_{ph},x_{ms})$ and a local minimum at $x_{min}\in (x_{ms},\infty)$, where $x_{max}$ and $x_{min}$ are uniquely determined by the equation $\lambda_{sph}(x) = \lambda$. Define the corresponding energy values $\varepsilon_{max} := \varepsilon_{sph}(x_{max})$ and $\varepsilon_{min} := \varepsilon_{sph}(x_{min})$. As $\lambda$ increases from $\lambda_{ms}$ to $\infty$, $x_{max}$ decreases from $x_{ms}$ to $x_{ph}$ and $\varepsilon_{max}$ increases from $\varepsilon_{ms} := \varepsilon_{sph}(x_{ms})$ to $\infty$ while $x_{min}$ increases from $x_{ms}$ to $\infty$ and $\varepsilon_{min}$ increases from $\varepsilon_{ms}$ to $1$. 
\end{enumerate}
\end{lemma}

\begin{figure}[ht]
\centerline{\resizebox{8.0cm}{!}{\includegraphics{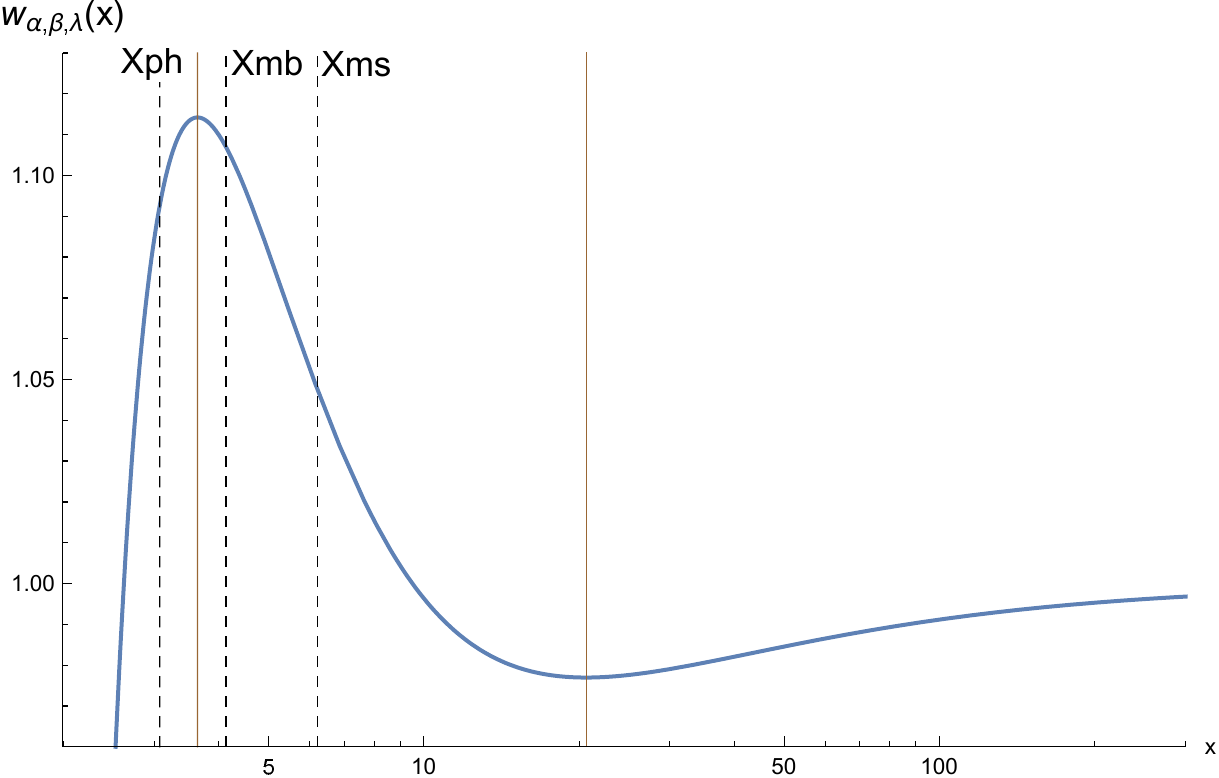}}
\resizebox{8.0cm}{!}{\includegraphics{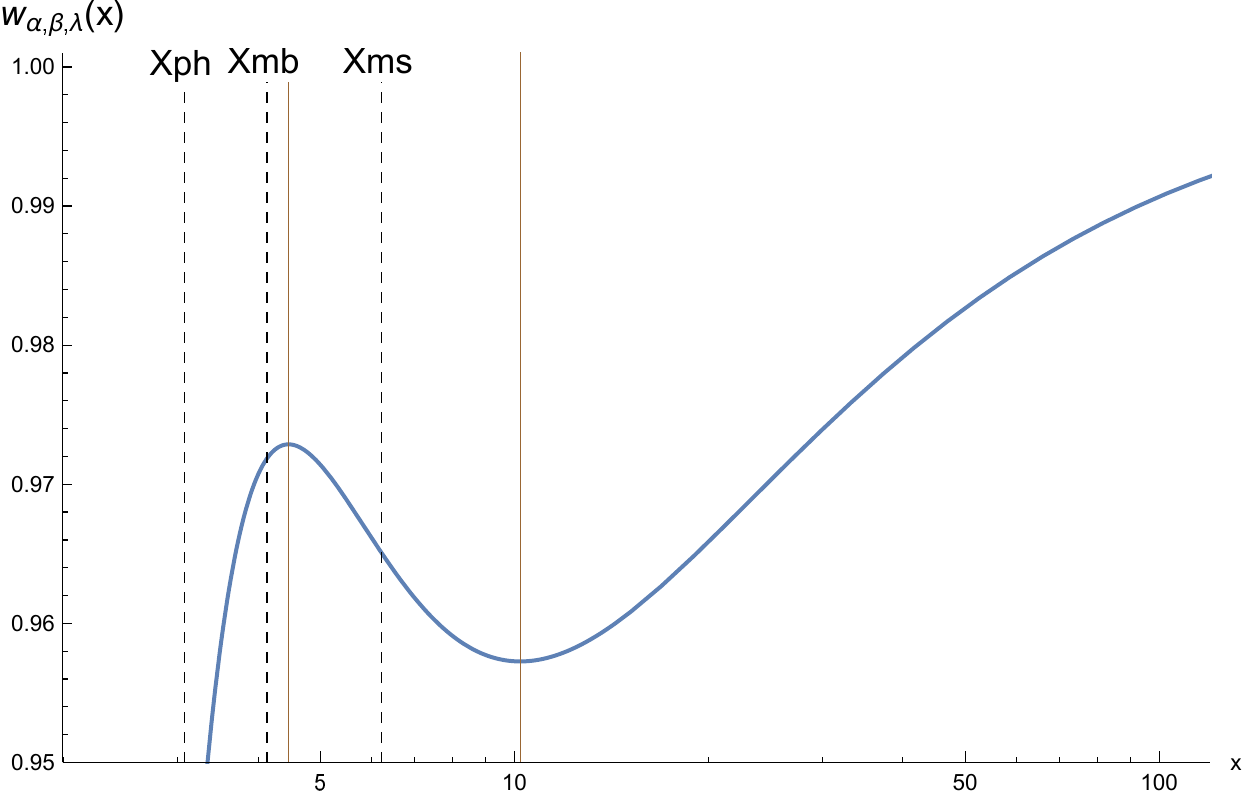}}
}
\caption{Plot of the effective potential $w_{\alpha,\beta,\lambda}$ for the parameter values $\alpha = 0.12$, $\beta = -0.61$ and $\lambda = 5$ (left panel) and $\lambda =3.9$ (right panel). Also shown (in vertical dashed lines) are the locations of the photon sphere, the marginally bound orbit, the marginally stable orbit and the locations (in vertical solid lines) of the local maxima and minima of $w_{\alpha,\beta,\lambda}$.}
\label{Fig:wEffective}
\end{figure}

Typical examples for the behavior of $w_{\alpha,\beta,\lambda}$ in case (ii) of the lemma are given in Fig.~\ref{Fig:wEffective}. To close this section, we summarize the properties of special orbits we have encountered so far and which play an important role throughout this article. Recall that in our parametrization, we fixed the parameter $\beta = \hat{L}_z/L$, such that the characteristic radii, energies and angular momentum of these orbits depend on $\alpha$ and $\beta$.

\begin{itemize}
\item {\bf spherical photon orbits}. These correspond to the limit of the unstable spherical orbits as the energy goes to infinity. Their radius $x = x_{ph}$ is determined by the unique zero of the function $h(x)$ defined in Eq.~(\ref{Eq:hDef}).
\item {\bf marginally stable orbits}. These are the spherical orbits with minimum angular momentum and energy. Their radius $x = x_{ms}$ is determined by the unique zero of the function $G$ defined in Lemma~\ref{Lem:ms}.
\item {\bf marginally bound orbits}. These are the unstable spherical orbits whose energy is equal to $1$, that is, equal to the asymptotic value of the effective potential. Their radius $x = x_{mb}$ corresponds to the unique root of $\varepsilon_{sph}(x) = 1$ in the interval $(x_{ph},x_{ms})$, and it also represents the minimum radius of parabolic-type orbits.
\item {\bf innermost stable orbits}. These are the unstable spherical orbits whose energy lies below $1$. Their radius $x = x_{max}$ represent the largest lower bound for the radii of bound orbits whose energy lies just below maximum of the effective potential.
\end{itemize}

\begin{figure}[ht]
\centerline{\resizebox{8.0cm}{!}{\includegraphics{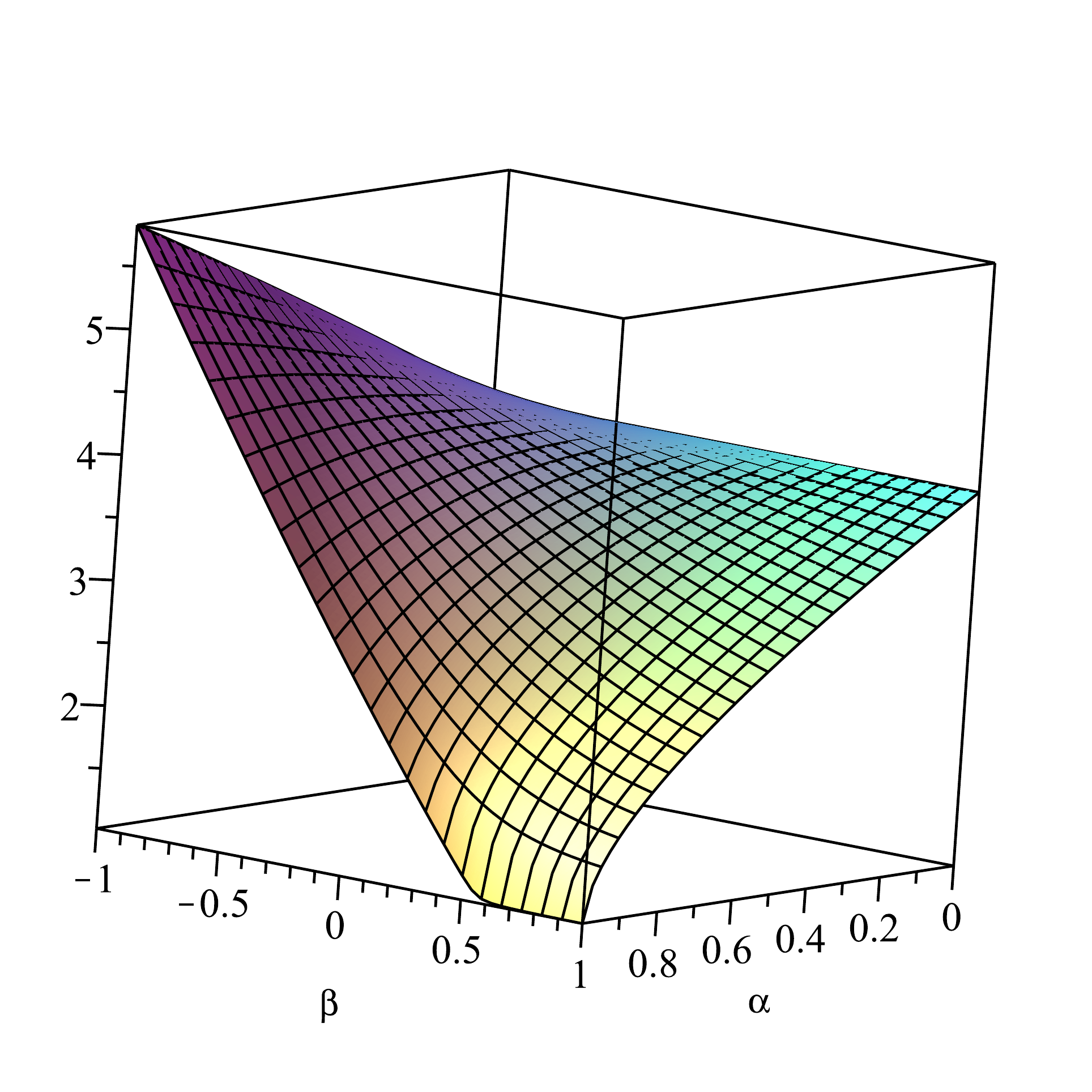}}
\resizebox{8.0cm}{!}{\includegraphics{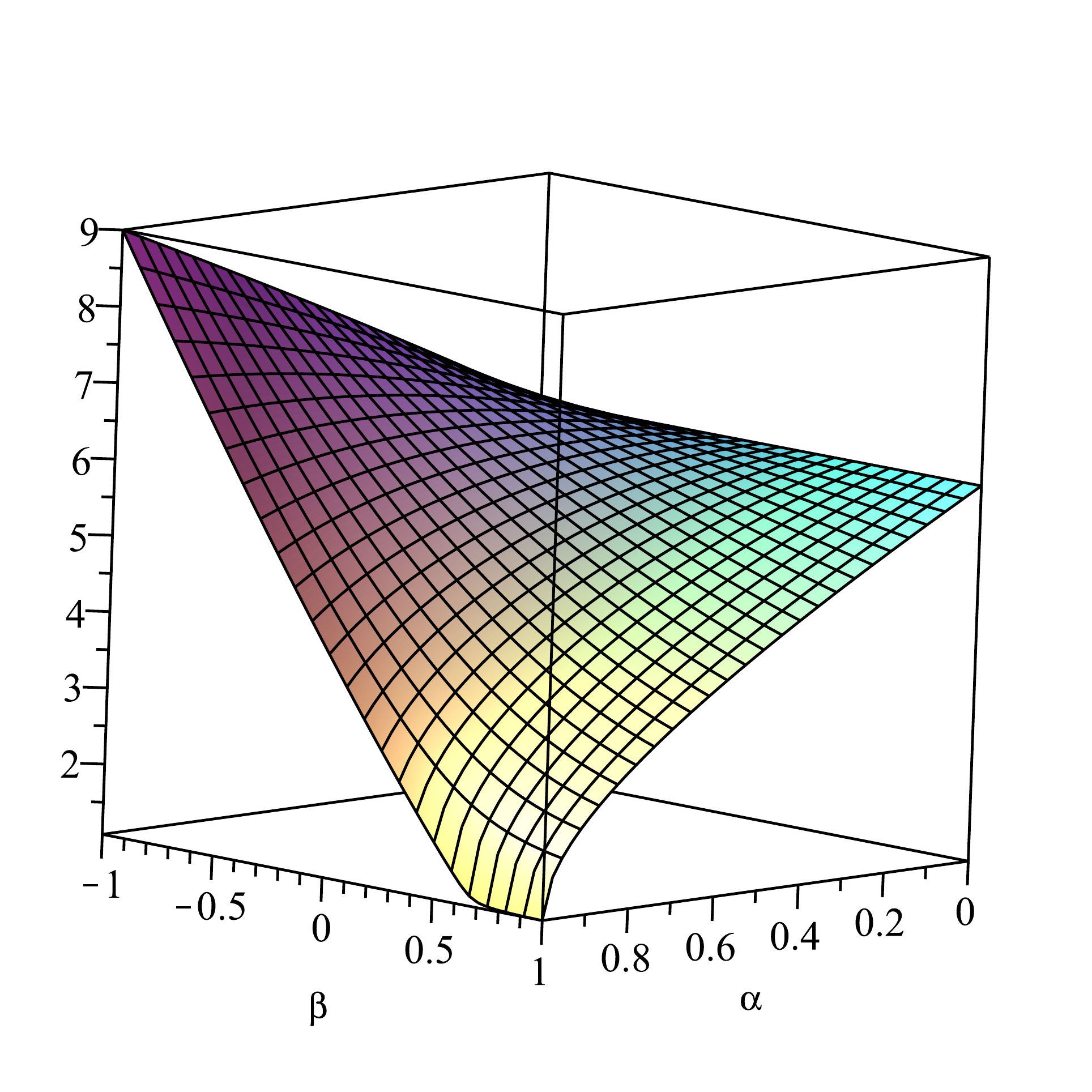}}
}
\caption{Plot of the functions $x_{mb}(\alpha,\beta)$ (left panel) and $x_{ms}(\alpha,\beta)$ (right panel).}
\label{Fig:xmbms}
\end{figure}

In some limiting cases it is possible to provide explicit expressions for the quantities $x_{ph}$, $x_{mb}$, $x_{ms}$, $\lambda_{mb}$, $\lambda_{ms}$ and $\varepsilon_{ms}$. In the non-rotating limit $\alpha = 0$ these quantities become independent of $\beta$, and they are given by
\begin{eqnarray}
x_{ph}(0,\beta) &=& 3,\quad
x_{mb}(0,\beta) = 4,\quad
x_{ms}(0,\beta) = 6,\\
\lambda_{mb}(0,\beta) &=& 4,\quad
\lambda_{ms}(0,\beta) = 2\sqrt{3},\quad
\varepsilon_{ms}(0,\beta) = \sqrt{8/9}.
\end{eqnarray}
For equatorial orbits and an arbitrary rotation parameter $\alpha\in (-1,1)$ one finds~\cite{jBwPsT72}
\begin{eqnarray}
x_{ph}(\alpha,\beta=\pm 1) &=& 2 + 2\cos\left[ \frac{2}{3}\arccos(-\alpha\beta) \right],\\
x_{mb}(\alpha,\beta=\pm 1) &=& \left( 1 + \sqrt{1-\alpha\beta} \right)^2 
 = 2 - \alpha\beta + 2\sqrt{1 - \alpha\beta},\\
x_{ms}(\alpha,\beta=\pm 1) &=& 3 + Z_2 - \beta\sqrt{(3 - Z_1)(3 + Z_1 + 2Z_2)},
\end{eqnarray}
with $Z_1 := 1 + (1-\alpha^2)^{1/3}[ (1+\alpha)^{1/3} + (1-\alpha)^{1/3}]$ and $Z_2 := \sqrt{Z_1^2 + 3\alpha^2}$. Using Eqs.~(\ref{Eq:lambdasph},\ref{Eq:epsilonsph},\ref{Eq:hms}) one also obtains from this
\begin{eqnarray}
&&
\lambda_{mb}(\alpha,\beta=\pm 1) = x_{mb}(\alpha,\beta),
\label{Eq:LambdaEpsMB}\\ 
&& \lambda_{ms}(\alpha,\beta=\pm 1) = \frac{x_{ms}(\alpha,\beta)}{\sqrt{3}},\qquad
\varepsilon_{ms}(\alpha,\beta=\pm 1) = \sqrt{1 - \frac{2}{3x_{ms}(\alpha,\beta)}}.
\label{Eq:LambdaEpsMS}
\end{eqnarray}
The behavior of $x_{mb}(\alpha,\beta)$ and $x_{ms}(\alpha,\beta)$ for arbitrary values of $\alpha\in [0,1)$ and $\beta\in (-1,1)$ are illustrated in Figs.~\ref{Fig:xmbms} and indicate that these function decrease monotonically for fixed $\alpha$ (although this property will not be used here).

\section{Parametrization of the bound orbits through the parameters $(\beta,p,e)$}
\label{App:OrbitParametrization}

It follows from Lemma~\ref{Lem:EffectivePotential} that the parameter range corresponding to bound orbits is described by the set
\begin{equation}
\mathcal{D}_\alpha := \left\{ (\beta,\lambda,\varepsilon) : -1 < \beta < 1, \lambda > \lambda_{ms}(\alpha,\beta), \varepsilon_{min}(\lambda) < \varepsilon < \min\{ 1, \varepsilon_{max}(\lambda) \}   \right\}.
\end{equation}
However, for many applications it is useful to parametrize the bound orbits in terms of the Kepler-type "semi-latus rectum" $p$ and eccentricity $e$, which are related to the turning points $x_1$ and $x_2$ of the radial motion through the transformation~\cite{wS02,jBmGtH15}
\begin{equation}
x_1 = \frac{p}{1+e},\qquad x_2 = \frac{p}{1-e}.
\label{Eq:x12pe}
\end{equation}
Here $e$ is restricted to the range $0 < e < 1$, the limits $e\to 0$ and $e\to 1$ corresponding to circular and parabolic-type orbits, respectively. Unlike the Kepler case in which $p$ is an arbitrary positive number, in the Kerr case the range of $p$ is restricted to $p > p_{ISO}(\alpha,\beta,e)$, the limit $p = p_{ISO}(\alpha,\beta,e)$ (determined below) corresponding to the innermost stable orbits, i.e. the orbits with minimal inner radius for given angular momentum. Therefore, the quantities $(\beta,p,e)$ are confined to the open set
\begin{equation}
\mathcal{E}_\alpha := \{ (\beta,p,e) : -1 < \beta < 1, 0 < e < 1, p > p_{ISO}(\alpha,\beta,e) \}.
\end{equation}

In this subsection, we describe the properties of the function $p_{ISO}(\alpha,\beta,e)$, and we prove that the map $T_\alpha: \mathcal{D}_\alpha\to \mathcal{E}_\alpha, (\beta,\lambda,\varepsilon)\mapsto (\beta,p,e)$ is invertible. To this purpose we first recall the fourth-order polynomial $R$ defined in Eq.~(\ref{Eq:rPlaneRestrictionBis}) and its relation with the effective potentials $W_\pm$ in Eq.~(\ref{Eq:RWpmRel}), which implies that on the interval $r > r_+$ all the roots of $W_+$ are also roots of $R$, with the same multiplicity. Furthermore, we see from the definition of $R$ in Eq.~(\ref{Eq:rPlaneRestrictionBis}) that in the region $r_- < r < r_+$ where $\Delta$ is negative, $R(r) > 0$, whereas $R(0) = -a^2(L^2 - \hat{L}_z^2)\leq 0$. Therefore, it follows that for $(\beta,\lambda,\varepsilon)\in \mathcal{D}_\alpha$ the function $R$ must have four real roots, two of which are given by the turning points $r_1 < r_2$ of $W_+(r) = E$ and the remaining roots $r_3 < r_4$ satisfying
\begin{equation}
0\leq r_3 < r_- < r_+ < r_4 < r_1 < r_2,
\end{equation}
with $r_3 = 0$ in the limiting case of Schwarzschild or equatorial orbits. In the limit $e\to 0$ the two roots $r_1 = r_2$ fall together and for ISOs the roots $r_4 = r_1$ are equal to each other, with $r_4 = r_1 = r_2$ for marginally stable orbits.

In order to get useful relations between the roots and the conserved quantities, we work again with dimensionless variables, setting $x_i = r_i/M_H$ for $i = 1,2,3,4$. Assuming that $(\beta,\lambda,\varepsilon)\in \mathcal{D}_\alpha$ and writing
\begin{equation}
\frac{R(r)}{(M_H^2 m)^2} = (\varepsilon x^2 - \alpha\beta\lambda)^2
 - (x^2 - 2x + \alpha^2)(x^2 + \lambda^2) = -(1-\varepsilon^2)(x-x_3)(x-x_4)(x-x_1)(x-x_2),
\label{Eq:PolyRRoots}
\end{equation}
one obtains the relations
\begin{eqnarray}
1 - \varepsilon^2 &=& \frac{2}{x_{1234}},
\label{Eq:ConsRootRel1}\\
\lambda^2 &=& \frac{x_1 x_2 x_3 + x_1 x_2 x_4 + x_1 x_3 x_4 + x_2 x_3 x_4}{x_{1234}},
\label{Eq:ConsRootRel2}\\
\alpha^2(1-\beta^2)\lambda^2 &=& \frac{2x_1 x_2 x_3 x_4}{x_{1234}},
\label{Eq:ConsRootRel3}\\
\alpha^2 + \lambda^2 + 2\alpha\beta\varepsilon\lambda 
 &=& 2\frac{x_1 x_2 + x_1 x_3 + x_1 x_4 + x_2 x_3 + x_2 x_4 + x_3 x_4}{x_{1234}},
\label{Eq:ConsRootRel4}
\end{eqnarray}
where we have introduced the abbreviation $x_{1234} := x_1 + x_2 + x_3 + x_4$. These equations allow one to determine the conserved quantities $(\varepsilon,\lambda,\beta)$ from the roots $x_1,x_2,x_3,x_4$. Since there are more equations than unknowns, the roots cannot be independent from each other. In fact, we will show that the knowledge of $\beta$, $x_1$ and $x_2$ is sufficient to determine the other roots $x_3$ and $x_4$ and the constants of motion $(\varepsilon,\lambda)$.

Before doing so, we note two immediate conclusions that can be drawn from Eqs.~(\ref{Eq:ConsRootRel1}--\ref{Eq:ConsRootRel4}). First, in the limit of marginally stable orbits, for which $x_4 = x_1 = x_2 = x_{ms}$, one obtains
\begin{equation}
1 - \varepsilon_{ms}^2 = \frac{2}{3x_{ms} + x_3},\qquad
\lambda_{ms}^2 = x_{ms}^2\frac{x_{ms} + 3 x_3}{3x_{ms} + x_3}.
\label{Eq:epslamms}
\end{equation}
For equatorial orbits, $x_3 = 0$, and these expressions reduce to  Eq.~(\ref{Eq:LambdaEpsMS}). Second, for marginally bound orbits, one has $x_2\to \infty$ with $x_4 = x_1 = x_{mb}$ remaining bounded, which yields $\varepsilon\to 1$ and
\begin{equation}
\lambda_{mb}^2 = x_{mb}(x_{mb} + 2x_3).
\end{equation}
This simplifies to the corresponding expression~(\ref{Eq:LambdaEpsMB}) when $x_3 = 0$.

With these observations at hand, one can prove the following general property of the effective potential $w_{\alpha,\beta,\lambda}(x)$ defined in Eq.~(\ref{Eq:wDef}):

\begin{lemma}
Let $\alpha\in [0,1)$, $\beta\in [-1,1]$ and $x\geq x_{mb}(\alpha,\beta)$ be fixed and set $\lambda_1 := \lambda_{sph}(x)$. Then, the function $(\lambda_1,\infty)\to \Real, \lambda\mapsto w_{\alpha,\beta,\lambda}(x)$ is increasing and unbounded.
\end{lemma}

\proof From Eq.~(\ref{Eq:wDef}) one easily finds
\begin{equation}
\frac{\partial}{\partial\lambda} w_{\alpha,\beta,\lambda}(x) 
 = \frac{ \lambda w_{\alpha,\beta,\lambda}(x) + \alpha\beta}{x^2 + \lambda^2}.
\label{Eq:dwdlambda}
\end{equation}
The right-hand side is manifestly positive for prograde orbits $\alpha\beta\geq 0$. However, in the retrograde case the right-hand side is certainly not positive for arbitrary values of $x > x_+$, since $w_{\alpha,\beta,\lambda}(x)$ becomes negative when $x\to x_+$.

We treat de retrograde as follows. First, it follows from the properties of the effective potential established in Lemma~\ref{Lem:EffectivePotential} that for each $\lambda > \lambda_1$ the point $x$ lies inside the potential well of $w_{\alpha,\beta,\lambda}$. Therefore, using Eq.~(\ref{Eq:epslamms}) one finds $w_{\alpha,\beta,\lambda}(x)^2\geq\varepsilon_{ms}^2\geq 1 - 2/(3x_{ms})$ and $\lambda^2\geq \lambda_{ms}^2\geq x_{ms}^2/3$ which imply
\begin{equation}
\lambda w_{\alpha,\beta,\lambda}(x)\geq \frac{x_{ms}}{\sqrt{3}}\sqrt{1 - \frac{2}{3x_{ms}}}.
\label{Eq:lambdawIneq}
\end{equation}
Next, we claim that for retrograde orbits, $x_{ms}(\alpha,\beta)\geq 4$, the equality being achieved for $\alpha = 1$ and $\beta = 0$. To prove this assertion, we recall that $x_{ms}$ is determined by the unique root of the function $G$ on the interval $x > x_{ph}$, and that  $G'(x) > 0$ at this root, see Lemma~\ref{Lem:ms}. Using Eqs.~(\ref{Eq:GDef},\ref{Eq:hIdentity}) a short computation reveals that
\begin{equation}
G(4) = 4\left\{ 8 + \alpha^2 - \left[ 4-\alpha^2(1-\beta^2) \right]\left[ \sqrt{4 - \alpha^2(1-\beta^2)} - \alpha\beta \right]^2 \right\} 
 \leq 4\left( 8 + \alpha^2 - 9 \right) = -4(1-\alpha^2)\leq 0,
\end{equation}
where we have used the assumption that $\alpha\beta\leq 0$ in the last step, and where we note that the equality holds if $\alpha = 1$ and $\beta=0$. Therefore, $x_{ms}\geq 4$ for retrograde orbits. Together with Eq.~(\ref{Eq:lambdawIneq}) this yields
\begin{equation}
\lambda w_{\alpha,\beta,\lambda}(x)\geq \frac{4}{3}\sqrt{\frac{5}{2}} > 1.
\end{equation}
This shows that the right-hand side of Eq.~(\ref{Eq:dwdlambda}) is also positive for retrograde orbits.

In order to prove the unboundedness property, consider the limit
\begin{equation}
\lim\limits_{\lambda\to \infty} \frac{x^2}{\lambda} w_{\alpha,\beta,\lambda}(x) 
 = \alpha\beta + \sqrt{x^2 - 2x + \alpha^2}.
\label{Eq:limw}
\end{equation}
The right-hand side is again manifestly positive when $\alpha\beta\geq 0$. When $\alpha\beta < 0$ we use the identity
\begin{equation}
h(x) = \overline{\Delta}(x) - \left[ \sqrt{x - \alpha^2(1-\beta^2)} - \alpha\beta \right]^2,
\end{equation}
to conclude that $h(2) \leq -(1-\alpha^2)\leq 0$, which shows that $x_{mb} > x_{ph}\geq 2$, such that the right-hand side of Eq.~(\ref{Eq:limw}) is again positive.
\qed

Coming back to the question regarding the invertibility of the map $T_\alpha: (\beta,\lambda,\varepsilon)\mapsto (\beta,p,e)$, for what follows we can fix $\beta\in (-1,1)$ since $\beta$ transforms trivially with respect to this map. Introducing the sets
\begin{equation}
\mathcal{D}_{\alpha,\beta} := \left\{ (\lambda,\varepsilon) : \lambda > \lambda_{ms}(\alpha,\beta), \varepsilon_{min}(\lambda) < \varepsilon < \min\{ 1, \varepsilon_{max}(\lambda) \}   \right\},
\label{Eq:DDef}
\end{equation}
and
\begin{equation}
\mathcal{E}_{\alpha,\beta} := \{(p,e) : 0 < e < 1, p > p_{ISO}(\alpha,\beta,e) \},
\label{Eq:EDef}
\end{equation}
our task is to determine the function $p_{ISO}(\alpha,\beta,e)$ and to prove that the map $T_{\alpha,\beta}: \mathcal{D}_{\alpha,\beta}\to \mathcal{E}_{\alpha,\beta}$ is invertible. Note first that this map is well defined if we replace $\mathcal{E}_{\alpha,\beta}$ with $(0,\infty)\times (0,1)$, since it follows from Lemma~\ref{Lem:EffectivePotential} that for given $(\lambda,\varepsilon)\in \mathcal{D}_{\alpha,\beta}$ there exist unique turning points $x_2 > x_1 > x_{mb}(\alpha,\beta)$ which are determined by the roots of the equation $w_{\alpha,\beta,\lambda}(x) = \varepsilon$. The key question is whether the converse holds as well: given $x_2 > x_1 > x_{mb}(\alpha,\beta)$, do there exist unique values $(\lambda,\varepsilon)\in \mathcal{D}_{\alpha,\beta}$ such that $x_1$ and $x_2$ are the turning point of $w_{\alpha,\beta,\lambda}(x) = \varepsilon$? The first observation is that this property does not hold without further restriction of $x_1$ and $x_2$. Indeed, when $x_1$ lies within the range $(x_{mb},x_{ms})$, $x_2$ cannot approach $x_1$ since $x_2 > x_{min} > x_{ms}$. As the next lemma shows, the closest $x_2$ can get to $x_1$ occurs when $\lambda = \lambda_{sph}(x_1)$ is minimized, such that $x_1$ corresponds to the inner radius of an ISO. For the following, denote for each $x_1\in (x_{mb},x_{ms})$ by $X_{2,ISO}(x_1) > x_{ms}$ the right turning point of $w_{\alpha,\beta,\lambda_{sph}(x_1)}(x) = \varepsilon_{sph}(x_1)$. Then, one can show:

\begin{proposition}
\label{Prop:TurningPointsInv}
Let $\alpha\in [0,1)$ and $\beta\in [-1,1]$ be fixed. Given $x_2 > x_1 > x_{mb}(\alpha,\beta)$ with the restriction $x_2 > X_{2,ISO}(x_1)$ if $x_1\in (x_{mb},x_{ms})$, there exist unique values $(\lambda,\varepsilon)\in \mathcal{D}_{\alpha,\beta}$ such that $x_1,x_2$ are the turning points of $w_{\alpha,\beta,\lambda}(x) = \varepsilon$. Conversely, given $(\lambda,\varepsilon)\in \mathcal{D}_{\alpha,\beta}$, it follows that the turning points of the orbit $w_{\alpha,\beta,\lambda}(x) = \varepsilon$ satisfy $x_2 > X_{2,ISO}(x_1)$ if $x_1\in (x_{mb},x_{ms})$.
\end{proposition}

\proof Suppose $x_2 > x_1 > x_{mb}(\alpha,\beta)$ with the restriction $x_2 > X_{2,ISO}(x_1)$ if $x_1\in (x_{mb},x_{ms})$ are given. Since $x_1$ must lie between the maximum and the minimum of $w_{\alpha,\beta,\lambda}$, the value of the angular momentum parameter must satisfy $\lambda\geq \lambda_1 := \lambda_{sph}(x_1)$. Next, introduce the smooth function
\begin{equation}
\varepsilon_1(\lambda) := w_{\alpha,\beta,\lambda}(x_1),\qquad \lambda\geq \lambda_1,
\label{Eq:epsilon1}
\end{equation}
According to the previous lemma this function is monotonously increasing and unbounded. 

Consider first the case $x_1 > x_{ms}$. Then, $\varepsilon_1(\lambda_1) < 1$ since $x_1$ is the location of the local minimum of $w_{\alpha,\beta,\lambda_1}$ when $\lambda = \lambda_1$. As $\lambda$ increases, the location of the local minimum moves to the right (see Lemma~\ref{Lem:EffectivePotential}); hence $x_1$ must be the left turning point of $w_{\alpha,\beta,\lambda}(x) = \varepsilon_1(\lambda)$. Since $\varepsilon_1(\lambda)\to \infty$ monotonically as $\lambda\to \infty$, there exists a unique $\lambda_2 > \lambda_1$ such that $\varepsilon_1(\lambda_2) = 1$. Next, denote for each $\lambda\in (\lambda_1,\lambda_2)$ by $X_2(\lambda)$ the location of the right turning point of $w_{\alpha,\beta,\lambda}(x) = \varepsilon_1(\lambda)$. Because $\varepsilon_1(\lambda)\to 1$ as $\lambda\to \lambda_2$ it follows that $X_2(\lambda)\to \infty$ as $\lambda\to \lambda_2$. Furthermore, $X_2(\lambda_1) = x_1$. Therefore, there exists $\lambda\in (\lambda_1,\lambda_2)$ such that $X_2(\lambda) = x_2$.

It remains to show that this $\lambda$ is unique. For this, it is sufficient to prove that the function $X_2(\lambda)$ is increasing. To prove this, differentiate both sides of the equation
$$
w_{\alpha,\beta,\lambda}(X_2(\lambda)) = w_{\alpha,\beta,\lambda}(x_1)
$$
with respect to $\lambda$, which yields
$$
\frac{\partial w_{\alpha,\beta,\lambda}}{\partial x}(X_2)\frac{dX_2}{d\lambda}
 = \frac{\partial w_{\alpha,\beta,\lambda}}{\partial\lambda}(x_1) 
 - \frac{\partial w_{\alpha,\beta,\lambda}}{\partial\lambda}(X_2).
$$
Using Eq.~(\ref{Eq:dwdlambda}) this can be rewritten as
\begin{equation}
\frac{\partial w_{\alpha,\beta,\lambda}}{\partial x}(X_2)\frac{dX_2}{d\lambda}
 = \frac{\lambda\varepsilon_1(\lambda) + \alpha\beta}{(\lambda^2 + x_1^2)
 \left[ \lambda^2 + X_2(\lambda)^2 \right]} \left[ X_2(\lambda)^2 - x_1^2 \right] > 0.
\end{equation}
Since the first factor on the left-hand side is positive (being $X_2$ the right turning point) we conclude that $X_2(\lambda)$ is an increasing function of $\lambda$ and hence the solution $X_2(\lambda) = x_2$ is unique.

Next, we analyze the case for which $x_1\in (x_{mb},x_{ms})$ and $x_2 > x_{2,ISO}(x_1)$. Now $x_1$ describes the location of the local maximum of $w_{\alpha,\beta,\lambda_1}$ and thus the range of the function $\varepsilon_1$ in Eq.~(\ref{Eq:epsilon1}) is restricted to the  interval $(\lambda_1,\lambda_{mb})$ with $\lambda = \lambda_1$ corresponding to the situation for which $x_1$ is the inner radius of the ISOs. As before, denote by $X_2(\lambda)$ the right turning point of $w_{\alpha,\beta,\lambda}(x) = \varepsilon_1(\lambda)$. As $\lambda$ increases from $\lambda_1$ to $\lambda_{mb}$, $\varepsilon_1$ increases monotonously to $1$ and hence $X_2(\lambda)$ increases monotonously from $X_{2,ISO}(x_1)$ to infinity. Therefore, there exists a unique $\lambda\in (\lambda_1,\lambda_{mb})$ such that $X_2(\lambda) = x_2$. 

Conversely, let $(\lambda,\varepsilon)\in \mathcal{D}_{\alpha,\beta}$ be given and denote by $x_2 > x_1$ the turning points of $w_{\alpha,\beta,\lambda}(x) = \varepsilon$. Suppose $x_1\in (x_{mb},x_{ms})$. Then, $\lambda\in (\lambda_1,\lambda_{mb})$, $x_2 = X_2(\lambda)$, and it follows from the monotonicity property of $X_2(\lambda)$ that $X_2(\lambda) > X_{2,ISO}(x_1)$. This concludes the proof of the proposition.
\qed

Before we proceed, we show the following property of the function $X_{2,ISO}(x_1)$:

\begin{lemma}
\label{Lem:ISOInv}
Let $\alpha\in [0,1)$ and $\beta\in [-1,1]$ be fixed. The function $(x_{mb},x_{ms})\to \Real$, $x\mapsto X_{2,ISO}(x)$ is a smooth monotonously decreasing function satisfying $X_{2,ISO}(x)\to \infty$ as $x\to x_{mb}$ and $X_{2,ISO}(x)\to x_{ms}$ as $x\to x_{ms}$.
\end{lemma}

\proof
The smoothness property and the limits is a consequence of the definition of $X_{2,ISO}$ and the fact that it is an isolated root of the polynomial~(\ref{Eq:PolyRRoots}). To prove the monotonicity property, differentiate both sides of the equation
\begin{equation}
w_{\alpha,\beta,\lambda_{sph}(x_1)}(X_{2,ISO}(x_1)) = \varepsilon_{sph}(x_1),\qquad
x_{mb} < x_1 < x_{ms},
\end{equation}
with respect to $x_1$, which yields (setting $x_2 := X_{2,ISO}(x_1)$):
$$
\left. \frac{\partial w_{\alpha,\beta,\lambda}}{\partial x}(x_2) \right|_{\lambda = \lambda_{sph}(x_1)} \frac{dX_{2,ISO}}{dx}(x_1) = \varepsilon_{sph}'(x_1) - \left. \frac{\partial w_{\alpha,\beta,\lambda}}{\partial\lambda}(x_2)\right|_{\lambda = \lambda_{sph}(x_1)} \lambda_{sph}'(x_1).
$$
Using Eqs.~(\ref{Eq:lambdasph},\ref{Eq:epsilonsph},\ref{Eq:epsilonlambdaDeriv},\ref{Eq:dwdlambda}) one can rewrite the right-hand side as
\begin{equation}
\left. \frac{\partial w_{\alpha,\beta,\lambda}}{\partial x}(x_2) \right|_{\lambda = \lambda_{sph}(x_1)} \frac{dX_{2,ISO}}{dx}(x_1) = \frac{\sqrt{x_1 - \alpha^2(1-\beta^2)}}{x_1^2}
\left[ 1 - \frac{x_1^2 + \lambda_{sph}(x_1)^2}{x_2^2 + \lambda_{sph}(x_1)^2}  \right] \lambda_{sph}'(x_1),\qquad
x_2 = X_{2,ISO}(x_1),
\label{Eq:ISODiff}
\end{equation}
for all $x_{mb} < x_1 < x_{ms}$. It follows from the fact that $x_1 < x_2$, $\lambda_{sph}'(x_1) < 0$ and that the first factor on the left-hand side of Eq.~(\ref{Eq:ISODiff}) is positive that $dX_{2,ISO}/dx(x_1) < 0$, and this concludes the proof of the lemma.
\qed

As a consequence of Proposition~\ref{Prop:TurningPointsInv} and the last lemma one has:

\begin{proposition}[Invertibility of the maps $T_{\alpha,\beta}$]
Let $\alpha\in [0,1)$ and $\beta\in [-1,1]$ be fixed. There exists a continuous function $p_{ISO}:[0,1]\to \Real$, $e\mapsto p_{ISO}(\alpha,\beta,e)$ satisfying $p_{ISO}(\alpha,\beta,0) = x_{ms}(\alpha,\beta)$ and $p_{ISO}(\alpha,\beta,1) = 2x_{mb}(\alpha,\beta)$ such that the map
\begin{equation}
T_{\alpha,\beta}: \mathcal{D}_{\alpha,\beta} \to \mathcal{E}_{\alpha,\beta}, (\lambda,\varepsilon)\mapsto (p,e),
\end{equation}
with $\mathcal{D}_{\alpha,\beta},\mathcal{E}_{\alpha,\beta}$ as defined in Eqs.~(\ref{Eq:DDef},\ref{Eq:EDef}) is a diffeomorphism.
\end{proposition}

\proof Introduce the open set
\begin{equation}
\mathcal{F}_{\alpha,\beta} := \left\{ (x_1,x_2) : x_2 > x_1 > x_{mb} \hbox{ and } x_2 > X_{2,ISO}(x_1) \hbox { if $x_1 < x_{ms}$} \right\}.
\end{equation}
It follows from the last proposition that the map
\begin{equation}
\Phi: \mathcal{D}_{\alpha,\beta} \to \mathcal{F}_{\alpha,\beta},\quad
(\varepsilon,\lambda) \mapsto (x_1,x_2)
\end{equation}
which maps $(\varepsilon,\lambda)\in \mathcal{D}_{\alpha,\beta}$ to the turning points $x_1,x_2$ of $w_{\alpha,\beta,\lambda}(x) = \varepsilon$, is invertible. Furthermore, since $x_1$ and $x_2$ are isolated roots of the polynomial~(\ref{Eq:PolyRRoots}) it follows that this map is smooth.

Next, consider the smooth map
\begin{eqnarray}
\Psi: \left\{ (x_1,x_2) : x_2 > x_1 > 0 \right\} &\to& \left\{ (p,e) : p > 0, 0 < e < 1 \right\},\\
(x_1,x_2) &\mapsto& (p,e)
 = \left( \frac{2x_1 x_2}{x_1 + x_2},\frac{x_2 - x_1}{x_1 + x_2} \right),
\end{eqnarray}
whose inverse is given by
\begin{equation}
\Psi^{-1}(p,e) = \left( \frac{p}{1+e}, \frac{p}{1-e} \right),\qquad
p > 0,\quad 0 < e < 1.
\end{equation}
Then, it follows that the map $T_{\alpha,\beta}: \mathcal{D}_{\alpha,\beta}\to \mathcal{E}_{\alpha,\beta} := \Psi(\mathcal{F}_{\alpha,\beta})$, $(\varepsilon,\lambda)\mapsto (p,e) := \Psi\circ\Phi(\varepsilon,\lambda)$ is a diffeomorphism.

It remains to prove that the set $\mathcal{E}_{\alpha,\beta} = \Psi(\mathcal{F}_{\alpha,\beta})$ has the form given in Eq.~(\ref{Eq:EDef}). For this, consider the image of the curve $x_2 = X_{2,ISO}(x_1)$ under the map $\Psi$:
\begin{equation}
\mathcal{B}: (x_{mb},x_{ms}) \to \Real^2: x_1
\mapsto \left( \mathcal{B}_1(x_1),\mathcal{B}_2(x_1) \right) := \Psi(x_1,X_{2,ISO}(x_1)). 
\end{equation}
Note that $\mathcal{B}(x_1)\to (2x_{mb},1)$ when $x_1\to x_{mb}$ and $\mathcal{B}(x_1)\to (x_{ms},0)$ when $x_1\to x_{ms}$. Furthermore, the image of $\mathcal{B}$ describes the boundary of $\mathcal{E}_{\alpha,\beta}$ in $(0,\infty)\times (0,1)$. In view of Lemma~\ref{Lem:ISOInv}, $X_{2,ISO}'(x_1)\leq 0$, and hence it follows that
$$
\frac{\partial\mathcal{B}_2}{\partial x_1}(x_1) 
 = \frac{\partial}{\partial x_1} \frac{X_{2,ISO}(x_1) - x_1}{X_{2,ISO}(x_1) + x_1} 
 = 2\frac{x_1 X_{2,ISO}'(x_1) - X_{2,ISO}(x_1)}{\left[ X_{2,ISO}(x_1) + x_1 \right]^2} < 0,
$$
which shows that $e$ decreases with respect to $x_1$. Hence, the boundary curve $\mathcal{B}$ can be re-parametrized in terms of $e$.
\qed

Since the roots $x_1$ and $x_2$ also depend smoothly on $\beta$, it follows immediately that the map $T_\alpha: \mathcal{D}_\alpha\to \mathcal{E}_\alpha$ is invertible and smooth for each $\alpha\in [0,1)$. The function $p_{ISO}(\alpha,\beta,e)$ and the inverse map $T_\alpha^{-1}$ can be determined in the following way. From Eq.~(\ref{Eq:x12pe}) one has $x_1 + x_2 = 2p/(1-e^2)$ and $x_1 x_2 = p^2/(1-e^2)$. Setting $w_+ := x_3 + x_4$ and $w_\times := x_3 x_4$ one obtains from Eqs.~(\ref{Eq:ConsRootRel2},\ref{Eq:ConsRootRel3}) a linear relation between $w_+$ and $w_\times$:
\begin{equation}
w_\times = \frac{\kappa^2 p}{2(p - \kappa^2)} w_+,
\label{Eq:wmx}
\end{equation}
where we have abbreviated $\kappa := \alpha\sqrt{1 - \beta^2}$. This relation allows one to express $w_\times$ in terms of $(\kappa,p,w_+)$. Note that $w_\times = 0$ for equatorial orbits.

For ISOs one has $x_4 = x_1 = p/(1+e)$. Thus, Eq.~(\ref{Eq:wmx}) implies that
\begin{equation}
\left.  x_3  \right|_{ISO}  = \left. \frac{\kappa^2}{2p - \kappa^2(3+e)} p  
\right|_{p = p_{ISO}(\alpha,\beta,e)}.
\end{equation}
This in turn, implies that $w_+ = x_3 + x_4$ is equal to
\begin{equation}
\left. w_+ \right|_{ISO} 
 = \left. \frac{2(p - \kappa^2)}{2p - \kappa^2(3 + e)}\frac{p}{1+e} 
 \right|_{p = p_{ISO}(\alpha,\beta,e)}
\label{Eq:w+ISO}
\end{equation}
for ISOs. The expression in Eq.~(\ref{Eq:w+ISO}) motivates the introduction of a new parameter $u_+$, defined through the relation
\begin{equation}
w_+ = \frac{2(p - \kappa^2)}{2p - \kappa^2(3 + e)}\frac{p}{1+e}
\frac{1}{u_+},
\label{Eq:u+Def}
\end{equation}
such that $u_+ = 1$ for ISOs. For general bound orbits, Eqs.~(\ref{Eq:ConsRootRel1}--\ref{Eq:ConsRootRel4}) imply the following algebraic equation for $u_+$:
\begin{eqnarray}
&& p^2(p-4) + \alpha^2 p(1-e^2) + \kappa^2\left[ p(3 + e^2) - \alpha^2(1-e^2) \right]
 - (p-\alpha^2)\left[ 2p - \kappa^2(3+e) \right] (1+e)u_+ 
\nonumber\\
&& \qquad +\, 2\alpha\beta p\sqrt{1+e}\sqrt{(1-e)p(p - \kappa^2) 
 + (p-1+e^2)\left[ 2p - \kappa^2(3+e) \right] u_+} = 0.
\label{Eq:u+}
\end{eqnarray}
Setting $u_+ = 1$, one obtains the ISO limit $p = p_{ISO}(\alpha,\beta,e)$ as a zero of the function
\begin{eqnarray}
H(\alpha,\beta,e,p) &:=& p^2(p-6-2e) + \alpha^2 p(3-e)(1+e)
 + 2\kappa^2\left[ p(3+2e+e^2) - 2\alpha^2(1+e) \right]
 \nonumber\\
 &+& 2\alpha\beta p\sqrt{1+e}\sqrt{(3-e)p^2 - 2p(1-e^2) - \kappa^2\left[ 4p - (3+e)(1-e^2) \right]}
\label{Eq:pIso}
\end{eqnarray}
for fixed $(\alpha,\beta,e)$. 

In the Schwarzschild limit $\alpha=0$, Eqs.~(\ref{Eq:u+},\ref{Eq:pIso}) simplify considerably and one obtains
\begin{equation}
u_+ = \frac{p-4}{2(1+e)},\qquad w_+ = \frac{2p}{p-4},\qquad 
p_{ISO}(e) = 6 + 2e,\qquad
\lim\limits_{\alpha\to 0}\frac{w_\times}{\alpha^2} = (1-\beta^2)\frac{p}{p-4}.
\label{Eq:u+Schwarzschild}
\end{equation}
To treat the case $\alpha > 0$ one rewrites Eq.~(\ref{Eq:u+}) in the form
\begin{equation}
a - b u_+ + 2\alpha\beta p\sqrt{(1-e^2)c + \nu bu_+} = 0,
\end{equation}
with the coefficients
\begin{eqnarray}
a &:=& p^2(p-4) + \alpha^2 p(1-e^2) + \kappa^2\left[ p(3 + e^2) - \alpha^2(1-e^2) \right],\\
b &:=& (1+e)(p-\alpha^2)\left[ 2p - \kappa^2(3+e) \right],\\
c &:=&p(p - \kappa^2),\\
\nu &:=& \frac{p-1+e^2}{p-\alpha^2}.
\end{eqnarray}
By squaring Eq.~(\ref{Eq:u+}) and solving the resulting quadratic equation, one obtains
\begin{equation}
u_+ = \frac{1}{b}\left[ a + 2\alpha^2\beta^2p^2\nu 
 + 2\alpha\beta p\sqrt{(1-e^2) c + a\nu + \alpha^2\beta^2 p^2\nu^2} \right],
\label{Eq:u+Sol}
\end{equation}
where the correct sign can be determined by taking the limits $\alpha\to 0$ or $p\to \infty$.

Eqs.~(\ref{Eq:x12pe},\ref{Eq:wmx},\ref{Eq:u+Def},\ref{Eq:u+Sol}) allow one to determine the four roots $x_1$, $x_2$, $x_3$ and $x_4$ explicitly in terms of the parameters $(\beta,p,e)$. The constants of motion $\varepsilon$ and $\lambda$ can be determined from these quantities using Eqs.~(\ref{Eq:ConsRootRel1},\ref{Eq:ConsRootRel3}), which yield
\begin{equation}
\varepsilon^2 = 1 - \frac{1-e^2}{p}\frac{1}{1 + \frac{1-e^2}{2p} w_+},\qquad
\lambda^2 = \frac{p}{1 + \frac{1-e^2}{2p} w_+}\frac{w_\times}{\kappa^2}.
\label{Eq:epsilonlambda2}
\end{equation}

\section{Details regarding the computation of the generalized action-angle variables}
\label{App:AngleVars}

In this appendix we provide the necessary details for the computation of the generalized action-angle variables $(J_\alpha,Q^\alpha)$. For this, recall the generating function $S(\gamma_x; I_\alpha)$ in Eq.~(\ref{Eq:GeneratingFunctionBisBis}) which depends on the two line integrals (see Eq.~(\ref{Eq:SrStheta}))
\begin{equation}
S_r(\gamma_r; I_\alpha) = \int\limits_{\gamma_r} V\frac{dr}{\Delta},\qquad
S_\vartheta(\gamma_\vartheta; I_\alpha) := \int\limits_{\gamma_\vartheta} p_\vartheta d\vartheta,
\label{Eq:SrSthetaBis}
\end{equation}
corresponding to the radial and polar motion, respectively. Also recall the relation between the quantities $I_\alpha$ defined in Eqs.~(\ref{Eq:DefI0}--\ref{Eq:DefI3}) which determine the generalized action variables and the original integrals of motion $P = (P_\alpha) = (m,E,L_z,L)$, such that $I_0(P) = m$, $I_1(P) = L_z$, $I_2(P) = S_\vartheta(\circ; I_\alpha)/(2\pi)$ and $I_3(P) = S_r(\circ; I_\alpha)/(2\pi)$ where the symbol $\circ$ indicates that one takes a closed loop in the line integrals~(\ref{Eq:SrSthetaBis}). Therefore, the strategy consists in first computing the functions $S_r$ and $S_\vartheta$ and their partial derivatives with respect to $P_\alpha$ (which have the form of open line integrals) and then to compute the action variables $I_\alpha$ and their derivatives with respect to $P_\alpha$ by ``closing" the integrals.

Partial differentiation of $S_\vartheta$ with respect to $(P_\alpha) = (m,E,L_z,L)$ and taking into account the relation $p_\vartheta^2 = L^2 - K(\vartheta)$ with the functions $K(\vartheta)$ defined in Eq.~(\ref{Eq:thetaPlaneRestriction}) yields
\begin{eqnarray}
\frac{\partial S_\vartheta}{\partial m} &=& -a_H^2 m\int\limits_{\gamma_\vartheta}
\cos^2 \vartheta \frac{d\vartheta}{p_\vartheta},
 \label{Eq:dSthetadm}\\
\frac{\partial S_\vartheta}{\partial E} &=& a_H\int\limits_{\gamma_\vartheta}
 (L_z - a_H E\sin^2 \vartheta)\frac{d\vartheta}{p_\vartheta}, 
\label{Eq:dSthetadE}\\
\frac{\partial S_\vartheta}{\partial L_z} &=& -\int\limits_{\gamma_\vartheta}
  (L_z - a_H E\sin^2 \vartheta)\frac{d\vartheta}{\sin^2 \vartheta p_\vartheta},
\label{Eq:dSthetadLz}\\
\frac{\partial S_\vartheta}{\partial L} &=& L\int\limits_{\gamma_\vartheta}
\frac{d\vartheta}{p_\vartheta}.
\label{Eq:dSthetadL}
\end{eqnarray}
Similarly, taking into account the relation $V^2 = R(r)$ with $R(r)$ defined in Eq.~(\ref{Eq:rPlaneRestrictionBis}), one obtains
\begin{eqnarray}
\frac{\partial S_r}{\partial m} &=& -m\int \limits_{(r,p_r)} \frac{r^2 dr}{V},
\label{Eq:dSrdm}\\
\frac{\partial S_r}{\partial E} &=& \int \limits_{(r,p_r)} 
\frac{(r^2 + a_H^2)(r^2 E - a_H\hat{L}_z)}{\Delta}\frac{dr}{V},
\label{Eq:dSrdE}\\
\frac{\partial S_r}{\partial L_z} &=& -a_H\int \limits_{(r,p_r)} 
\frac{r^2 E - a_H\hat{L}_z}{\Delta}\frac{dr}{V},
\label{Eq:dSrdLz}\\
\frac{\partial S_r}{\partial L} &=& -L\int \limits_{(r,p_r)} \frac{dr}{V}.
\label{Eq:dSrdL}
\end{eqnarray}
For the following, we express these integrals in terms of Legendre's elliptic integrals. We find it convenient to work in terms of the dimensionless quantities defined in Eq.~(\ref{Eq:DimLessVariables}).

\subsection{Polar integrals}

We start with the computation of the polar integrals, determined by the generating function $S_\vartheta(\vartheta; P_\alpha)$ and its partial derivatives with respect to $(P_\alpha) = (m,E,L_z,L)$. For this we set $\zeta := \cos\vartheta$ and write
\begin{equation}
p_\vartheta^2 = L^2 - K(\vartheta) = \left( \frac{M_H m}{\sin\vartheta} \right)^2 q(\zeta),
\label{Eq:ptheta2}
\end{equation}
with the polynomial
\begin{equation}
q(\zeta) :=  \alpha^2(1-\varepsilon^2)\zeta^4 - \left( \lambda^2 + 2\alpha\beta\varepsilon\lambda + \alpha^2 \right)\zeta^2 + \lambda^2(1-\beta^2).
\label{Eq:qDef}
\end{equation}
Since $q(0) > 0$ and $q(1) = -\lambda_z^2 < 0$, this polynomial has two positive roots $\zeta_1,\zeta_2$ satisfying $0 < \zeta_1 < 1 < \zeta_2$, and the motion is restricted to the interval $[-\zeta_1,\zeta_1]$ which corresponds to $[\vartheta_1,\pi-\vartheta_1]$ in terms of the polar angle. Writing $q(\zeta) = \alpha^2(1-\varepsilon^2)(\zeta^2 - \zeta_1^2)(\zeta^2 - \zeta_2^2)$ one finds that the roots satisfy the relations
\begin{eqnarray}
\zeta_1^2 + \zeta_2^2 
&=& \frac{\lambda^2 + 2\alpha\beta\varepsilon\lambda + \alpha^2}
{\alpha^2(1-\varepsilon^2)},
\label{Eq:z12Rel1}\\
\zeta_1^2 \zeta_2^2 &=& \frac{\lambda^2(1-\beta^2)}{\alpha^2(1-\varepsilon^2)},
\label{Eq:z12Rel2}
\end{eqnarray}
when $\alpha\neq 0$, while in the non-rotating limit $\alpha=0$ the polynomial $q$ has the single root $\zeta_1 = \sqrt{1-\beta^2}$. In terms of the angle $\phi$ introduced in Eqs.~(\ref{Eq:thetaphi},\ref{Eq:pthetaphi}) one finds the following expressions for the generating function $S_\vartheta$:
\begin{equation}
S_\vartheta(\gamma_\vartheta; P_\alpha) = M_H m\frac{\lambda\sqrt{1-\beta^2}}{\zeta_1^3}
\left[ \zeta_1^2 \EE_*(\phi,k_1) - k_1^2(1-\zeta_1^2)\FF_*(\phi,k_1)
 - (1-\zeta_1^2)(\zeta_1^2 - k_1^2)\Pi_*(\phi,\zeta_1^2,k_1) \right],
\end{equation}
where we recall that $k_1 = \zeta_1/\zeta_2$ and where the notation $\EE_*(\phi,k)$ refers to the same integral as in Eq.~(\ref{Eq:EllipticEE}) with the lower integration limit replaced with $-\pi/2$, such that $\EE_*(\phi,k_1) := \EE(\phi,k_1) + \EE(k_1)$, and similarly for $\FF_*(\phi,k_1)$ and $\Pi_*(\phi,\zeta_1^2,k_1)$.

Similarly, one obtains from Eqs.~(\ref{Eq:dSthetadm}--\ref{Eq:dSthetadL}),
\begin{equation}
\frac{\partial S_\vartheta}{\partial m} = M_H\GG^0(\phi),\quad
\frac{\partial S_\vartheta}{\partial E} = M_H\GG^1(\phi),\quad
\frac{\partial S_\vartheta}{\partial L_z} = \GG^2(\phi),\quad
\frac{\partial S_\vartheta}{\partial L} = \GG^3(\phi),
\label{Eq:PartialStheta}
\end{equation}
with the functions $\GG^\alpha(\phi)$ defined by
\begin{eqnarray}
\GG^0(\phi) &:=& -\frac{\alpha^2\zeta_1^3}{\lambda\sqrt{1-\beta^2}}\DD_*(\phi, k_1),
\label{Eq:GG0Def}\\
\GG^1(\phi) &:=& \frac{\alpha\beta\zeta_1}{\sqrt{1-\beta^2}}\FF_*(\phi, k_1) 
 - \varepsilon\GG^0(\phi),
\label{Eq:GG1Def}\\
\GG^2(\phi) &:=& \frac{\zeta_1}{\lambda\sqrt{1-\beta^2}}
 \left[ \alpha\varepsilon \FF_*(\phi,k_1) - \lambda_z\Pi_*(\phi,\zeta_1^2,k_1) \right],
\label{Eq:GG2Def}\\
\GG^3(\phi) &:=& \frac{\zeta_1}{\sqrt{1-\beta^2}} \FF_*(\phi, k_1).
\label{Eq:GG3Def}
\end{eqnarray}
In the limit $\beta\to \pm 1$ of equatorial orbits it follows from Eqs.~(\ref{Eq:z12Rel1},\ref{Eq:z12Rel2}) that $\zeta_1\to 0$ while $\zeta_2$ and $\zeta_1/\sqrt{1-\beta^2}$ have finite limits, such that $k_1 = 0$, and in this case one finds that $\GG^0 = 0$ vanishes, while $\GG^1(\phi) = -\alpha\GG^2(\phi)$, $\GG^3(\phi) = -\beta\GG^2(\phi)$ and $\GG^2(\phi) = -\hat{\lambda}_z(\phi + \pi/2)/\sqrt{\lambda_z^2 + \alpha^2(1-\varepsilon^2)}$.

In the non-rotating limit $\alpha\to 0$ one has $\zeta_1\to \sqrt{1-\beta^2}$ and $\zeta_2\to \infty$ such that $k_1\to 0$ which implies that $\GG^0(\phi)$ and $\GG^1(\phi)$ vanish identically, while $\GG^2(\phi) = -\beta\Pi_*(\phi,1-\beta^2,0)$ and $\GG^3(\phi) = \phi + \pi/2$.

\subsection{Radial integrals}

Next, we compute the generating function $S_r(\gamma_r; P_\alpha)$ and its partial derivatives with respect to $P_\alpha$. For this, recall that
\begin{equation}
V^2 = R(r) = (E r^2 - a\hat{L}_z)^2 - \Delta(r)(m^2 r^2 + L^2).
\end{equation}
As discussed in the previous appendix, for $(m,E,L_z,L)\in \Omega$ the polynomial on the right-hand side has four real roots $r_i$, $i=1,2,3,4$, satisfying the inequalities
\begin{equation}
0\leq r_3 < r_- < r_+ < r_4 < r_1 < r_2.
\label{Eq:RootsRel}
\end{equation}
Recalling the definitions of the elliptic integrals in Eqs.~(\ref{Eq:EllipticFF}--\ref{Eq:EllipticDD}) (see also Eqs.~(\ref{Eq:R1}--\ref{Eq:R4}) and (\ref{Eq:R5},\ref{Eq:R6})), the integrals in Eqs.~(\ref{Eq:dSrdm}--\ref{Eq:dSrdL}) can be expressed in terms of the $\pi$-periodic angle $\chi$ introduced in Eqs.~(\ref{Eq:rchi},\ref{Eq:Vchi}) and the quantities $b$, $k$ and $C$ defined in Eqs.~(\ref{Eq:bkC}) as follows:
\begin{equation}
\frac{\partial S_r}{\partial m} = M_H\HH^0(\chi),\quad
\frac{\partial S_r}{\partial E} = M_H\HH^1(\chi),\quad
\frac{\partial S_r}{\partial L_z} = \HH^2(\chi),\quad
\frac{\partial S_r}{\partial L} = \HH^3(\chi),
\label{Eq:PartialSr}
\end{equation}
with the functions $\HH^\alpha(\chi)$ defined by\footnote{These functions are related to the functions $\HH_0$, $\HH_1$ and $\HH_2$ introduced in~\cite{pRoS18} for equatorial orbits, in which case $x_3 = 0$ and $x_4$ was denoted by $x_0$. In fact, when $S_r$ is restricted to $L = \pm\hat{L}_z = L_z - aE$, the partial derivatives of $S_r$ with respect to $(m,E,L_z)$ obtained from Eq.~(\ref{Eq:PartialSr}) are found to agree with the corresponding expressions in Eq.~(A7) of Ref.~\cite{pRoS18}.}
\begin{eqnarray}
\HH^0(\chi) &:=& -\frac{C}{2}\left\{ (x_4 x_{124} - x_1 x_2)\FF(\chi,k)
 + (x_1-x_3)(x_2-x_4)\EE(\chi,k) + (x_1-x_4)x_{1234} \Pi(\chi,b^2,k) \right.
 \nonumber\\
 && \left.  \qquad\qquad - \frac{1}{2}(x_1-x_3)(x_2-x_1)\frac{\sin(2\chi)}{1 - b^2\sin^2\chi} 
 \right\},
\label{Eq:HH0Def}\\
\HH^1(\chi) &:=& 2C\left\{ \left[ \frac{x_4(\varepsilon x_4^2 - \alpha\beta\lambda)}{(x_4 - x_+)(x_4 - x_-)} - \frac{1}{2}\alpha\beta\lambda \right]\FF(\chi,k)
 + \varepsilon(x_1 - x_4)\Pi(\chi,b^2,k) \right. 
\nonumber\\
 && \left. \qquad
 - \frac{x_1 - x_4}{x_+ - x_-} \left[ \frac{x_+(\varepsilon x_+^2 - \alpha\beta\lambda)}{(x_1 - x_+)(x_4 - x_+)} \Pi(\chi,b_+^2,k) - (+ \leftrightarrow -) \right]
 \right\} - \varepsilon\HH^0(\chi),
\label{Eq:HH1Def}\\
\HH^2(\chi) &:=& -\alpha C\left\{
 \frac{\varepsilon x_4^2 - \alpha\beta\lambda}{(x_4 - x_+)(x_4 - x_-)}\FF(\chi,k)
 - \frac{x_1 - x_4}{x_+ - x_-} \left[ \frac{\varepsilon x_+^2 - \alpha\beta\lambda}{(x_1 - x_+)(x_4 - x_+)} \Pi(\chi,b_+^2,k) - (+ \leftrightarrow -) \right] \right\},
\label{Eq:HH2Def}\\
\HH^3(\chi) &:=& -\lambda C\FF(\chi,k),
\label{Eq:HH3Def}
\end{eqnarray}
where we have set
\begin{equation}
b_\pm := \sqrt{\frac{x_4 - x_\pm}{x_1 - x_\pm}} b.
\end{equation}
Note that in the non-rotating limit $\alpha\to 0$ or the limit of equatorial orbits $\beta\to \pm 1$ it follows that $x_3\to 0$. It is also possible to compute the generating functions $S_r$ itself. Based on the observation that
\begin{equation}
S_r(\gamma_r;\mu m, \mu E, \mu L_z, \mu L) = \mu S_r(\gamma_r; m,E,L_z,L)
\end{equation}
for all $\mu > 0$, which is a manifestation of the weak equivalence principle, one obtains the relation
\begin{equation}
S_r(\gamma_r; m,E,L_z,L) = m\frac{\partial S_r}{\partial m} + E\frac{\partial S_r}{\partial E} 
 + L_z\frac{\partial S_r}{\partial L_z} + L\frac{\partial S_r}{\partial L},
\end{equation}
which allows one to compute $S_r$ from Eqs.~(\ref{Eq:PartialSr},\ref{Eq:HH0Def}--\ref{Eq:HH3Def}). After some simplifications based on the identity~(\ref{Eq:PolyRRoots}) one finds
\begin{eqnarray}
S_r(\gamma_r,m,E,L_z,L) &=& \frac{M_H m C}{x_{1234}}\left\{ 
(x_1 - x_4)(x_2 - x_4)\FF(\chi,k) - (x_1 - x_3)(x_2 - x_4)\EE(\chi,k) 
\right. 
\nonumber\\
 && \qquad\qquad
 + (x_1 - x_4)(x_{1234} - 4)\Pi(\chi,b^2,k)
 + \frac{1}{2}(x_1 - x_3)(x_2 - x_1)\frac{\sin(2\chi)}{1 - b^2\sin^2\chi}
 \nonumber\\
 && \left. \qquad\qquad
 - 2\frac{x_1 - x_4}{x_+ - x_-}\left[ (x_+ - x_3)(x_2 - x_+)\Pi(\chi,b_+^2,k) 
 - (+ \leftrightarrow -) \right]
 \right\}.
\end{eqnarray}

\section{List of relevant elliptic integrals}
\label{App:EllipticIntegrals}

In this appendix we briefly summarize some of the elliptic integral expressions that are relevant for the results in this article. The starting point is the fourth-order polynomial
$$
{\cal R}(x) = (x_2 - x)(x - x_1)(x - x_3)(x - x_4),
$$
where we assume that the four roots are real and ordered such that $x_3 < x_4 < x_1 < x_2$. Introducing the constants
$$
k := \sqrt{\frac{x_4 - x_3}{x_1 - x_3}} b,\qquad b := \sqrt{\frac{x_2 - x_1}{x_2 - x_4}},
$$
satisfying $0 < k < b < 1$, the variable substitution
$$
\chi := \arcsin\left( \sqrt{\frac{x - x_1}{x_2 - x_1}\frac{x_2 - x_4}{x - x_4}} \right),\qquad x_1 < x < x_2,
$$
leads to
$$
\frac{dx}{\sqrt{{\cal R}(x)}} = \tilde{C}\frac{d\chi}{\sqrt{1 - k^2\sin^2\chi}}.
\qquad \tilde{C} := \frac{2}{\sqrt{(x_1 - x_3)(x_2 - x_4)}}.
$$
From this, one obtains the following integral expressions (see~\cite{aE53,mAiS84,rFwH09} for more details) which are valid for $x_1 < x < x_2$:
\begin{eqnarray}
\int\limits_{x_1}^x \frac{dy}{\sqrt{{\cal R}(y)}} &=& \tilde{C} \FF(\chi,k),
\label{Eq:R1}\\
\int\limits_{x_1}^x \frac{y dy}{\sqrt{{\cal R}(y)}} &=& 
 \tilde{C}\left[ x_4\FF(\chi,k) + (x_1 - x_4)\Pi(\chi,b^2,k) \right],
\label{Eq:R2}\\
\int\limits_{x_1}^x \frac{y^2 dy}{\sqrt{{\cal R}(y)}} &=& \frac{\tilde{C}}{2}\left[
 (x_4 x_{124} - x_1 x_2)\FF(\chi,k) + (x_1 - x_3)(x_2 - x_4)\EE(\chi,k) \right. \nonumber\\
 && \left. + (x_1 - x_4)x_{1234}\Pi(\chi,b^2,k) - (x_2 - x_1)(x_1 - x_3)\frac{\cos\chi\sin\chi}{1 - b^2\sin^2\chi}\sqrt{1 - k^2\sin^2\chi} \right],
\label{Eq:R3}\\
\int\limits_{x_1}^x \frac{1}{y - x_0}\frac{dy}{\sqrt{{\cal R}(y)}} &=& \frac{\tilde{C}}{x_4 - x_0} \left[
 \FF(\chi,k) - \frac{x_1 - x_4}{x_1 - x_0}\Pi(\chi,b_0^2,k) \right],
\label{Eq:R4}
\end{eqnarray}
where here we have abbreviated $x_{124} := x_1 + x_2 + x_4$ and $x_{1234} := x_1 + x_2 + x_3 + x_4$, $x_0$ is an arbitrary real constant smaller than $x_4$,
$$
b_0 := \sqrt{\frac{x_4 - x_0}{x_1 - x_0}} b,
$$
and where Legendre's integrals $\FF$, $\EE$ $\Pi$ and $\DD$ are defined by~\cite{DLMF}
\begin{eqnarray}
\FF(\phi,k) &:=& \int \limits_{0}^{\phi} \frac{d\theta}{\sqrt{1 - k^2\sin^2\theta}} ,\qquad \KK(k):=  \FF(\pi/2,k),
\label{Eq:EllipticFF}\\
\EE(\phi,k) &:=& \int\limits_0^\phi \sqrt{1 - k^2\sin^2\theta} d\theta,\qquad
\EE(k) := \EE(\pi/2,k),
\label{Eq:EllipticEE}\\
\Pi(\phi,b^2,k) &:=& \int\limits_0^\phi 
\frac{d\theta}{\sqrt{1 - k^2\sin^2\theta}(1 - b^2\sin^2\theta)},\qquad
\Pi(k) := \Pi(\pi/2,k),
\label{Eq:EllipticPi}\\
\DD(\phi,k) &:=& \int\limits_0^\phi \frac{\sin^2\theta d\theta}{\sqrt{1 - k^2\sin^2\theta}},\qquad
\DD(k) := \DD(\pi/2,k).
\label{Eq:EllipticDD}
\end{eqnarray}

From Eq.~(\ref{Eq:R4}) one also obtains the integrals
\begin{eqnarray}
\int\limits_{x_1}^x \frac{1}{(y - x_+)(y - x_-)}\frac{dy}{\sqrt{{\cal R}(y)}} 
 &=& \tilde{C}\left\{ \frac{\FF(\chi,k)}{(x_4 - x_+)(x_4 - x_-)} - \frac{x_1 - x_4}{x_+ - x_-}
\left[ \frac{\Pi(\chi,b_+^2,k)}{(x_1 - x_+)(x_4 - x_+)} - (+ \leftrightarrow -) \right] \right\},
\label{Eq:R5}\\
\int\limits_{x_1}^x \frac{y}{(y - x_+)(y - x_-)}\frac{dy}{\sqrt{{\cal R}(y)}} 
 &=& \tilde{C}\left\{ \frac{x_4\FF(\chi,k)}{(x_4 - x_+)(x_4 - x_-)} - \frac{x_1 - x_4}{x_+ - x_-}
\left[ \frac{x_+\Pi(\chi,b_+^2,k)}{(x_1 - x_+)(x_4 - x_+)} - (+ \leftrightarrow -) \right] \right\},
\label{Eq:R6}
\end{eqnarray}
for arbitrary $0 < x_- < x_+ < x_4$.

\bibliographystyle{unsrt}
\bibliography{../References/refs_kinetic}

\begin{thebibliography}{10}

\bibitem{hA11}
H.~Andr{\'{e}}asson.
\newblock The {E}instein-{V}lasov system/kinetic theory.
\newblock {\em Living Reviews in Relativity}, 14(4), 2011.

\bibitem{bC68}
B.~Carter.
\newblock Global structure of the {K}err family of gravitational fields.
\newblock {\em Phys. Rev.}, 174:1559--1571, 1968.

\bibitem{mWrP70}
M.~Walker and R.~Penrose.
\newblock On quadratic first integrals of the geodesic equations for type [22]
  spacetimes.
\newblock {\em Commun.Math.Phys.}, 18:265--274, 1970.

\bibitem{oStZ14b}
O.~Sarbach and T.~Zannias.
\newblock The geometry of the tangent bundle and the relativistic kinetic
  theory of gases.
\newblock {\em Class. Quantum Grav.}, 31:085013, 2014.

\bibitem{hB52}
H.~Bondi.
\newblock On spherically symmetrical accretion.
\newblock {\em Monthly Notices Roy Astronom. Soc.}, 112:195--204, 1952.

\bibitem{fM72}
F.C. Michel.
\newblock Accretion of matter by condensed objects.
\newblock {\em Astrophysics and Space Science}, 15:153--160, 1972.

\bibitem{fHrL39}
F.~Hoyle and R.A. Lyttleton.
\newblock {The effect of interstellar matter on climatic variation}.
\newblock {\em Proceedings of the Cambridge Philosophical Society}, 35:405,
  1939.

\bibitem{hBfH44}
H.~Bondi and F.~Hoyle.
\newblock {On the Mechanism of Accretion by Stars}.
\newblock {\em Monthly Notices of the Royal Astronomical Society},
  104(5):273--282, 1944.

\bibitem{pRoS16}
P.~Rioseco and O.~Sarbach.
\newblock Accretion of a relativistic, collisionless kinetic gas into a
  {S}chwarzschild black hole.
\newblock {\em Class. Quantum Grav.}, 34(9):095007, 2017.

\bibitem{pRoS17}
P.~Rioseco and O.~Sarbach.
\newblock Spherical steady-state accretion of a relativistic collisionless gas
  into a {S}chwarzschild black hole.
\newblock {\em J. Phys. Conf. Ser.}, 831(1):012009, 2017.

\bibitem{pMoA21a}
P.~Mach and A.~Odrzywo\l{}ek.
\newblock Accretion of the relativistic {V}lasov gas onto a moving
  {S}chwarzschild black hole: Exact solutions.
\newblock {\em Phys. Rev. D}, 103:024044, 2021.

\bibitem{pMoA21b}
P.~Mach and A.~Odrzywo\l{}ek.
\newblock Accretion of dark matter onto a moving {S}chwarzschild black hole: An
  exact solution.
\newblock {\em Phys. Rev. Lett.}, 126:101104, 2021.

\bibitem{pMaO22}
P.~Mach and A.~Odrzywo\l{}ek.
\newblock Accretion of the relativistic {V}lasov gas onto a moving
  {S}chwarzschild black hole: low-temperature limit and numerical aspects.
\newblock {\em Acta Phys. Pol. B Proc. Suppl.}, 15(1-A7), 2022.
\newblock Presented at the 7th conference of the {P}olish {S}ociety on
  {R}elativity, \L{}\'od\'z, {P}oland, 20-23 september 2021.

\bibitem{aGetal21}
A.~Gamboa, C.~Gabarrete, P.~Dom\'{\i}nguez-Fern\'andez, D.~N\'u{\~n}ez, and
  O.~Sarbach.
\newblock Accretion of a {V}lasov gas onto a black hole from a sphere of finite
  radius and the role of angular momentum.
\newblock {\em Phys. Rev. D}, 104:083001, 2021.

\bibitem{aCpMaO22}
A.~Cie\'slik, P.~Mach, and A.~Odrzywolek.
\newblock {Accretion of the relativistic Vlasov gas in the equatorial plane of
  the Kerr black hole}.
\newblock {\em Phys. Rev. D}, 106:104056, 2022.

\bibitem{jLoP73}
J.L. Lebowitz and O.~Penrose.
\newblock Modern ergodic theory.
\newblock {\em Phys. Today}, 26(2):23--29, 1973.

\bibitem{CornfeldFominSinai-Book}
I.P. Cornfeld, S.V. Fomin, and Ya.G. Sinai.
\newblock {\em Ergodic Theory}.
\newblock Springer-Verlag, New York, 1982.

\bibitem{dL62}
D.~Lynden-Bell.
\newblock The stability and vibrations of a gas of stars.
\newblock {\em Monthly Notices Roy Astronom. Soc.}, 124:279--296, 1962.

\bibitem{dL67}
D.~Lynden-Bell.
\newblock Statistical mechanics of violent relaxation in stellar systems.
\newblock {\em Monthly Notices Roy Astronom. Soc.}, 136:101--121, 1967.

\bibitem{sTmHdL86}
S.~Tremaine, M.~H\'enon, and D.~Lynden-Bell.
\newblock {H}-functions and mixing in violent relaxation.
\newblock {\em Monthly Notices Roy Astronom. Soc.}, 219:285--297, 1986.

\bibitem{dM99}
D.~Merritt.
\newblock Elliptical galaxy dynamics.
\newblock {\em Publications of the Astronomical Society of the Pacific},
  111(756):129--168, February 1999.

\bibitem{sT99}
S.~Tremaine.
\newblock The geometry of phase mixing.
\newblock {\em Monthly Notices Roy Astronom. Soc.}, 307:877--883, 1999.

\bibitem{cMcV11}
C.~Mouhot and C.~Villani.
\newblock On {L}andau damping.
\newblock {\em Acta Math.}, 207:29--201, 2011.

\bibitem{bY16}
B.~Young.
\newblock Landau damping in relativistic plasmas.
\newblock {\em J. Math. Phys.}, 57:021502, 2016.

\bibitem{rMeT17}
R.~Mathew and E.~Tiesinga.
\newblock Phase-space mixing in dynamically unstable, integrable few-mode
  quantum systems.
\newblock {\em Phys. Rev. A}, 96:013604, 2017.

\bibitem{tDaKeKyS02}
T.V. Dudnikova, A.~I. Komech, E.~A. Kopylova, and Y.~M. Suhov.
\newblock On convergence to equilibrium distribution, {I}. {T}he
  {K}lein-{G}ordon equation with mixing.
\newblock {\em Comm. Math. Phys.}, 225:1--32, 2002.

\bibitem{tDaKnRyS02}
T.V. Dudnikova, A.~I. Komech, N.E. Ratanov, and Y.~M. Suhov.
\newblock On convergence to equilibrium distribution, {II}. {T}he wave equation
  in odd dimensions, with mixing.
\newblock {\em J. Stat. Phys.}, 108:1219--1253, 2002.

\bibitem{cM19}
C.~Mitchell.
\newblock Weak convergence to equilibrium of statistical ensembles in
  integrable {H}amiltonian systems.
\newblock {\em J. Math. Phys.}, 60:052702 (1)--(15), 2019.

\bibitem{pRoS20}
P.~Rioseco and O.~Sarbach.
\newblock Phase space mixing in external gravitational central potentials.
\newblock {\em Class. Quantum Grav.}, 37(19):195027, 2020.

\bibitem{pRoS18}
P.~Rioseco and O.~Sarbach.
\newblock Phase space mixing in the equatorial plane of a {K}err black hole.
\newblock {\em Phys. Rev. D}, 98(12):124024, 2018.

\bibitem{lApBjS18}
L.~Andersson, P.~Blue, and J.~Joudioux.
\newblock Hidden symmetries and decay for the {V}lasov equation on the {K}err
  spacetime.
\newblock {\em Comm. Partial Differential Equations}, 43:47--65, 2018.

\bibitem{lB20}
L.~Bigorgne.
\newblock Decay estimates for the massless {V}lasov equation on {S}chwarzschild
  spacetimes.
\newblock 2020.
\newblock arXiv:2006.03579.

\bibitem{hA21}
H.~Andr\'easson.
\newblock Existence of steady states of the massless {E}instein-{V}lasov system
  surrounding a {S}chwarzschild black hole.
\newblock {\em Annales Henri Poincare}, 22(12):4271--4297, 2021.

\bibitem{fJ22}
F.E. Jabiri.
\newblock Stationary axisymmetric {E}instein-{V}lasov bifurcations of the
  {K}err spacetime.
\newblock 2 2022.
\newblock arXiv:2202.10245 [math.AP].

\bibitem{wS02}
W.~Schmidt.
\newblock Celestial mechanics in {K}err space-time.
\newblock {\em Class. Quantum Grav.}, 19:2743--2764, 2002.

\bibitem{tHeF08}
T.~Hinderer and E.E. Flanagan.
\newblock {Two timescale analysis of extreme mass ratio inspirals in Kerr. I.
  Orbital Motion}.
\newblock {\em Phys. Rev. D}, 78:064028, 2008.

\bibitem{rFwH09}
R.~Fujita and W.~Hikida.
\newblock Analytical solutions of bound timelike geodesic orbits in {K}err
  spacetime.
\newblock {\em Class. Quant. Grav.}, 26:135002, 2009.

\bibitem{eFgGgS03}
E.~Fiorani, G.~Giachetta, and G.~Sardanashvily.
\newblock The {L}iouville-{A}rnold-{N}ekhoroshev theorem for non-compact
  invariant manifolds.
\newblock {\em J. Phys. A}, 36:L101--L107, 2003.

\bibitem{jBmGtH13}
J.~Brink, M.~Geyer, and T.~Hinderer.
\newblock Orbital resonances around black holes.
\newblock {\em Phys. Rev. Lett.}, 114:081102, 2015.

\bibitem{jBmGtH15}
J.~Brink, M.~Geyer, and T.~Hinderer.
\newblock Astrophysics of resonant orbits in the {K}err metric.
\newblock {\em Phys. Rev. D}, 91(8):083001, 2015.

\bibitem{rAcGoS22}
R.~Acu{\~n}a{-}C{\'a}rdenas, C.~Gabarrete, and O.~Sarbach.
\newblock An introduction to the relativistic kinetic theory on curved
  spacetimes.
\newblock {\em General Relativity and Gravitation}, 54(23), February 2022.

\bibitem{MTW-Book}
C.W. Misner, K.S. Thorne, and J.A. Wheeler.
\newblock {\em Gravitation}.
\newblock W. H. Freeman, 1973.

\bibitem{Arnold-Book}
V.I. Arnold.
\newblock {\em Mathematical Methods of Classical Mechanics}.
\newblock Springer-Verlag, New York, 1989.

\bibitem{Zehnder-Book}
E.~Zehnder.
\newblock {\em Lectures on Dynamical Systems: Hamiltonian Vector Fields and
  Symplectic Capacities}.
\newblock European Mathematical Society, Zurich, 2010.

\bibitem{sM20}
B.S. Mityagrin.
\newblock The zero set of a real analytic function.
\newblock {\em Mathematical Notes}, 107:529--530, 2020.

\bibitem{dL62a}
D.~Lynden-Bell.
\newblock Stellar dynamics. {O}nly isolating integrals should be used in
  {J}eans theorem.
\newblock {\em Monthly Notices Roy Astronom. Soc.}, 124:1--9, 1962.

\bibitem{MoBoschWhite-Book}
H.~Mo, F.~van~den Bosch, and S.~White.
\newblock {\em Galaxy Formation and Evolution}.
\newblock Cambridge University Press, Cambridge, U.K., 2010.

\bibitem{PoissonWill-Book}
E.~Poisson and C.M. Will.
\newblock {\em Gravity}.
\newblock Cambridge University Press, United Kingdom, 2014.

\bibitem{sCjL21}
S.~Chaturvedi and J.~Luk.
\newblock Phase mixing for solutions to 1d transport equation in a confining
  potential.
\newblock 2021.

\bibitem{mMpRhB22}
M.~Moreno, P.~Rioseco, and H.~Van~Den Bosch.
\newblock Mixing in an anharmonic potential well.
\newblock {\em J. Math. Phys.}, 63:071502, 2022.

\bibitem{cGoS22c}
C.~Gabarrete and O.~Sarbach.
\newblock Axisymmetric, stationary collisionless gas configurations surrounding
  {S}chwarzschild black holes.
\newblock {\em Class. Quant. Grav.}, 40(5):055012, 2023.

\bibitem{cGoS22a}
C.~Gabarrete and O.~Sarbach.
\newblock Kinetic gas disks surrounding {S}chwarzschild black holes.
\newblock {\em Acta Phys. Pol. B Proc. Suppl.}, 15(1-A10), January 2022.
\newblock Presented at the 7th conference of the {P}olish {S}ociety on
  {R}elativity, \L{}\'od\'z, {P}oland, 20-23 september 2021.

\bibitem{Chandrasekhar-Book}
S.~Chandrasekhar.
\newblock {\em The Mathematical Theory of Black Holes}.
\newblock Oxford University Press, Great Clarendon Street, Oxford 0X2 6DP,
  1992.

\bibitem{ONeill-Book}
B.~O'Neill.
\newblock {\em The Geometry of the {K}err black holes}.
\newblock Dover Publications, Inc., Mineola, New York, 2015.

\bibitem{eTpTjM13}
E.~Tejeda, P.A. Taylor, and J.C. Miller.
\newblock An analytic toy model for relativistic accretion in {K}err spacetime.
\newblock {\em Mon. Not. Roy. Astron. Soc.}, 429:925, 2013.

\bibitem{jBwPsT72}
J.~Bardeen, W.~Press, and S.~Teukolsky.
\newblock Rotating black holes: Locally nonrotating frames, energy extraction,
  and scalar synchrotron radiation.
\newblock {\em Astrophysical Journal}, 178:347--370, 1972.

\bibitem{aE53}
Editor A.~Erd{\'e}lyi.
\newblock {\em Higher Transcendental Functions. Vol. II}.
\newblock McGraw-Hill, The University of Michigan, 1953.

\bibitem{mAiS84}
M.~Abramowitz and I.A. Stegun.
\newblock {\em Pocketbook of Mathematical Functions}.
\newblock Harri Deutsch, Thun, 1984.

\bibitem{DLMF}
Digital library of mathematical functions.
\newblock \url{http://dlmf.nist.gov/}.

\end{thebibliography}

\end{document}